\documentclass{aa}  

\usepackage{amsmath}    
\usepackage{amssymb}    
\usepackage{array}
\usepackage{bm}
\usepackage{booktabs}
\usepackage[font=small,labelfont=bf]{caption}
\usepackage{comment}
\usepackage{float}
\usepackage{graphicx}
\usepackage[colorlinks, citecolor={blue}]{hyperref} 
\usepackage{layouts}
\usepackage{pdflscape}
\usepackage{placeins}
\usepackage{ragged2e}
\usepackage{subcaption}
\usepackage{txfonts}
\def\as{$^{\prime\prime}$}

\def\kst{KS test}
\def\logh{$\log(\mathrm{hardness})$}
\def\logL{$\log(L_{\mathrm{tot}})$}

\def\LX{$L_{\mathrm{2-10~keV}}$}

\def\chandra{\textit{Chandra}}
\def\swift{\textit{Swift}}
\def\xmm{\textit{XMM-Newton}}

\def\swiftcat{\swift\ X-ray \& UV catalog}
\def\xmmcat{\xmm\ simultaneous catalog}

\newcommand{\bi}[1]{\textit{\textbf{#1}}}

\begin{document} 

\title{Do radio active galactic nuclei reflect X-ray binary spectral states?}

\author{Emily Moravec\inst{1,2}
    \and
    Ji\v r\'i Svoboda\inst{1}
    \and 
    Abhijeet Borkar\inst{1}
    \and 
    Peter Boorman\inst{1}
    \and 
    Daniel Kynoch\inst{1}
    \and
    Francesca Panessa\inst{3}
    \and
    Beatriz Mingo\inst{4}
    \and
    Matteo Guainazzi\inst{5}
    }
      
\institute{Astronomical Institute of the Czech Academy of Sciences, Bo\v cn\'i II 1401/1A, 14100 Praha 4, Czech Republic, \email{emoravec@nrao.edu}
    \and
    Green Bank Observatory, P.O. Box 2, Green Bank, WV 24944, USA
    \and
    INAF - Istituto di Astrofisica e Planetologia Spaziali, via Fosso del Cavaliere 100, I-00133 Roma, Italy
    \and
    School of Physical Sciences, The Open University, Walton Hall, Milton Keynes, MK7 6AA, UK
    \and
    European Space Agency, ESTEC, Keplerlaan 1 2201AZ Noordwijk, The Netherlands 
    }

\date{Received 09 December 2021; accepted 16 February 2022; Published XX}
 
\abstract
{Over recent years there has been mounting evidence that accreting supermassive black holes in active galactic nuclei (AGNs) and stellar mass black holes have similar observational signatures: thermal emission from the accretion disk, X-ray coronas, and relativistic jets. Further, there have been investigations into whether or not AGNs have spectral states similar to those of X-ray binaries (XRBs) and what parallels can be drawn between the two using a hardness-intensity diagram (HID).}
{To address whether AGN jets might be related to accretion states as in XRBs, we explore whether populations of radio AGNs classified according to their (a) radio jet morphology, Fanaroff-Riley classes I and II  (FR I and II), (b) excitation class, high- and low-excitation radio galaxies (HERG and LERG), and (c) radio jet linear extent, compact to giant, occupy different and distinct regions of the AGN HID (total luminosity vs. hardness).}
{We do this by cross-correlating 15 catalogs of radio galaxies with the desired characteristics from the literature with \xmm\ and \swift\ X-ray and ultraviolet (UV) source catalogs. We calculate the luminosity and hardness from the X-ray and UV photometry, place the sources on the AGN HID, and search for separation of populations and analogies with the XRB spectral state HID.}
{We find that (a) FR Is and IIs, (b) HERGs and LERGs, and (c) FR I-LERGs and FR II-HERGs occupy distinct areas of the HID at a statistically significant level (p-value < 0.05), and we find no clear evidence for population distinction between the different radio jet linear extents. The separation between FR I-LERG and FR II-HERG populations is the strongest in this work. }
{Our results indicate that radio-loud AGNs occupy distinct areas of the HID depending on the morphology and excitation class, showing strong similarities to XRBs.}


\maketitle
%

\section{Introduction}\label{sect:intro}
Accreting black holes are some of the most luminous astronomical objects in the sky and are interesting laboratories with which to study physical processes happening under extreme conditions of gravity, ultra-dense matter, and particle acceleration. Observations have revealed a variety of black hole flavors as reflected by their masses. On the low end of the mass spectrum, there are stellar-mass black holes (SBHs), which are typically found in X-ray binaries (XRBs) and have masses that range from a few $M_{\odot}$ to a few tens of $M_{\odot}$ \citep{Fender04,McClintock06,Dunn10,Zhang13}. On the high end, there are supermassive black holes (SMBHs; $10^5\lesssim M<10^{10}M_{\odot}$) found in active galactic nuclei (AGNs). A question that is actively being investigated is whether accreting SBH and SMBH systems are analogous to each other, differing only in mass scale. 

\subsection{X-ray binary evolution}
\begin{figure}
    \centering
    \includegraphics[width=\linewidth]{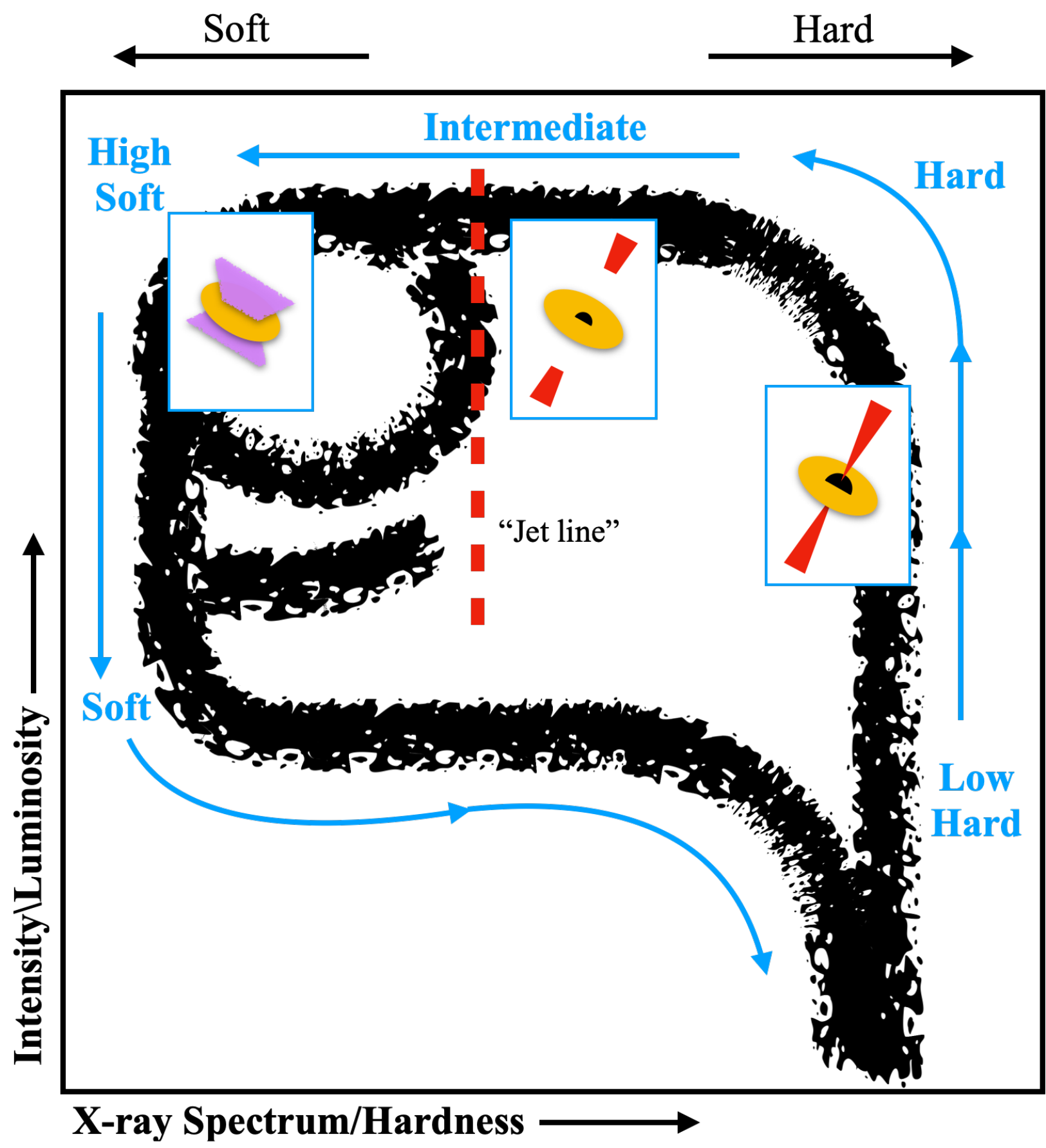}
    \caption{Cartoon representing the ``q'' diagram for XRBs, adapted from \cite{Fender04} and \cite{Fender12}, which depicts the progression of an XRB outburst through the different spectral states. An XRB starts an outburst in the low-hard state, progresses to the hard state through intermediate states to the high-soft state, then down to the soft state, and then back to the low-hard state to begin again (described in more detail in Sect. \ref{sect:intro}). }
    \label{fig:XRB_state_diagram}
\end{figure}

One of the most well-known and established properties of accreting SBHs primarily found in XRBs is their cyclical and evolutionary progression through certain accretion states. The progression of XRBs through these accretion states can be tracked in a hardness versus intensity (or luminosity) diagram in which the XRBs typically trace out a ``q'' shape \citep{Fender12}. The hardness for XRBs is defined as the X-ray color (the ratio of flux between different X-ray bands), and the intensity is typically defined as the X-ray luminosity or the ratio of the X-ray luminosity to the Eddington luminosity. As the name hardness-intensity diagram (HID) suggests, the various states are defined by the hardness of the X-ray spectrum (hard and soft states) and the luminosity (low and high) of the source. Based on \cite{Fender12}, we separate the movement of a source through these spectral states into four phases (see Fig. \ref{fig:XRB_state_diagram}). 
\begin{enumerate}
    \item At the beginning of an outburst, the source increases in luminosity by several orders of magnitude. Its X-ray spectrum remains hard and is dominated by the emission due to the thermal Comptonization of lower-energy seed photons on hot electrons \citep[see, e.g.,][]{Zdziarski1985}. The source in the hard state is often associated with relatively steady radio emission at gigahertz frequencies originating from a jet \citep[see, e.g.,][]{Corbel00, Gallo03}.
    \item The source then moves from the hard state through intermediate stages to the high-soft state. During this transition, the X-ray spectrum changes from hard to soft as the blackbody-like component attributed to the accretion disk brightens and eventually dominates in the soft state, resulting in a softening of the X-ray spectrum. As the source transitions from the hard state to the high-soft state (still at high luminosity), the jets of the source change as well. The source will progress from having steady radio jets (hard state) to producing discrete injections and flares (intermediate states). Eventually, the jets are quenched and disappear in the high-soft state. This evolution of the jet is depicted in the HID by the source crossing a ``jet line'' in the intermediate states. When sources cross this jet line either during (a) the initial crossing of the source as it moves from the hard to soft state or (b) a recrossing of the line in small cycles, it can produce series of temporary blob injections known as ballistic jets \citep{Mirabel94,Narayan12}.
    \item Still in the soft state, the source will then decrease in luminosity, the radio emission will fade away, the accretion disk will dominate the X-ray spectrum, and the accretion rate typically drops. This phase is typically the longest.
    \item The source transitions from the soft state back to the hard state (at lower luminosities) and fades into quiescence until the next outburst.
\end{enumerate}

The XRB state transitions can in reality  be much more complex than described here, often showing failed outbursts or rare transitions. Some sources went through an outburst cycle multiple times, such as GX 339-4 \citep{Corbel13, Zdziarski04, Belloni05, Homan05, Barnier22}, whereas others, such as Cyg X-1, never dropped to quiescence and are continuously fed by accretion \citep{Esin1998, Grinberg14, Cechura15}. Overall, the cyclical outbursts of XRBs typically last from months to years \citep{Fender04}. We refer the reader to \cite{Fender04}, \cite{Remillard06}, \cite{Done07}, \cite{Dunn10}, \cite{Fender12}, \cite{Zhang13}, and \cite{Fender16} for more in depth analyses and variations of the progression of an XRB through the HID.

\subsection{Comparing AGN to XRBs}
We wish to determine how similar the accretion processes of SMBHs are to those of SBHs and whether AGNs exhibit accretion states as XRBs do. It is already known that black holes of both types of systems follow the same fundamental plane of black hole activity of correlated radio and X-ray emission when black hole mass is taken into account \citep{Merloni03,Falcke04,Kording06a}. There are similarities in other shared characteristics of these systems, such as the presence of the accretion disk and energetic corona in both systems \citep{Arcodia20}, evidence for a jet line \citep{Zhu20}, and their timing properties \citep{McHardy06}. And though the power density spectra show a characteristic break frequency that is scaled by mass \citep{McHardy06}, it is more difficult to compare the duration of an XRB outburst to the duration of AGN activity. 

The length of an outburst in an XRB is weeks to months \citep{Dunn10}. Furthermore, each particular phase (or spectral state) of this outburst lasts for a particular amount of time, ranging from minutes to days \citep[see Figure 8 in][]{Dunn10}. If XRBs and AGNs are analogs, then the AGN duty cycle (the fraction of its lifetime that the galaxy spends in the AGN phase) would take millions of years to be the mass-scaled analogy to an XRB outburst. Radio active galaxies are useful to trace the ``ejection active phase''; however, it is not known for how long the AGNs maintain their active phase or which part of them is in a radio active phase. Further, the precise value of the AGN duty cycle is unknown, but estimates for the lifetimes of an AGN ``outburst'' range from $10^{5}$-$10^{8}$ yr based on the analysis of the spectral age of radio galaxies \citep{Konar13,Turner18,Brienza20}, radio galaxy properties and models \citep{Shabala08,Shabala20,Maccagni20}, and X-ray observations of AGNs \citep{Vantyghem14,Schawinski15}. Determining the duty cycle and lifetime of AGNs is complicated as it seems to depend on host galaxy mass \citep{Best05,Shabala08,Sabater19} and AGN power \citep{Parma99}. However, when the XRB outburst duration is scaled by mass and compared to current estimates of the AGN lifetime, they are roughly consistent.

Clearly, AGNs operate on much longer timescales than XRBs, making it difficult to directly compare their transitions dynamically and on similar timescales as a whole. However, one possible avenue for comparing XRB and AGN accretion state changes is with a class of AGNs named changing-look AGNs (CLAGNs). These sources have been serendipitously detected to have changed from one Seyfert type to the other, and sometimes back to the original type. The observational manifestation of this change is that their broad lines have been found to be appearing and/or disappearing \citep[see, e.g.,][]{Tohline76,Anderson71,Cromwell70,Denney14,LaMassa15,Shappee14} or the X-ray spectra of the sources are seen to be changing from reflection-dominated, Compton-thick absorption to Compton-thin spectra, or vice versa \citep[][and others]{Guainazzi02,Matt03,Risaliti09}. There have been concerted efforts to look for such sources in the archival observations of large-scale catalogs such as the Sloan Digital Sky Survey (SDSS), and several dozen have been identified \citep[e.g.,][]{MacLeod19}. The timescales involved with the change in AGN type are typically on the order of months to years, which is much shorter than the expected AGN dynamical timescale of >$10^5$ years. Observations have also shown that the change in CLAGNs occurs around Eddington ratios of $\sim$\,$10^{-2}$ and is accompanied by changes in luminosity, as also seen in XRBs \citep{MacLeod19,Ruan19}. Thus, these CLAGNs are an interesting set of objects that can be used to compare the accretion state transitions between AGNs and XRBs, but such analysis is beyond the scope of this work. Beyond CLAGNs, the typical way to compare AGNs to XRBs is to use populations of AGNs.

Aside from comparing the fundamental properties of each system type (mass, luminosity, disk, corona, timescale, and duty cycle), \cite{Kording06b} and \cite{Sobolewska11} suggest that different classes and properties of AGNs might correspond to specific spectral states of XRBs. In order to investigate whether AGNs have similar spectral states to XRBs, \cite{Kording06b} assembled an AGN HID for the first time, called the ``disk-fraction/luminosity diagram'' in their work. In \cite{Kording06b} and following similar works, the hardness and intensity are defined differently for AGNs than for XRBs due to the complexities in determining the level of thermal emission from the accretion disk in AGNs.  For XRBs, the total luminosity, the power-law component, and the disk component can all be measured from an X-ray spectrum. But for AGNs, ultraviolet (UV) data are also needed in addition to an X-ray spectrum because the thermal emission from the accretion disk peaks in the UV. \cite{Kording06b} define the luminosity as the total luminosity of the system (the sum of the disk and power-law component of a source) and the hardness as the relative strength of the X-ray corona power-law component compared to the disk component of a source (using optical observations to determine the disk component). 

Using the disk-fraction/luminosity diagram, \cite{Kording06b} postulated that different spectral states may explain the radio-loud (RL) and radio-quiet (RQ) AGN dichotomy \citep{Kording06b}\footnote{We note that the exact definition of radio-loudness often varies according to the source, but these different definitions should not lead to substantial differences. In order of citation in this introduction, \cite{Kording06b} define radio-loudness as $L_{\rm 1.4\,GHz}/L_{\mathrm{optical~B-band}}$, \cite{Nipoti05} and \cite{Zhu20} as $L_{\mathrm{5\,GHz}}$/$L_{\mathrm{4400\,\AA}}$, \cite{Svoboda17} as $L_{\mathrm{4.8\,GHz}}$/$L_{\mathrm{UV+X}}$, and \cite{Fernandez21} as $F_{\mathrm{8.4\,GHz}}$/$F_{\mathrm{2-10\,keV}}$.}. \cite{Kording06b} also found that low-luminosity Faranoff-Riley type I radio galaxies are associated with the hard state, RL quasars are associated with the hard intermediate state, and RQ quasars are associated with either the soft intermediate state or simply the soft state. Similar previous work by \cite{Nipoti05} (without an AGN HID) found that RL and RQ AGNs can be associated with specific XRB states, where RL AGNs might be the analog of XRB high-intermediate states and RQ AGNs the analogs of the non-flaring high-soft state. They point out that a typical XRB flares a few per cent of the time, which is similar to the fraction of quasars that are RL. If SMBH systems are similar to SBH systems, perhaps SMBH systems cycle through RL and RQ phases as part of a particular quasar-triggering event \citep{Nipoti05}.

Further work investigating whether RL and RQ AGNs reflect XRB accretion states using an AGN HID was done by \citet{Svoboda17}. Similar to \cite{Kording06b}, they define the luminosity as the total luminosity of the system (the sum of the disk and power-law component of a source) and the hardness as the relative strength of the power-law component compared to the disk component of a source. However, \cite{Svoboda17} used (a) UV observations instead of optical to determine the accretion disk component in order to be closer to the UV peak of thermal emission and (b) X-ray observations from a wider and harder X-ray band ($2-12$\,keV compared to $0.1-2.4$\,keV) to avoid any effects of X-ray absorption and to more accurately determine the X-ray power-law emission. Additionally, the UV and X-ray observations in their sample are simultaneous to eliminate possible effects of AGN variability. Their final sample contains 1522 unique sources, but only 175 AGNs could be classified as either RL or RQ. It is clear from their HID that the RL sources have on average higher hardnesses, which confirms the idea that the AGN radio dichotomy could indeed be related to the evolution of AGN accretion states. On a related note, \cite{Zhu20} studied jetted RL quasars and non-jetted RQ quasars and found that jetted RL quasars are harder than RQ quasars, suggesting the presence of a jet line in the AGN HID\footnote{\cite{Zhu20} define hardness as a normalized $L_{\mathrm{2\,keV}}$/$L_{\mathrm{2500\,\AA}}$ and intensity as $L_{\mathrm{2500\,\AA}}$.} akin to the XRB HID.

The most recent work to use an AGN HID to investigate the existence of AGN spectral states is by \cite{Fernandez21}, who used a sample of 167 nearby Seyfert 1s, Seyfert 2s, and low-ionization nuclear emission-line regions (LINERs). They created an HID equivalent by defining the luminosity as the ratio of the rescaled total line luminosity (mid-infrared and optical lines) to the Eddington luminosity and defining the hardness as the Lyman hardness that uses the ratio of mid-infrared lines. Similar to \cite{Kording06b} and \cite{Svoboda17}, \cite{Fernandez21} find that RL sources are associated with the hard state and RQ sources are associated with the soft state. Further, \cite{Fernandez21} place the Seyferts and LINERs on the AGN HID and find, similar to \cite{Sobolewska11}, that different classes of AGNs reflect specific spectral states of XRBs, and they recover the characteristic q-shaped morphology of XRB HIDs. Specifically, they find that the (a) broad-line Seyferts and about half of the Seyfert 2 population, which both have highly excited gas and RQ cores consistent with disk-dominated nuclei, are associated with the soft state and (b) the remaining half of the Seyfert 2 nuclei and the bright LINERs are associated with the bright hard and intermediate states. 

Overall, using the AGN HID, \cite{Nipoti05}, \cite{Kording06b}, \cite{Svoboda17}, \cite{Zhu20}, and \cite{Fernandez21} provide evidence that RL and RQ AGNs reflect XRB spectral states, which is consistent with the picture that AGNs might be similar to XRBs in having accretion states. The results of \cite{Fernandez21}, that different classes of AGNs reflect spectral states of XRBs, support this idea.

\section{Radio-AGN properties and comparing them to XRBs}
One of the manifestations of the different accretion states of SBHs in XRBs is the presence or absence of radio jets. As mentioned previously, XRBs are often seen to evolve from the low-hard state to the high-hard state accompanied by the launching and presence of radio jets, which eventually get quenched as the state transitions to the high-soft state. Similarly, SMBHs in AGNs can also launch radio jets showing current or past radio activity. Most AGNs are RQ and are typically non-jetted whereas only 10-20\% are RL and jetted \citep{Kellermann89,Kellermann16}. However, it has been found that some RQ sources have low-power jets in their core \citep{Panessa13,Harrison15,Panessa19,Jarvis19}. Active galactic nuclei can also produce short-lived, powerful jets in the super-Eddington regime \citep{Begelman78,Abramowicz88,Sadowski14}. For the RL sources with easily observable jets in particular, it is useful to investigate whether different radio-AGN jet morphologies and properties correlate with specific spectral states of XRBs, particularly when the XRBs are in the outburst phase and launch jets. 

Radio AGNs with jets typically display two lobes. Double-lobed radio AGNs are historically divided into two morphological classes: Fanaroff-Riley classes I and II \citep[FR I and FR II;][]{FR74}. FR I sources are ``edge-darkened'' in that the emission is brighter near the radio core and becomes fainter radially outward. FR II sources are ``edge-brightened'' in that two well-separated lobes end in distinctive areas of brightest emission (i.e., ``hotspots''). Historically, there was thought to be a relatively clean divide in power between the two morphologies with FR IIs having higher radio powers, but \cite{Mingo19} showed that radio luminosity does not reliably predict whether a source is FR I or FR II based on high sensitivity survey data. Currently, the FR I--II morphological difference is primarily explained as a difference in jet dynamics in the two systems where the edge-brightened FR IIs are thought to have jets that remain relativistic throughout, terminating in a hotspot, while the edge-darkened, center-brightened FR Is are believed to disrupt on kiloparsec  scales \citep[e.g.,][]{Bicknell95,Tchekhovskoy16}. It has also long been suggested that the structural difference between FR Is and FR IIs is caused by the interplay of the jet and the environmental density on the host-scale, such that jets in a rich environment will be disrupted and become FR I more easily than jets in a poor environment \citep{Ledlow96,Bicknell95,Kaiser07}. There still remains considerable debate about the link between accretion mode and jet morphology for FR morphologies \citep[e.g.,][]{Hardcastle07,Best12,Mingo14,Ineson15,Hardcastle18} and this work aims to contribute to this debate.

Beyond the FR I--II classification, one can classify radio galaxies based on the extent of their radio jets, which ranges from compact to giant. Compact radio sources exhibit jets within their host galaxy and there are many compact radio galaxy classifications: FR0s, gigahertz peaked spectrum radio sources, high-frequency peaker (HFP) radio sources, compact steep spectrum (CSS) radio sources, and compact symmetric objects (CSOs). Deep sensitive surveys at radio frequencies have revealed a population of radio sources that are associated with AGNs and have similar core luminosity to those of FR I sources, but lack the substantial extended radio emission that FR I--II sources contain and are typically at the resolution limit of the surveys \citep[weaker by a factor of $\sim$100, see][]{Baldi09,Baldi15,Baldi18}. These sources have been named FR0s and have extents $\lesssim$ 5 kpc typically (see \citealt{Baldi18,Odea21} for a more in depth review). Gigahertz peaked spectrum (GPS) sources are selected to have their radio spectra dominated by a peak in the flux density around 1 GHz \citep{Odea21}, whereas HFP sources peak above 5 GHz \citep{Dallacasa00,Odea21}. Sources that peak at frequencies below 400 MHz are called CSS sources. Such sources are not selected specifically on the basis of the location of the spectral peak \citep{Fanti90,Odea21} as GPS and HFPs and are thought to be young FR II radio galaxies \citep{Odea98}. Both GPS and HFP sources tend to have projected linear sizes less than 500 pc, while CSS sources tend to have sizes between 500 pc and 20 kpc \citep{Odea98,Fanti90,Odea21}. CSOs have been defined to be those with symmetric double-lobe radio emission and an overall size less than about 1 kpc \citep{Odea21}. We note that it is possible for there to be overlap of FR0s with other compact classifications. For example, FR0s can be CSS or GPS sources \citep{Sadler14,Whittam16,Odea21}.

On the other end of the scale of AGN radio extent are giant radio galaxies (GRGs), which are typically defined as radio AGNs with linear extents > 0.7 Mpc \citep{Dabhade20}. The linear size of radio AGNs with a classical FR I--II morphology can extend from less than a few tens of parsecs to several megaparsecs. In the past 60 years, thousands of radio galaxies have been found, but only $\sim$800 radio galaxies have been discovered that exhibit megaparsec scale sizes \citep{Dabhade20}.

Another property of radio AGNs that has the potential to reflect XRB spectral states is excitation class. Radio AGNs can be classified according to the optical emission lines produced in the narrow-line region ([OIII]$\lambda$5007, [NII]$\lambda$6584, [SII]$\lambda$6716, and [OI]$\lambda$6364). High-excitation radio galaxies (HERGs) show strong high-excitation broad and narrow lines similar to those in Seyfert galaxies (diagnosed typically with the [OIII]$\lambda$5007 line), whereas low-excitation radio galaxies (LERGs) exhibit weak or no line emission (spectroscopic indicators of low excitation are [NII]$\lambda$6584, [SII]$\lambda$6716, and [OI]$\lambda$6364). The different excitation modes are associated with different accretion rates and radiative efficiencies \citep[see][and references within]{Best12,Heckman14}. On one hand, HERGs have higher accretion rates ($L$/$L_\mathrm{Edd}\sim0.1$, typically $\sim$0.1-0.2, see \citealt{Best12,Mingo14}) and accrete efficiently (advection processes and the potential energy of the gas accreted by the SMBH is efficiently converted into radiation). On the other hand, LERGs have lower accretion rates ($L$/$L_{\mathrm{Edd}}\leq0.1$, typically $\sim$0.01, \citealt{Mingo14}) and accrete inefficiently (the jet carries the bulk of the AGN energy output). Interestingly, \cite{Best12} and \cite{Mingo14} find evidence for an approximate division between the Eddington ratios of low- and high-excitation objects at $L/L_{\mathrm{Edd}}\sim0.01-0.1$ and \cite{Maccarone03} find that $L/L_{\mathrm{Edd}}\sim0.01$ represents the division between different accretion states in XRBs (from the low/hard to the high/soft states).

Building upon the work of and methods used by \cite{Svoboda17}, the goal of this work is to investigate whether different radio-AGN properties (morphology, extent, and excitation class) beyond radio-loudness correlate with specific XRB spectral/accretion states. This paper is organized as follows. In Sect. \ref{sect:cats}, we describe the radio, UV, and X-ray source catalogs that we use for our analysis. In Sect. \ref{sect:methods} we describe the methods for cross-matching the catalogs and calculating the luminosities of interest. In Sect. \ref{sect:results} we present the results of placing radio AGNs with different properties (morphology, extent, excitation class) on the HID. In Sect. \ref{sect:disc} we discuss our results. Lastly, in Sect. \ref{sect:concl} we summarize our results and conclusions. In this work, we use a concordance cosmology with $H_0$=67.7 km s$^{-1}$ Mpc$^{-1}$, $\Omega_{\Lambda}$=0.69, and $\Omega_{\mathrm{m}}$=0.31 \citep{Planck18}.

\section{Source catalogs}\label{sect:cats}
In order to investigate whether AGNs with different radio properties lie in distinct areas of the HID, it is necessary to obtain three quantities concerning the radio galaxy: the (a) radio property (jet morphology, excitation class, or linear extent), (b) X-ray luminosity, $L_{X}$, for which we need an X-ray flux measurement, and (c) UV luminosity, $L_{UV}$, for which we need a UV flux measurement. In Sect. \ref{sect:rad_cats}, we describe the catalogs that we use in this work that provide classifications (morphology, excitation class, or linear extent) of a sample of radio galaxies. To obtain $L_X$ and $L_{UV}$, we created two source catalogs that contain both X-ray and UV measurements for sources, which we describe in Sects. \ref{sect:xmm_simult} and \ref{sect:swift_UX}. In Sect. \ref{sect:xmm_simult}, we describe the creation of an \xmm\ source catalog of simultaneous X-ray and UV observations and in Sect. \ref{sect:swift_UX} we describe the creation of a source catalog of X-ray and UV observations made by the \textit{Neil Gehrels Swift Observatory} (\swift).

\subsection{Radio catalogs}\label{sect:rad_cats}
In this work, we use 15 individual catalogs that classify a sample of radio galaxies according to radio morphology (FR I or II), excitation class (HERG or LERG), or linear extent (compact to giant). Some of the radio catalogs contain several classifications (e.g., morphology and excitation class). We detail these catalogs in this section. The name that is used to refer to each catalog in this work is indicated by \textit{italics} in the following description.

\subsubsection{Radio morphology catalogs}
We created a catalog of low redshift ($z\lesssim$ 0.15) radio galaxies, referenced in this work as \textit{FRXCAT}, by compiling the following individual catalogs: FRIICAT \citep{FRIICAT}, FRICAT \citep{FRICAT}, small FR Is from \cite{FRICAT}, FR0CAT \citep{Baldi18}, and COMP2CAT \citep{JM19}. FRIICAT is a catalog of 122 FR II radio galaxies that were selected from a published sample obtained by combining observations from the National Radio Astronomy Observatory VLA Sky Survey (NVSS), the Faint Images of the Radio Sky at Twenty centimetres (FIRST) survey, and the SDSS. The catalog includes sources with an edge-brightened radio morphology, $z\lesssim$ 0.15, and at least one of the emission peaks located at radius larger than 30 kpc from the center of the host \citep{FRIICAT}. FRICAT is a catalog of 219 FR I radio galaxies that were identified in the same way as FRIICAT except that FRICAT is a catalog of sources with an edge-darkened radio morphology and extending to a radius larger than 30 kpc from the center of the host. In addition, \cite{FRICAT} selected a sample (sFRICAT) of 14 smaller (10 < r < 30 kpc) FR Is, limiting to $z<$ 0.05. FR0CAT is a sample of 108 compact radio sources with $z\lesssim$ 0.05, a radio size $\lesssim$5 kpc, and an optical spectrum characteristic of low-excitation galaxies \citep{Baldi18}. Lastly, COMP2CAT is a catalog of 32 compact double-lobed radio galaxies that are edge-brightened radio sources whose projected linear size does not exceed 60 kpc with $z\lesssim$ 0.15 \citep{JM19}. We cross-matched all individual catalogs within FRXCAT with \cite{Best12} in \verb|topcat|~\citep{topcat} using the SDSS name of the source to identify whether these sources were HERGs or LERGs. We did not cross-match COMP2CAT with \cite{Best12} because \cite{JM19} already contained excitation information and all sources within COMP2CAT were LERGs except one.

\textit{Gendre+10} \citep{Gendre10} is a catalog of NVSS-FIRST galaxies with morphological classifications from all three NVSS, FIRST, and follow-up Very Large Array (VLA) observations. It is a compilation of the Combined NVSS–FIRST Galaxies (CoNFIG) catalogs 1, 2, 3, and 4 for a total of 859 unique sources that were classified as FR I, FR II, compact, or uncertain. Sources with size smaller than 3\as\ were classified as compact: ``C'' or ``C$^*$,'' depending on whether or not the source was confirmed compact from the Very Long Baseline Array calibrator list or the Pearson–Readhead survey. We excluded the sources that no not have redshift information (221). 

\textit{GRG\_catalog} \citep{Dabhade20} is a catalog of 820 GRGs that is a part of the Search and Analysis of Giant Radio Galaxies with Associated Nuclei (SAGAN)\footnote{\href{https://sites.google.com/site/anantasakyatta/sagan}{https://sites.google.com/site/anantasakyatta/sagan}}. This catalog is a database of all known GRGs from the literature to date. This catalog has the following morphological classifications: FR I, FR II, hybrid morphology radio source, and double-double radio galaxy. In addition, \cite{Dabhade20} has classified a subset of the GRGs as HERGs or LERGs using a \textit{Wide-field Infrared Survey Explorer} color-color analysis of four mid-infrared bands (\textit{W1}, \textit{W2}, \textit{W3}, and \textit{W4}). 

\textit{Macconi+20} \citep{Macconi20} is a sample of 79 radio galaxies that are sources from the revised Third Cambridge Catalogue of Radio Sources (3CR) that are at $z <$ 0.3 and are classified both in the optical (HERGs and LERGs) and radio bands (FR Is vs. FR IIs). Of the total 79, 30 are FR II-HERGs, 17 FR II broad-line radio galaxies (which are classified as HERGs according to their narrow-line-region emission and in this work we classify them as HERGs), 19 are FR II-LERGs, and 13 are FR Is (only 12 of the FR Is have X-ray data). \cite{Macconi20} preformed X-ray analysis to compute the $L_{\mathrm{2-10~keV}}$ and $\Gamma$ (photon index) for the FR II-HERGs and FR II-LERGs using \xmm/\chandra\ data. Due to poor statistics and/or the complexity of the emission, \cite{Macconi20} fixed $\Gamma$=1.7 in 7 out of 19 FR II-LERGs and in 27 out of 32 FR II-HERGs. The $L_{\mathrm{2-10~keV}}$ and $\Gamma$ for the FR Is used in \cite{Macconi20} originate from \cite{Balmaverde06} and were also calculated using spectral fitting. From \cite{Balmaverde06}, 10 out of the 12 FR Is that had X-ray data had constrained $L_{\mathrm{2-10~keV}}$ and $\Gamma$ values. We restricted the Macconi+20 sample used in this work to those sources that have a constrained or fixed $\Gamma$ value.

\textit{Mingo+19} \citep{Mingo19} is a catalog of 5805 radio galaxies identified in the Low Frequency Array Two-Metre Sky Survey Data Release 1 \citep[DR1;][]{Shimwell19,Williams19} that are morphologically classified as FR I, FR II, hybrid, or unresolved using an automated classification algorithm LoMorph\footnote{\href{https://github.com/bmingo/LoMorph/}{https://github.com/bmingo/LoMorph/}} (see \citealt{Mingo19} for details). \cite{Mingo19} identify ``small'' sources (which are the smallest identified sources) for which they can classify the morphology, but state that the classification of these sources is less reliable. In addition, some sources were given the additional morphological classification of wide-angle tail or narrow-angle tail. In order to obtain redshifts for these sources, we cross-matched the \cite{Mingo19} sample with the photometric redshifts for the entire DR1 catalog from \cite{Duncan19} using the \verb|Source_name| in \verb|topcat|. We restricted the Mingo+19 sample used in this work to those that are classified (i.e., we removed those with `Indeterminate' = True). We removed those with `Small' = True for the FR final catalog as they do not have a reliable FR classification due to their small extent (see Sect. \ref{sect:hid_fr}), but we leave these sources in for the extent catalog (see Sect. \ref{sect:hid_ext}). 

\textit{Miraghaei+17} \citep{Miraghaei17} is a sample of $\sim$1300 1.4-GHz-selected extended radio sources from \cite{Best12} that were visually classified primarily as FR I, FR II, hybrid, or unclassified using FIRST and NVSS images. We refer to this sample as \textit{Miraghaei+17\_FR}. We restricted the Miraghaei+17\_FR sample used in this work to those that have a morphological classification (``FRclass'' < 400, where 400 is the morphological code). In addition, \cite{Miraghaei17} identified a sample of compact radio sources that correspond to those sources identified as single-component FIRST sources by \cite{Best12} (see \citealt{Miraghaei17} for details). From the combined sample of both extended and compact sources, 245 had HERG or LERG classifications (Table 4 in \citealt{Miraghaei17}), and we refer to this sample as \textit{Miraghaei+17\_FR\_HL}.

Radio Sources Associated with Optical Galaxies and Having Unresolved or Extended Morphologies I (\textit{ROGUE I}; \citealt{ROGUEI20}) is a catalog of 32,616 spectroscopically selected galaxies whose radio morphology have been visually classified using FIRST and NVSS images. The main morphological classifications of interest for this work are those with FR I, FR II, hybrid, one-sided FR I, one-sided FR II, double-double radio galaxy, wide-angle tail, narrow-angle tail, and head-tail. We removed the sources whose ``Finalclass'' was blended, halo, not clear, not detected, star-forming region, compact, or extended from the ROGUE I sample for the purpose of this work. 

\subsubsection{Radio morphology catalogs: Compact radio sources}
\textit{Chandola+20} \citep{Chandola20} used the Giant Metrewave Radio Telescope (GMRT) to observe 27 low- and intermediate-luminosity radio AGNs that were classified as either LERG or HERGs. If the linear projected size of the radio emission in the GMRT data was $\lesssim$ 20 kpc, the source was classified as compact and if the linear projected size was $\gtrsim$ 20 kpc, the source was classified as extended. The positions of the radio sources are defined by their SDSS counterparts.

\textit{Kosmaczewski+20} \citep{Kosmaczewski20} is a sample of 29 objects that are in the earliest phase of radio galaxy evolution.\ They are classified as GPS and/or CSOs and have X-ray data. \cite{Kosmaczewski20} calculated the $L_{\mathrm{2-10~keV}}$, but $\Gamma$ was not calculated or recorded in this work.

\textit{Liao+20\_I} \citep{Liao20a} collected a sample of 545 young radio sources from the literature classified as GPS, CSS, HFP, or CSO. They then removed blazars and searched for the SDSS spectroscopic counterparts within 2\as\ of the NASA/IPAC Extragalactic Database (NED) position, which resulted in a final sample of 126 young radio sources with optical counterparts.

\textit{Liao+20\_II} \citep{Liao20b} started with the parent sample of 468 young radio AGNs from \cite{Liao20a} that were classified as compact sources in the literature and were not blazar-type objects. Then the sample was cross-matched with \chandra\ and \xmm\ X-ray archives to find X-ray detections within 2 and 5\as\ of the NED source positions. The final sample of young radio sources with $L_{\mathrm{2-10~keV}}$ and $\Gamma$ values contains 91 sources. We restricted the Liao+20\_II sample used in this work to those that have a ``logL\_X'' and ``Gamma'' not equal to 0.

\textit{Sobolewska+19} \citep{Sobolewska19} is a sample of 24 CSOs that have either \chandra\ or \xmm\ observations.

\subsubsection{HERG--LERG catalogs}
\textit{Best+12} \citep{Best12} is a sample of 18,286 RL AGNs constructed by combining the seventh data release of the SDSS with the NVSS and FIRST surveys. Using this sample, the authors label the radio AGNs as HERGs or LERGs by calculating the ``excitation index'' \citep{Buttiglione10} using measurements from the SDSS spectra. We restricted the Best+12 sample used in this work to those that are either HERGs or LERGs (i.e., L=1 or H=1 in the table, which is a total of 10,344 objects).

\textit{Ching+17} \citep{Ching17} is a sample of 12,329 radio sources from the Large Area Radio Galaxy Evolution Spectroscopic Survey (LARGESS) that are identified via FIRST and have optical identifications via SDSS, WiggleZ, or Galaxy And Mass Assembly (GAMA). Of this sample, 10,856 have reliable spectroscopic redshifts. \cite{Ching17} classify sources as HERGs or LERGs based on measurements of the [O$_{\mathrm{III}}$] line using similar cutoffs to \cite{Best12}. We restricted the Ching+17 sample used in this work to those sources that are either HERGs or LERGs and have a redshift (a total of $\sim$6,700 objects).

\subsection{Simultaneous \xmm\ X-ray and UV source catalog}\label{sect:xmm_simult}
One catalog that we used to obtain the X-ray and UV measurements of the radio galaxies of interest is a catalog of X-ray and UV observations that were taken simultaneously. We followed \cite{Svoboda17} and created a catalog of simultaneous \xmm\ X-ray and UV observations specifically in order to minimize the effect of variable X-ray absorption on the source luminosity estimates.\footnote{Because the UV and X-ray data trace the central engine and the radio emission from the catalogs described in Sect. \ref{sect:rad_cats} trace the jet activity on much larger scales, we do not require simultaneity of radio, UV, and X-ray observations.} We created this catalog by first cross-matching the 4XMM-DR10 \xmm\ Serendipitous Source Catalogue \citep[4XMM-DR10;][]{Webb20} with the XMM-OM-SUSS 5.0, which is the 2020 release of the XMM Optical Monitor (OM) Serendipitous Ultraviolet Source Survey (SUSS) catalog \citep{Page12}, within a 5\as\ cross-matching radius using \verb|topcat|. The 5\as\ radius is approximately equal to or slightly larger than the nominal accuracy of the astrometric reconstruction. To ensure the simultaneity of the observations, we selected the X-ray and UV observations that have the same OBSID. We note that it is possible to have several X-ray and UV observations for a single source. 

4XMM-DR10 contains source detections that are drawn from 11,647 \xmm\ EPIC observations made between 2000 February 3 and 2019 December 14 in an energy interval of 0.2\,--\,12 keV. All data sets were publicly available by 2020 December 10. XMM-OM-SUSS 5.0 is the fifth release of the catalog and contains source detections by the OM\ instrument on board \xmm\ spanning the period of observations from \xmm\ revolution 34 (February 2000) to revolution 3704 (February 2020). The \xmm\ OM UV filters are UVW2 ($\lambda_{\mathrm{eff}}$ = 2120 \AA), UVM2 ($\lambda_{\mathrm{eff}}$ = 2310 \AA), and UVW1 ($\lambda_{\mathrm{eff}}$ = 2910 \AA), and the optical filters are U ($\lambda_{\mathrm{eff}}$ = 3440 \AA), B ($\lambda_{\mathrm{eff}}$ = 4500 \AA), and V ($\lambda_{\mathrm{eff}}$ = 5430 \AA).

The observed photon index $\Gamma$ can be estimated from the flux measurements in two neighboring X-ray bands in the 4XMM-DR10 catalog. Using $F_{0.5-2\mathrm{keV}}$, $F_{2-12\mathrm{keV}}$, and monochromatic flux ratios, we define $\Gamma$ as
\begin{equation}\label{eq:gamma}
    \Gamma = 2 - 2\log\left( \frac{F_{0.5-2\mathrm{keV}}}{F_{2-12\mathrm{keV}}} \right)\frac{1}{\log\left( \frac{(r_1+1)(r_1-1)}{(r_2+1)(r_2-1)} \right)}
,\end{equation}
where $r_1$ is a ratio of the boundaries of the soft band (0.5 keV/2 keV) and $r_2$ is a ratio of the boundaries of the hard band (12 keV/2 keV). Equation \ref{eq:gamma} reduces to $\Gamma \approx 2 + 3.144 \times \log\frac{F_{0.5-2\mathrm{keV}}}{F_{2-12\mathrm{keV}}}$ \citep{Svoboda17}. We used $\Gamma$ for the quality cuts described in Sect. \ref{sect:quality_cuts} and for flux extrapolation to obtain the coronal luminosity in Sect. \ref{sect:LX}.

The X-ray flux for various data releases of \xmm\ Serendipitous Source Catalog in different energy bands is calculated from a measured count rate assuming a power-law spectral model with $\Gamma$=1.7 and
a cold absorbing column density of $N_H$ = 3$\times$10$^{20}$cm$^{-2}$ (see \footnote{\href{https://xmmssc-www.star.le.ac.uk/Catalogue/2XMM/UserGuide\_xmmcat.html\#EmldetFit}{https://xmmssc-www.star.le.ac.uk/Catalogue/2XMM\\/UserGuide\_xmmcat.html\#EmldetFit}}, \citealt{Rosen16}, and \citealt{Webb20}).

\subsection{\swift\ X-ray and UV source catalog}\label{sect:swift_UX}
To supplement the \xmmcat\ and be able to obtain more X-ray and UV measurements for the radio galaxy samples described in Sect. \ref{sect:rad_cats}, we also cross-matched the \swift-X-Ray Telescope (XRT) Point Source (2SXPS) catalog \citep{Evans20} and the \swift\ Ultraviolet/Optical Telescope Serendipitous Source Catalogue \citep[UVOTSSC;][]{Page14,Yershov14} UV catalog within a 5\as cross-matching radius using \verb|topcat|. This \swift\ UV and X-ray source catalog contains all X-ray and UV sources that are within 5\as\ of one another and there could be multiple UV observations associated with one X-ray source. 

The 2SXPS catalog contains the sources detected by the \swift\ XRT in the 0.3-10 keV energy range. The X-ray flux is calculated using the measured count rate and one of the following three methods: (a) a fixed spectrum with a power law with $\Gamma$=1.7 and $N_H$ is ``GalacticNH,'' (b) absorbed power-law spectral values derived from the hardness ratios, and (c) absorbed power-law spectral values taken from a fit to a custom-built spectrum (see \footnote{\href{https://www.swift.ac.uk/2SXPS/docs.php\#sources\_flux}{https://www.swift.ac.uk/2SXPS/docs.php\#sources\_flux}} and \citealt{Evans20}). We follow the flux measurements preference built into the catalog where the order is c, b, and a. If the flux comes from method b or c, we chose the unabsorbed flux.

The UVOTSSC catalog was compiled from 23,059 \swift\ data sets taken within the first five years of observations with the \swift\ UVOT (from the beginning of the mission in 2005 until 2010 October 1) and uses the UVW1 ($\lambda_{\mathrm{eff}}$ = 1928 \AA), UVM1 ($\lambda_{\mathrm{eff}}$ = 2246 \AA), UVW2 ($\lambda_{\mathrm{eff}}$ = 2600 \AA), U ($\lambda_{\mathrm{eff}}$ = 3465 \AA), B ($\lambda_{\mathrm{eff}}$ = 4392 \AA), and V ($\lambda_{\mathrm{eff}}$ = 5468 \AA) filters. We note that it is possible to have several UV observations of the UV source associated with a single X-ray source. In the 2SXPS catalog, the entries are created by stacking observations at the source location, and thus we did not require simultaneity of the X-ray and UV observations as with the \xmmcat. 

\begin{figure*}
    \centering
    \includegraphics[width=0.84\linewidth]{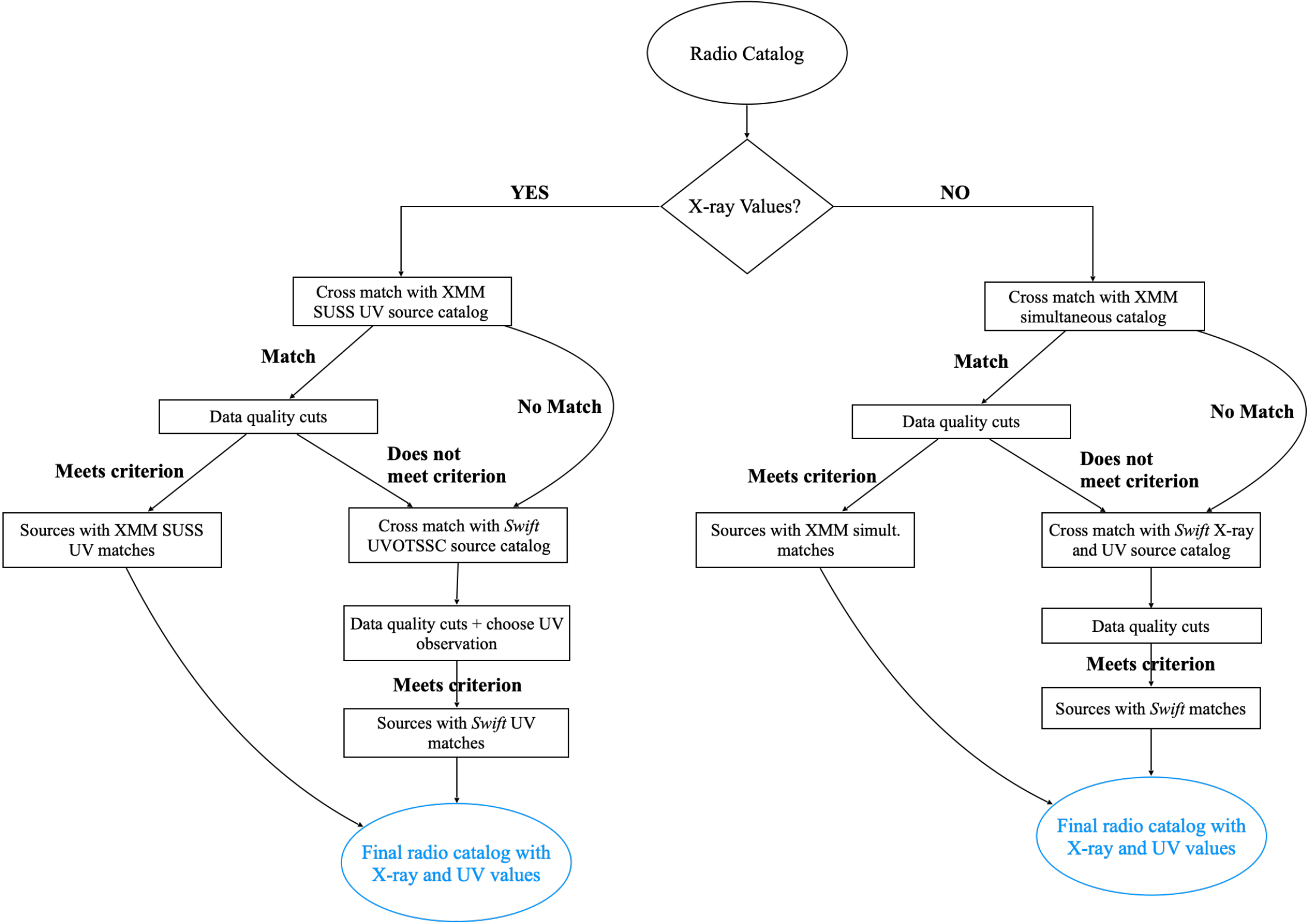}
    \caption{Flowchart describing the process of cross-matching the radio catalogs with \xmm\ and \swift\ source catalogs.}
    \label{fig:flowchart}
\end{figure*}

\subsection{Comparison samples}\label{sect:comparison_samples}
We compare our sources with a specific radio property to two larger samples of AGNs to provide context: (a) 3,632 bright quasars and AGNs from the \xmmcat\ (see Sect. \ref{sect:xmm_simult}) and (b) 292 local Seyfert AGNs from the Swift/BAT AGN Spectroscopy Survey (BASS; \citealt{Koss17}). A detailed description and analysis of the results of these catalogs will be discussed in Borkar et al., \textit{in prep.}, but here we give a brief description of the creation of these catalogs.

The \xmmcat\ provides the largest number of simultaneous X-ray and UV observations of AGNs. In order to obtain bona fide AGN sources with measured redshifts, the \xmmcat\ was cross-matched with the SDSS DR14 AGN catalog \citep{Paris18} and the Veron-Cetty \& Veron catalog of AGNs \citep{Veron10}. Further quality cuts and calculations similar to those listed Sect. \ref{sect:methods} were performed (including a $\Gamma$ cut that matches that described in Sect. \ref{sect:quality_cuts}). The final catalog results in a sample of 3,632 individual observations. This catalog can contain multiple observations of the same object, which are treated as independent data points for comparison in this work. The catalog consists primarily of bright quasars ($L > 10^{44}$ ergs$^{-1}$) that have redshifts up to 3, but most sources have redshift $z < 2$. The sources have a SMBH mass of $10^7 - 10^{10} M_{\odot}$ with the median value of $10^{8.6} M_{\odot}$. We cross-matched the sources with the VLA Sky Survey catalog to obtain their radio luminosity and find that about 11\% of the sources are RL. The RL sources are predominantly located in the hard part of the HID, and the loudest sources are found in the top-right part of the HID. This is similar to the observed radio jets in the hard state in XRBs, providing confirmation to the similarity between XRB and AGN accretion states.

To complement this sample of bright quasars, we also compiled a sample from BASS, which consists of nearby ($z < 0.1$) local Seyfert AGNs. The BASS sample is a sample of hard X-ray selected AGNs and thus is a mixture of sources -- both high and low luminosity and a distribution of Eddington ratios \citep{Panessa15,Koss17,Ricci17}. We follow the same procedure to obtain the multiwavelength data, as discussed above for XMM-Newton sample. The final sample consists of 292 sources with luminosities ranging from $10^{39}-10^{46}$ ergs$^{-1}$. Their SMBH masses lie between $10^5 - 10^{10}~M_{\odot}$ with the median value of $10^{7.8} M_{\odot}$. We exclude obscured sources (those with $\log (N_H)$ > 21.9) from this sample and thus have a final sample size of 163 BASS sources.

For this work, we note that the XMM-Newton sample contains distant AGNs and quasars, which are accreting efficiently and are found in the ``high-soft'' state. In contrast, the BASS sample consists primarily of local Seyfert AGNs ($z<0.1$), which, compared to the XMM-Newton sample, represent harder sources and lower luminosities, including those in the ``low-hard'' state. In relevant figures, we show the XMM comparison sample as black solid outlined and/or blue shaded contours and the BASS comparison sample as dashed-dotted outlined and/or gray shaded contours.

\bgroup
\def\arraystretch{1.25}
\begin{table*}
\centering
\caption{Number of sources in the radio catalogs after each step of the cross-matching and filtering processes.}
\label{tab:catalog_num}
\begin{tabular}{c|c|c|ccc|ccc|c}
    \hline\hline 
    & & & \multicolumn{3}{c|}{\xmm} & \multicolumn{3}{c|}{\swift\ Only} & \\
    \textbf{Catalog} &     \textbf{N} & \textbf{Source} &  \textbf{X+UV} &  \textbf{UV} &  \textbf{Qual.} &  \textbf{X+UV} &  \textbf{UV} &  \textbf{Qual.} &  \textbf{Total} \\
    \hline
             Best+12 & 18286 &         Cat. &         110 &  $-$ &         58 &    59 &     $-$ &           18 &     76 \\
         Chandola+20 &    27 &         Cat. &           1 &  $-$ &          1 &     3 &     $-$ &            0 &      1 \\
            Ching+17 &  6720 &         Cat. &          38 &  $-$ &         16 &    18 &     $-$ &            4 &     20 \\
              FRXCAT &   507 &         Cat. &          26 &  $-$ &         14 &     7 &     $-$ &            4 &     18 \\
           Gendre+10 &   636 &         Cat. &          40 &  $-$ &         20 &    35 &     $-$ &           19 &     39 \\
        GRG\_catalog &   820 &         Cat. &          23 &  $-$ &          8 &    10 &     $-$ &            5 &     13 \\
     Kosmaczewski+20 &    29 &         Cat. &           7 &  $-$ &          1 &     5 &     $-$ &            3 &      4 \\
          Liao+20\_I &   126 &         Cat. &           9 &  $-$ &          1 &     4 &     $-$ &            3 &      4 \\
         Liao+20\_II &    91 &         Lit. &         $-$ &   16 &          9 &   $-$ &       7 &            4 &     13 \\
          Macconi+20 &    63 &         Lit. &         $-$ &   26 &         23 &   $-$ &       2 &            2 &     25 \\
            Mingo+19 &  5805 &         Cat. &          15 &  $-$ &          8 &    11 &     $-$ &            5 &     13 \\
    Miraghaei+17\_FR &  1329 &         Cat. &          16 &  $-$ &          7 &     9 &     $-$ &            3 &     10 \\
Miraghaei+17\_FR\_HL &   245 &         Cat. &           7 &  $-$ &          3 &     6 &     $-$ &            1 &      4 \\
              ROGUEI & 32616 &         Cat. &          34 &  $-$ &         21 &    17 &     $-$ &            8 &     29 \\
      Sobolewska+19a &    25 &         Cat. &          11 &  $-$ &          2 &     5 &     $-$ &            2 &      4 \\
    \hline
\end{tabular}
\tablefoot{\bi{Column 1:} Catalog name as defined in Sect. \ref{sect:cats}. \bi{Column 2:} Number of sources in the literature radio catalog. \bi{Column 3:} Source of the X-ray data. ``Cat.'' means that the source is either the \xmmcat\ or the \swiftcat. ``Lit.'' means that the X-ray values ($\Gamma$ and $L_X$) were from the literature reference associated with the catalog name. \bi{Columns 3-8:} Number of sources left after filtering, cross-matching the radio catalogs with \xmm\ and \swift\ X-ray and UV catalogs, and data quality cuts. If the source of the X-ray data is a source catalog, X+UV is the number of sources left in the radio catalog after applying filtering to the sources that matched with the simultaneous \xmmcat\ and the \swiftcat\ under the respective observatory columns (see Sect. \ref{sect:cross_matching}). In this case, we did not cross-match with the UV-only catalogs, resulting in a $``-$'' value in the UV column. If the source of X-ray data is from the literature reference, UV is the number of sources after filtering that matched with the XMM-OM-SUSS 5.0 and \swift\ UVOTSSC UV source catalogs (see Sect. \ref{sect:cross_matching}). In this case, we did not cross-match with the X-ray+UV catalogs and only cross-matched with UV catalogs, resulting in a $``-$'' value in X+UV column. ``Qual.'' is the number of sources left after performing the data quality cuts described in Sect. \ref{sect:quality_cuts}. \bi{Column 9:} Total amount of sources for the catalog produced by combining the sources left in the radio catalogs with either \xmm\ or \swift\ matches after cross-matching and data quality cuts.}
\end{table*}
\egroup

\section{Methods}\label{sect:methods}
The final catalogs of radio sources that have morphology, excitation, or linear extent classifications with X-ray and UV measurements were created by executing the following six steps. First, in order to obtain X-ray and UV fluxes of the radio galaxies, we cross-matched the individual radio catalogs with the \xmm\ simultaneous (see Sect. \ref{sect:xmm_simult}) and then the \swiftcat\ (see Sect. \ref{sect:swift_UX}). During this step, we applied filtering to the samples detailed in the previous section (require redshifts, remove unwanted radio morphologies, etc.; see Sect. \ref{sect:cross_matching}). 

Second, in some cases there can be several UV observations of one source that match with one X-ray observation. In this case, we selected one UV observation for each source (see Sect. \ref{sect:select_UV}). 

Third, we applied data quality cuts to the UV and X-ray data (see Sect. \ref{sect:quality_cuts}).
Fourth, we applied a Galactic extinction correction (see Sect. \ref{sect:galactic_NH}). Fifth, we calculated the $L_{UV}$, $L_X$, and hardness (see Sects. \ref{sect:LUV}, \ref{sect:LX}, and \ref{sect:hardness}). Finally, we created final catalogs according to specific radio properties (FR I--II, compact--giant, HERG--LERG), where duplicate sources have been taken into account and resolved (see Sect. \ref{sect:final_cats}).

\subsection{Cross-matching}\label{sect:cross_matching}
In order to place the radio galaxies of interest on the HID, we required the $L_X$ and $L_{UV}$ of these galaxies. Thus, we cross-matched the radio galaxy catalogs with both \xmm\ and \swift\ source catalogs to find radio sources with UV and X-ray observations and flux measurements (see Fig. \ref{fig:flowchart} for a flowchart of this process).

For a majority of the catalogs (Best+12, Chandola+20, Ching+17, FRXCAT, Gendre+10, GRG\_catalog, Kosmaczewski+20, Liao+20\_I, Mingo+19, both Miraghaei+17, ROGUE I, and Sobolewska+19a), we first cross-match the radio coordinates of the sources with the X-ray coordinates in the \xmmcat\ and find the closest match within 5\as, then do the same with the UV source coordinates. There may be multiple source observations associated with one UV source, and in this case we chose the closest UV observation to the radio source.

Then for the sources that did not have a match in the \xmmcat\ or sources that do not make it through the data quality cuts described in Sect. \ref{sect:quality_cuts}, we cross-match these remaining sources with the \swiftcat\ (see Sect. \ref{sect:swift_UX}). Similarly to the above, we cross-match the radio coordinates for the sources given in the literature with the X-ray coordinates in the \swiftcat\ and find the closest match within 5\as, then do the same with the UV source coordinates. There could be several UV observations of the closest UV source. We describe the process of choosing a UV observation in Sect. \ref{sect:select_UV}. We cross-matched with the \xmmcat\ first to prioritize simultaneous observations and the more sensitive \xmm\ X-ray observations.

There were two radio catalogs that already had $L_X$ and $\Gamma$ values available. For Macconi+20, spectral fitting was done to acquire these values, and for Liao+20\_II the values were obtained from the literature references within and are typically a result of spectral fitting. In this case, it was necessary to acquire only UV flux measurements. In this case, we followed the preference order of \xmm\ then \swift\ and cross-matched the catalog first with XMM-OM-SUSS 5.0. Then, for the sources that do not have an associated simultaneous \xmm\ UV observation or did not make it through the quality cuts, we cross-matched those with the \swift\ UVOTSSC catalog. Again, we find the closest UV matches within 5\as\ of the radio source, and in the case of the \swift\ UVOTSSC sources, we chose a UV observation according to Sect. \ref{sect:select_UV}.

\subsection{Selection of UV \swift\ observations}\label{sect:select_UV}
In some cases, there may be several UV observations of the closest UV source in the \swift\ catalogs. For the \swift\ UV observations, multiple observations of one source have the same RA and Dec, and thus we used a different method than with \xmm. 

To calculate $L_{UV}$, we used the UV flux from one filter from one observation. Since it is standard to use $F_{\mathrm{UWV1}}$ to calculate the $L_{UV}$ of the accretion disk (see reasoning in Sect. \ref{sect:LUV}), we chose to use the UV observation that has the highest significant detection and a filter $\lambda$ that is (a) closest to $\lambda_{\mathrm{UVW1}}$ in the observed frame and (b) $\geq1240~\AA$ as measurements of the UV flux at wavelengths lower than 1240 $\AA$ could be contaminated by Ly$\alpha$ emission.

We also needed to choose a UV observation with which to calculate the UV spectral index, $\beta$ (see Sect. \ref{sect:LUV} for exact details). We did this by using the two filters that have $\lambda$ closest to the filter that represents UVW1 in the observed frame (including itself) of the object. Similar to the above method, we chose the UV observation that had the highest average significance in the two filters that have $\lambda$ closest to the filter that represents UVW1 in the observed frame (including itself) of the object. The de-reddened fluxes are used in this selection. We note that the UV observation selected for $L_{UV}$ could be different than the one selected for the UV slope calculation.

\subsection{Data quality cuts}\label{sect:quality_cuts}
We apply similar data quality cuts as \cite{Svoboda17}: To ensure a significant detection, the UV flux is required to be > 3$\sigma$ detection.
To avoid underexposed observations, we ensured that (a) the X-ray exposure time is greater than 10 ks, and (b) the uncertainty in a UV or X-ray flux measurement does not exceed 100\%.
Finally, we removed sources with (a) a flat $\Gamma$ ($\leq$ 1.5), which is indicative of significantly absorbed AGNs (see Sect. \ref{sect:errors} for a more detailed discussion), or (b) a steep $\Gamma$ ($\geq$ 3.5), which is physically far from what is seen in AGN X-ray slopes.

The number of sources that were eliminated for each cut is given in Appendix \ref{A:rejects}. Additionally, after correcting for Galactic extinction (see Sect. \ref{sect:galactic_NH}), we discard any UV filters for further use that have $\lambda\leq1240~\AA$ as measurements of the UV flux at lower wavelengths than 1240$\AA$ could be contaminated by Ly$\alpha$ emission. If a source only has filters with $\lambda\leq1240~\AA$, then the source is discarded.

\subsection{Galactic $N_H$}\label{sect:galactic_NH}
The UV and optical fluxes are affected by Galactic extinction. For the de-reddening, we used the relation by \cite{Guver09},
\begin{equation}
    A_v = \frac{N_H}{2.2\times10^{21}}
,\end{equation}
where $N_H$ is the column density in a Galactic HI map in combination with \verb|astropy.dust_extinction| assuming $R_v$=3.1. For the sources using \xmm\ simultaneous data or UVOTSSC for UV, we determine the Galactic $N_H$ in the direction of the source using the Galactic HI map by \cite{Kalberla05}. For the sources using \swift\ X-ray data, the Galactic $N_H$ is included as a parameter in the catalog and was calculated using \cite{Willingale13}.

\subsection{Thermal disk luminosity}\label{sect:LUV}
The goal is to obtain a measurement of the thermal disk emission whose spectral energy distribution (SED) usually peaks in the UV (see Sect. 2.4 of \citealt{Svoboda17}). To calculate the luminosity, it is necessary to choose a flux from the available filters (UVW2, UVM2, UVW1, U, B, V). Since our sample spans a wide range in redshift, we chose to use the flux from the nearest UV or optical filter to the observed-frame wavelength of a reference filter $\lambda_{\mathrm{ref,obs}} = (1 + z) \times \lambda_{ref}$. Following the logic of \cite{Svoboda17}, we chose $\lambda_{ref}$ to be the central, rest-frame wavelength of the UVW1 filter (\xmm\ OM 2910 \AA\ and \swift\ 2600 \AA) because the UVW1 filter has the (a) highest throughput for \xmm\ and (b) most flux values of the six \xmm\ filters in many catalogs (e.g., the sample in \citealt{Svoboda17}, the \xmmcat\ in this work, and the XMM SUSS 5.0 catalog). In the case of \swift\ in order to be consistent in methodology, we also used UVW1 as the reference filter. Thus for the disk luminosity calculation, we use the flux ($F_{\lambda 2910~\AA}$) from the filter that is (a) closest to the reference wavelength in the observed-frame, and (b) passed the quality cuts described in Sect. \ref{sect:quality_cuts}. 

\bgroup
\def\arraystretch{1.25}
\begin{table*}
\centering
\caption{Basic information about the final catalogs.}
\label{tab:classes_tbl}
\begin{tabular}{p{2.0cm} | >{\RaggedRight}p{4.25cm} | p{0.8cm} | >{\RaggedRight}p{9.0cm}}
    \hline\hline 
    \textbf{Type} & \textbf{Rad. Class} &  \textbf{Total} & \textbf{Ref.} \\
    \hline
   Morphology & FR I (26), FR II (38) &     64 & FRXCAT, GRG\_catalog, Gendre+10, Macconi+20, Mingo+19, Miraghaei+17, ROGUEI \\
   Excitation & HERG (26), LERG (94) &   120 &  Best+12, Ching+17, GRG\_catalog, Liao+20\_I, Macconi+20 \\
   Morph.+Excit. &FR I LERG (12), FR II HERG (19), FR II LERG (8) &     39 &  FRXCAT, GRG\_catalog, Macconi+20, Miraghaei+17\\
   Extent & C (11), CO (20), CSS (5), FR0 (8), FRIIC (1), FRS (8), G (13), Norm (52) &   118 & Chandola+20, FRXCAT, GRG\_catalog, Gendre+10, Kosmaczewski+20, Liao+20\_I, Liao+20\_II, Macconi+20, Mingo+19, Miraghaei+17, ROGUEI, Sobolewska+19a \\
\hline
\end{tabular}
\tablefoot{\bi{Column 1:} Type of radio characteristic that defines the table. See Sect. \ref{sect:results} for definitions. \bi{Column 2:} radio classifications within the table plus the number of sources with that classification (in parentheses). \bi{Column 3:} Total number of sources in that table. \bi{Column 4:} Names of the catalogs that have sources with these radio classes. See Sect. \ref{sect:rad_cats} for references. }
\end{table*}
\egroup

If the flux was not from the UVW1 equivalent in the observed frame, we converted the flux to a UVW1 flux by multiplying the flux by a factor $(\lambda_{\mathrm{ref,obs}}/\lambda_{\mathrm{obs}})^\beta$, where $\lambda_{\mathrm{ref,obs}}$ is the observed wavelength of the reference filter, $\lambda_{\mathrm{obs}}$ is the observed wavelength of the filter that has flux, and $\beta$ is the UV slope. The $\beta$ in the wavelength domain is calculated by
\begin{equation}
    \beta = \frac{\log F_a/F_b}{\log \lambda_a/\lambda_b}
,\end{equation}
where $F_a$ and $F_b$ are observed flux densities in the nearest filters to $\lambda_{\mathrm{ref}}$ in the observed frame, and $\lambda_a$ and $\lambda_b$ are the mean wavelengths of the corresponding filters. Following \cite{Svoboda17}, when only a single filter had flux, a default of $\beta=-1.5$ is used, based on previous UV studies of quasars \citep{Scott04,Richards06}. Additionally, to ensure physical and realistic values, we restricted $\beta$ to be $-2.8<\beta<0$ based on the work of \citet{Svoboda17}, who found that $\langle \beta \rangle$ = -1.4 with $\sigma$=1.4 for a sample of 1522 AGNs. If $\beta$ was found to be outside this interval, a default of $\beta=-1.5$ was used for flux extrapolation. Lastly, we multiplied the observed UV flux by a K-correction factor to get the source UV flux at the rest wavelength $\lambda_{\mathrm{rest}}$.

We used the redshift and Galactic-extinction-corrected UV flux to estimate the disk luminosity ($L_D$), which can be defined as
\begin{equation}
    L_D = A \times 4\pi D_L^2 \lambda F_{\lambda, 2910~\AA}
,\end{equation}
where $D_L$ is the luminosity distance constrained from the redshift measurement $z$ and $A$ is an empirical factor that is chosen such that the sum of the disk and the power-law luminosity, $L_{\mathrm{tot}}$, roughly corresponds to the bolometric luminosity. \cite{Svoboda17} cross-matched their sample of AGNs with simultaneous UV and X-ray observations with those of \cite{Vasudevan09} and determined that $\langle A \rangle=2.4$ with $\sigma=$0.6. Although $A$ will vary with the black hole mass and accretion rate, for simplicity and consistency with \cite{Svoboda17}, we applied a factor $A=2.4$ when estimating $L_D$ for all sources in our sample. We note that there is a relatively small black hole mass range ($8 \leq M_{BH} \leq 10$) in this sample so we do not expect this to affect the results presented in Sect. \ref{sect:results}. Though a one-size-fits-all scaling will introduce some error, determining $A$ for each individual source is beyond the scope of this work and will not change substantially the results presented in this paper.

\subsection{Coronal (power-law) luminosity}\label{sect:LX}
Following the methods of \cite{Svoboda17}, we define the coronal (power-law) luminosity as the extrapolated X-ray luminosity in the energy interval 0.1-100 keV. The power-law luminosity can therefore be written as 
\begin{equation}
    L_P = 4\pi D_L^2 F_{0.1-100 \mathrm{keV}}
        ,\end{equation}
where $D_L$ is the luminosity distance constrained from the redshift measurement $z$ and $F_{0.1-100 \mathrm{keV}}$ is the X-ray flux in the 0.1-100 keV energy range. $F_{0.1-100 \mathrm{keV}}$ is calculated by an extrapolation of the observed 2-10 keV flux, 
\begin{equation}
    F_{0.1-100\,\mathrm{keV}} = F_{a-b\,\mathrm{keV}}\left( \frac{100^{-\Gamma+2}-0.1^{-\Gamma+2}}{b^{-\Gamma+2}-a^{-\Gamma+2}} \right)
,\end{equation}
where the photon index $\Gamma$ is either (a) that which is described in Sect. \ref{sect:xmm_simult} for the \xmmcat\, or (b) taken from the literature in the case of a literature catalog. Here, $a$ is the lower limit of the energy range for which the flux was calculated and $b$ is the upper limit of the energy range for which the flux was calculated; they are both instrument-specific. For \xmm, $a$=2 and $b$=12, for \swift, $a$=0.3, and $b$=10, and for the catalogs with literature X-ray luminosity and spectral index, $a$=2 and $b$=10. 

Additionally, for the X-ray values taken from the literature, $L_{2-10\mathrm{keV}}$. We converted this luminosity into $F_{2-10\mathrm{keV}}$ using the standard $L = 4\pi D_L^2F$ relationship and the cosmological parameters from \cite{Planck18} via \verb|astropy.cosmology.Planck18|. We used \cite{Planck18} for all calculations as different cosmologies will change the values in the HID $\leq$1\%.

We applied the standard K-correction to the X-ray fluxes from the \xmm\ simultaneous and \swift\ catalogs before converting the flux to a luminosity using $F_{intrinsic} = F_{observed} \times (1+z)^{\Gamma-2}$ where $\Gamma$ is the observed photon index described in Sect. \ref{sect:xmm_simult}. Since the sources were selected to have photon indices that are typically associated with low levels of intrinsic obscuration, we assume the observed photon index is a good approximation of the intrinsic value.

\begin{figure}
    \centering
    \includegraphics[width=\linewidth]{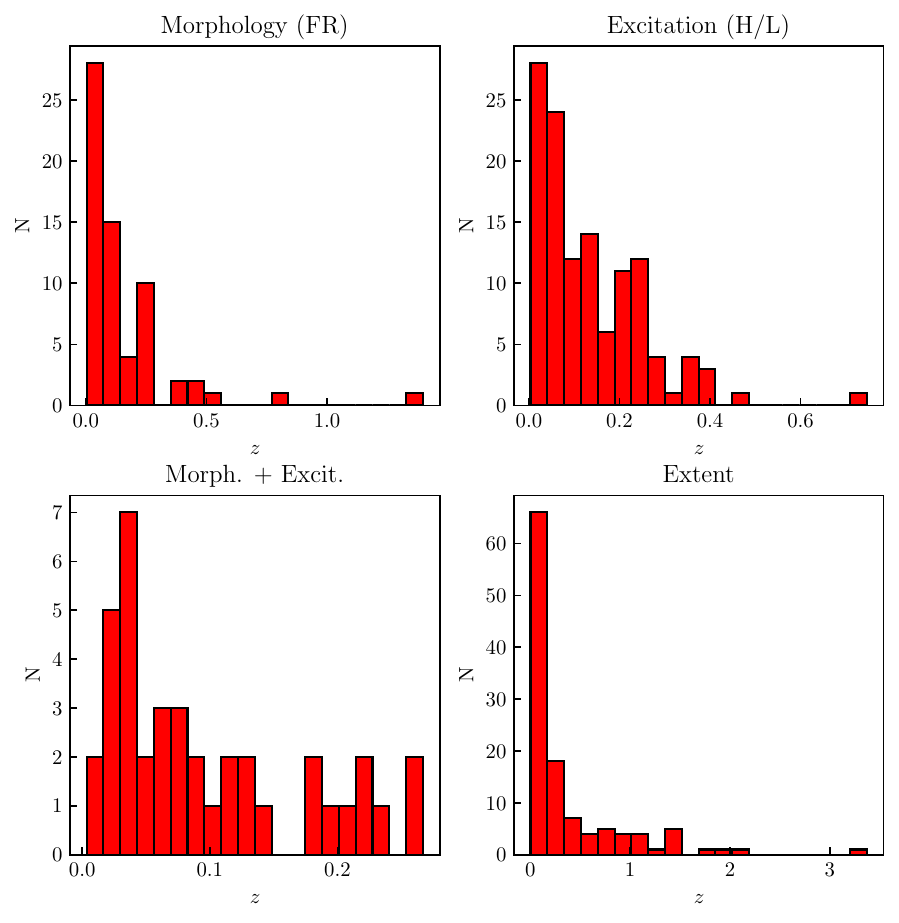}
    \caption{Redshift distribution for the sources in the morphology (FR I and FR II: upper left), excitation class (HERG and LERG: upper right), morphology plus excitation (lower left), and linear radio extent (lower right) catalogs.}
    \label{fig:z_dist}
\end{figure}

\subsection{Spectral hardness}\label{sect:hardness}
For XRBs, the HID is widely used to track the spectral state evolution of black holes (e.g.,\ \citealt{Fender04}) with the spectral hardness on the abscissa and the X-ray intensity on the ordinate. For XRBs, both the thermal emission and the hard X-ray emission are measured in X-rays. However, for AGNs the thermal emission peaks in the UV instead, and this needs to be accounted for in the definition of hardness. Thus following \cite{Svoboda17}, we define hardness ($H$) as a ratio of the power-law luminosity ($L_P$) against the total luminosity (a sum of the corona power-law $L_P$ and disk $L_D$ luminosity):
\begin{equation}
    H = \frac{L_P}{L_{tot}} = \frac{L_P}{L_P+L_D}
.\end{equation}
To verify the applicability of this method of calculating $L_{tot}$, we calculated $L_{bol}$ for a large sample of AGNs via empirical relations relating $L_{bol}$ to \LX\ \citep{Netzer19} and find that it agrees with the value of $L_{tot}$ calculated by combining $L_P$ and $L_D$ ($L_{tot} = L_P + L_D$) within a factor of 2-3, which will not affect our overall results.

\begin{figure*}
    \begin{subfigure}{.5\textwidth}
      \centering
      \includegraphics[width=0.99\linewidth]{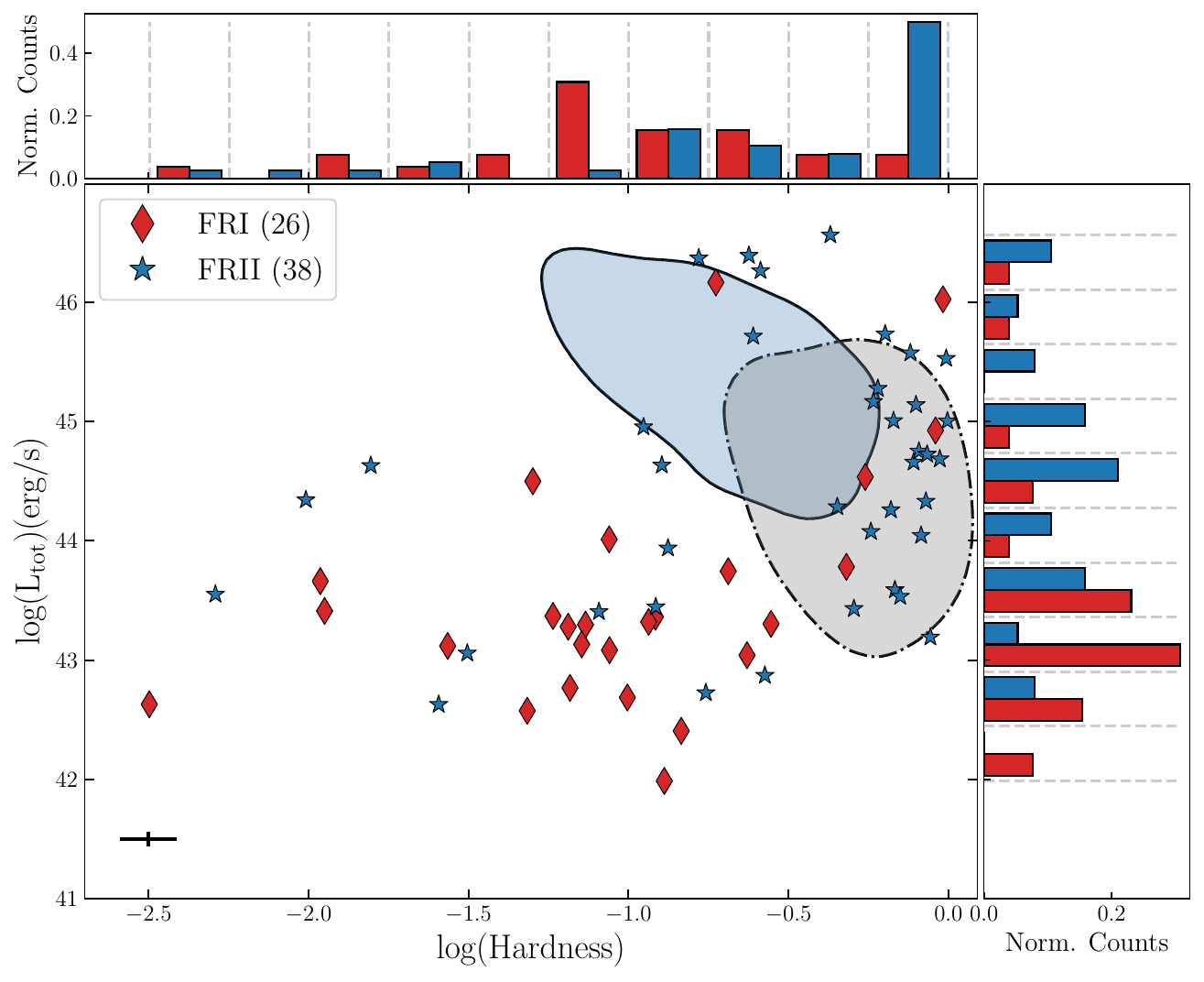}
      \caption{}
      \label{fig:hid_fr_a}
    \end{subfigure}
    \begin{subfigure}{.5\textwidth}
      \includegraphics[width=0.99\linewidth]{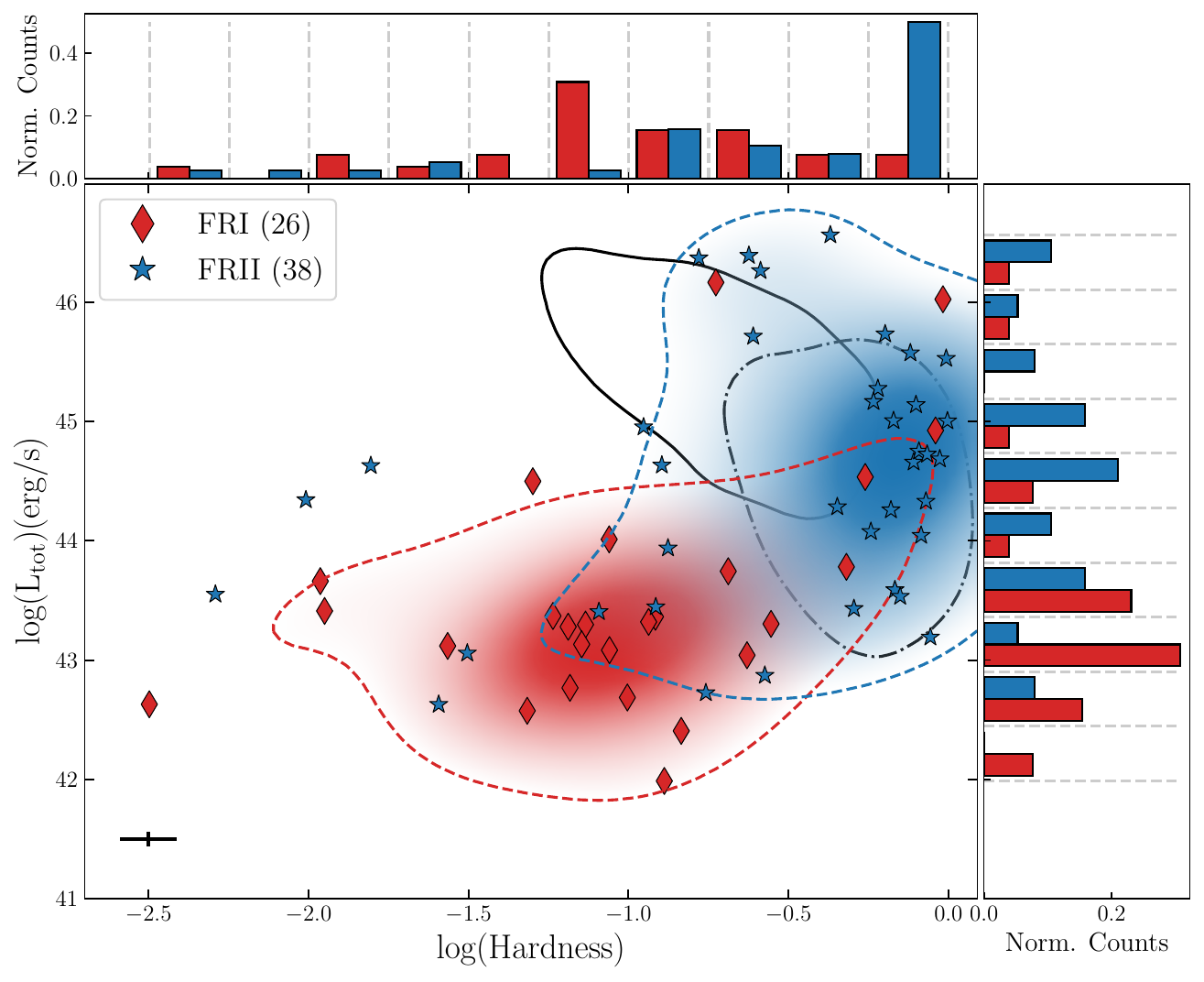}
      \caption{}
      \label{fig:hid_fr_b}
    \end{subfigure}
   \caption{HID for FR I (red diamonds) and FR II (blue stars) sources. The histograms show the distribution of the normalized number of counts for the FR I and II sources in hardness and $L_{\mathrm{tot}}$. The mean error is shown in the lower-left-hand corner. All contours show a smoothed Gaussian KDE enclosing 68\% of the data points. \textbf{(Panel a):} Emphasis placed on the two comparison samples (described in Sect. \ref{sect:comparison_samples}), with the XMM AGN sample depicted by the solid outlined contour with blue filling and the BASS sample depicted by the dashed-dotted outlined contour with gray filling. \textbf{(Panel b):} Plot (a) but with emphasis placed on the difference between the FR I and FR II populations and their placement with respect to the comparison samples, showcased with corresponding contours and shading. We see that FR Is and FR IIs occupy different areas of the HID.}
   \label{fig:hid_fr}
\end{figure*}

\begin{figure}
    \centering
    \includegraphics[width=0.84\linewidth]{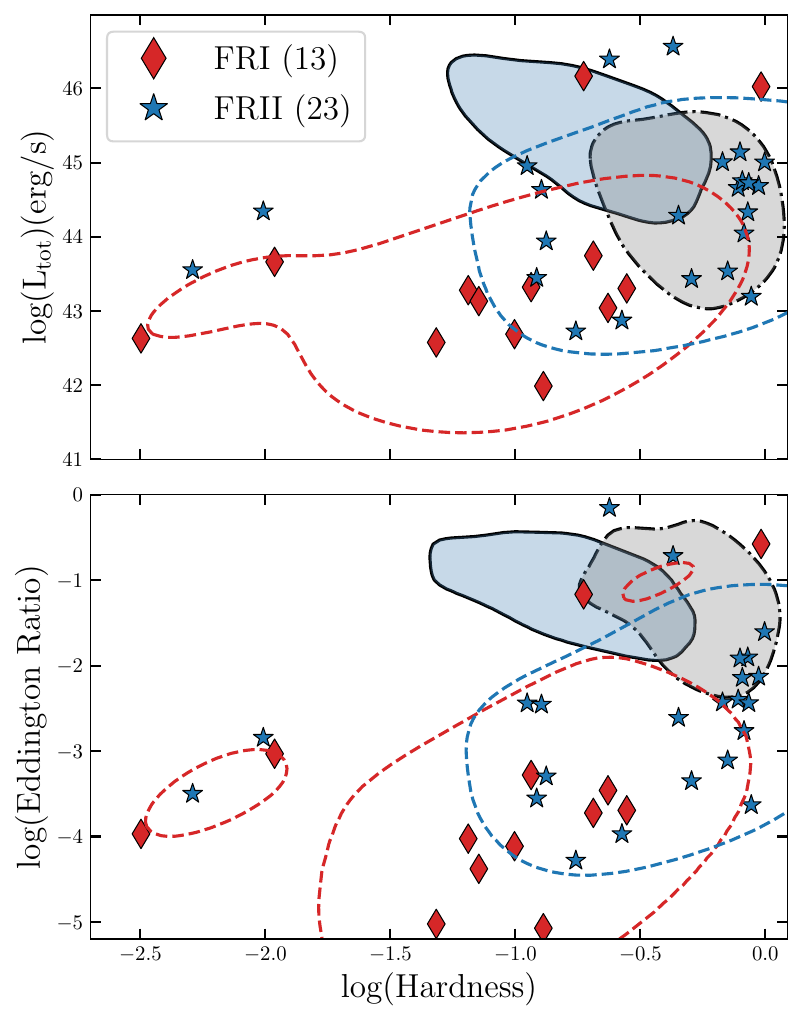}
    \caption{Investigating the effect of black hole mass on the placement of FR I and FR II sources on the HID. \textit{Top panel:} Fig. \ref{fig:hid_fr} but only for sources that have black hole mass measurements. \textit{Bottom panel:} Fig. \ref{fig:hid_fr} but with the Eddington ratio instead of luminosity on the ordinate for the sources that have black hole mass measurements.}
    \label{fig:edd_fr}
\end{figure}

\subsection{Final catalogs}\label{sect:final_cats}
After all the radio catalogs described in Sect. \ref{sect:rad_cats} went through the cross-matching process, quality cuts, and calculations, we sorted them into ``final catalogs'' that contain all sources that have a specific radio property such as morphology (FR I and II), excitation class (HERG and LERG), and extent (compact, normal, and giant). Because the radio catalogs used in this work (described in Sect. \ref{sect:rad_cats}) are independent from one another, one radio galaxy could appear in multiple catalogs. Thus when creating these final catalogs, we must account for duplicate entries.

To account for duplicate entries, we first chose a radio property of interest (morphology, excitation class, extent) and isolated all sources that have this classification. Then we identified any sources in this new catalog that matched to the same X-ray source based on the catalog X-ray name (i.e., for the \xmm\ catalog: 4XMM JRA+Dec, and for the \swift\ catalog: 2SXPS JRA+Dec). Then we do the same based on the UV catalog name (i.e., for the \xmm\ catalog: XMMOM JRA+Dec, and for the \swift\ catalog: \textit{Swift} UVOT JRA+Dec). And lastly, we identify any sources that have radio coordinates within 2\as\ of one another. We note that the number of matches does not change if we use a radio matching radius of 2\as\ or 5\as.\

At each duplicate check point (X-ray name, UV name, and radio RA and Dec) when duplicates are found, we performed a mean aggregate for numerical values and we treat the radio property aggregate in the following way. If there are more than two sources to combine into one, we took the majority classification. If there are exactly two classifications, we took both. If a source with multiple source morphology classifications contains a hybrid classification (e.g., FR I--II), we chose the classification of the other classifications (e.g., if a source has hybrid and FR I classifications, we chose FR I). Additionally, some radio catalogs have certainty assessments as to how confident the authors are in their morphology classification \citep[e.g.,][]{Miraghaei17,ROGUEI20}. If an uncertain source is one of two classifications, we took the more confident classification as the final classification. We do this duplicate resolution procedure for the final catalogs of morphology (FR I and II), excitation class (HERG and LERG), both morphology and excitation class, and extent (compact, normal, giant) sources. The final catalog counts are detailed in Table \ref{tab:classes_tbl} and their redshift distributions are shown in Fig. \ref{fig:z_dist}.

\section{Results}\label{sect:results}
We explored whether jet morphologies (FR I and II), excitation class (HERG and LERG), and radio extent (compact to giant) reflect specific XRB spectral states through inspecting their placement on the HID. 

\begin{figure*}
    \begin{subfigure}{.5\textwidth}
      \centering
      \includegraphics[width=0.99\linewidth]{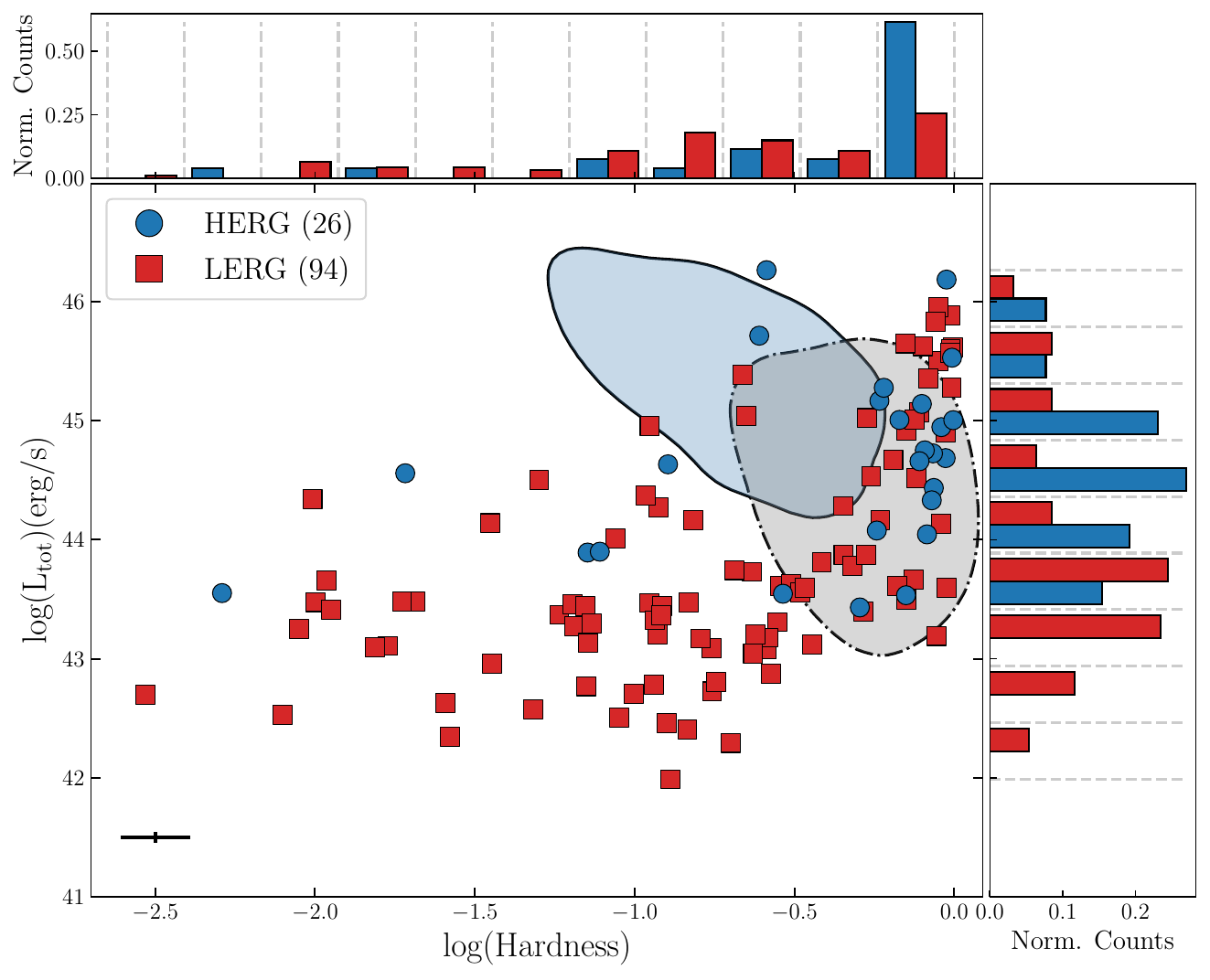}
      \caption{}
      \label{fig:hid_hl_a}
    \end{subfigure}
    \begin{subfigure}{.5\textwidth}
      \includegraphics[width=0.99\linewidth]{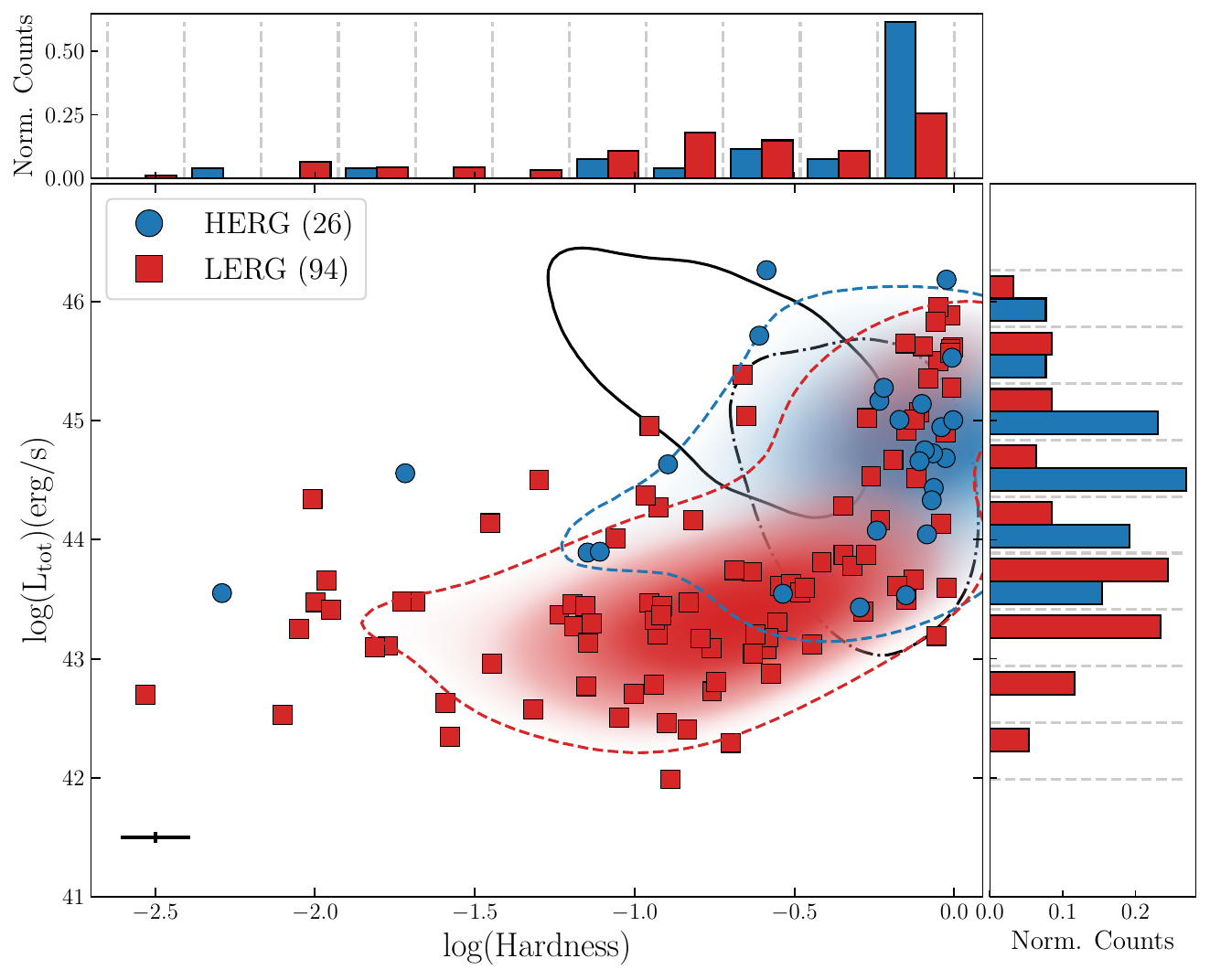}
      \caption{}
      \label{fig:hid_hl_b}
    \end{subfigure}
   \caption{HID for HERG (red squares) and LERG (blue circles) sources. The histograms show the distribution of the normalized number of counts for the HERGs and LERGs in hardness and $L_{\mathrm{tot}}$. The mean error is shown in the lower-left-hand corner. All contours show a smoothed Gaussian KDE enclosing 68\% of the data points. \textbf{(a):} Emphasis placed on the two comparison samples (described in Sect. \ref{sect:comparison_samples}), with the XMM AGN sample depicted by the solid outlined contour with blue filling and the BASS sample depicted by the dashed-dotted outlined contour with gray filling. \textbf{(b):} Plot (a) but with emphasis placed on the difference between the HERG and LERG populations and their placement with respect to the comparison samples, showcased with corresponding contours and shading. We see that HERGs and LERGs occupy different areas of the HID.}
   \label{fig:hid_hl}
\end{figure*}

\begin{figure}
    \centering
    \includegraphics[width=0.83\linewidth]{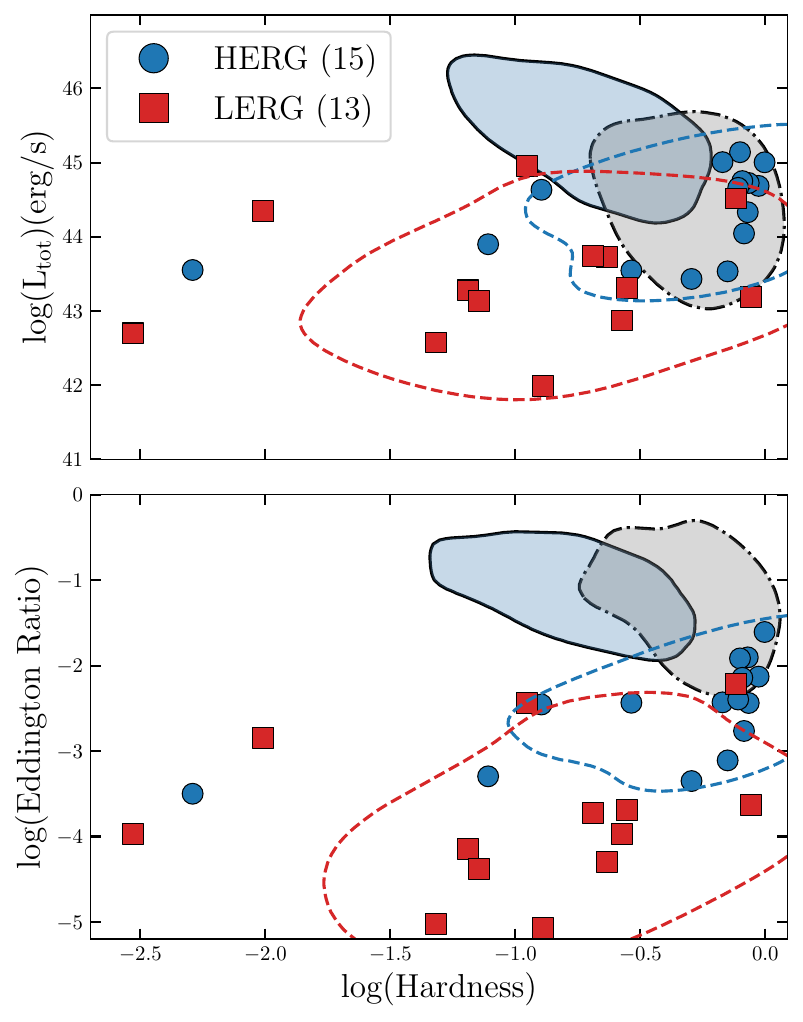}
    \caption{Investigating the effect of black hole mass on the placement of HERG and LERG sources on the HID. \textit{Top panel:} Fig. \ref{fig:hid_hl} but only for sources that have black hole mass measurements. \textit{Bottom panel:} Fig. \ref{fig:hid_hl} but with the Eddington ratio instead of luminosity on the ordinate for the sources that have black hole mass measurements.}
    \label{fig:edd_hl}
\end{figure}

\begin{figure*}
    \begin{subfigure}{.5\textwidth}
      \centering
      \includegraphics[width=0.99\linewidth]{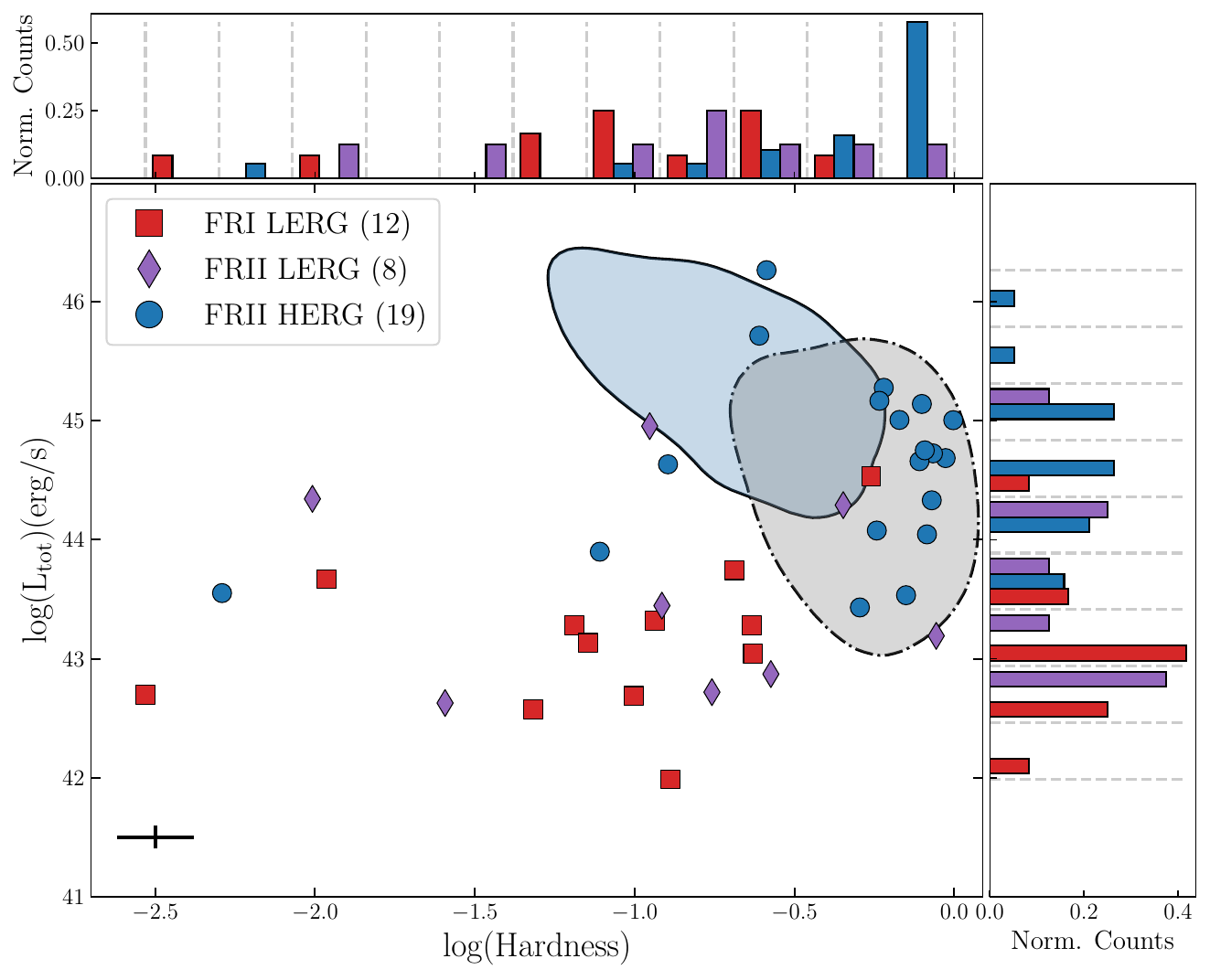}
      \caption{}
      \label{fig:hid_fr_hl_a}
    \end{subfigure}
    \begin{subfigure}{.5\textwidth}
      \includegraphics[width=0.99\linewidth]{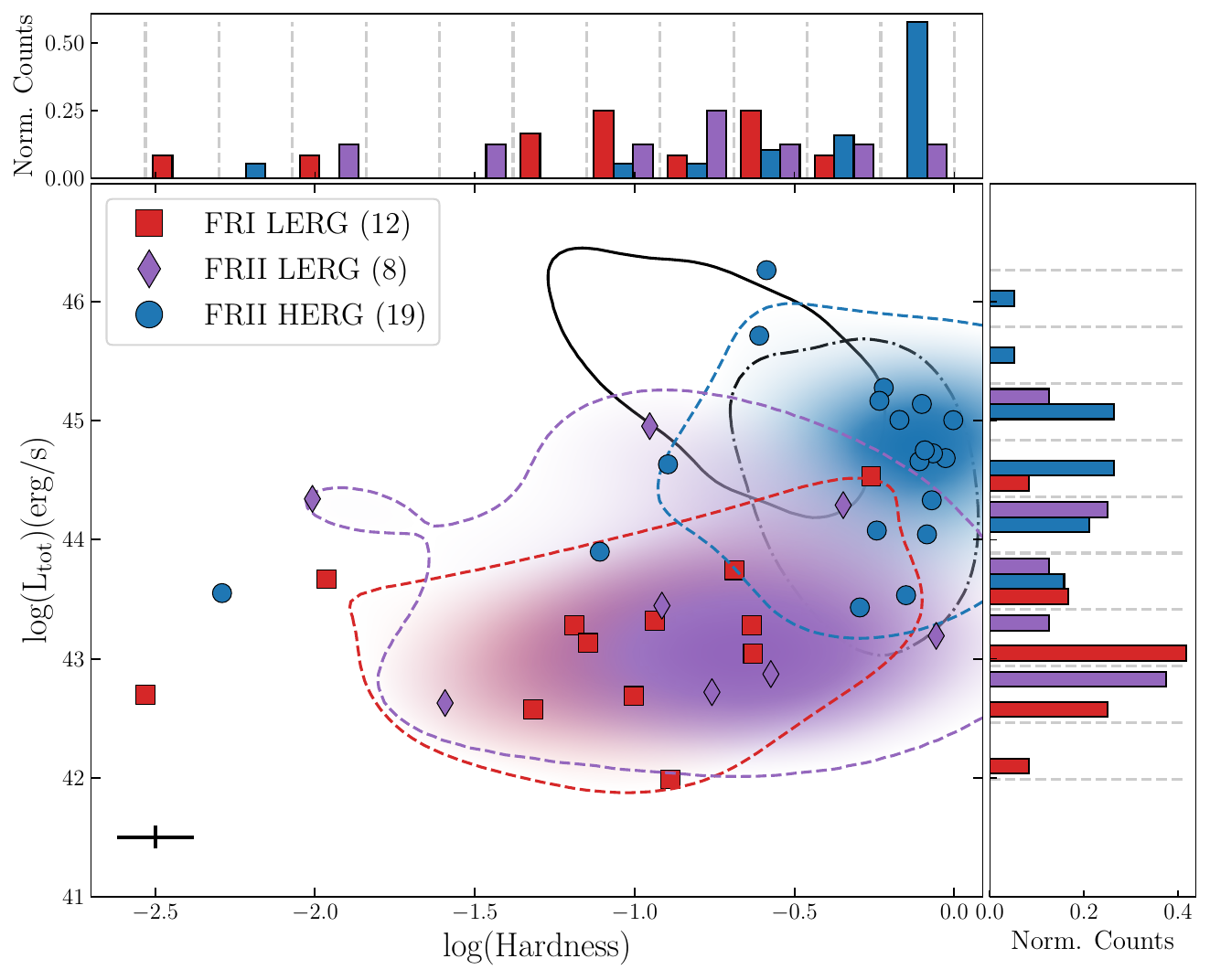}
      \caption{}
      \label{fig:hid_fr_hl_b}
    \end{subfigure}
   \caption{HID for FR I--II HERG--LERG sources. In both plots, FR I-LERGs are plotted as red squares, FR II-LERGs are purple diamonds, and FR II-HERGs are blue circles. The histograms show the distribution of the normalized number of counts for different classes in hardness and $L_{\mathrm{tot}}$. All contours show a smoothed Gaussian KDE enclosing 68\% of the data points. The mean error is shown in the lower-left-hand corner. \textbf{(a):} Emphasis placed on the two comparison samples, with the XMM AGN sample depicted by the solid outlined contour with blue filling and the BASS sample depicted by the dashed-dotted outlined contour with gray filling. \textbf{(b):} Plot (a) but with emphasis placed on the difference between the FR I-LERG, FR II-LERG, and FR II-HERG populations and their placement with respect to the comparison samples, showcased with corresponding contours and shading. We see a clear separation between FR I-LERGs and FR II-HERGs.}
   \label{fig:hid_fr_hl}
\end{figure*}

In Figs. \ref{fig:hid_fr}-\ref{fig:hid_ext} we display smoothed Gaussian kernel density estimation (KDE) contours that enclose 68\% of the data points to guide the eye. These contours were created using \verb|scipy.stats.kde.gaussian_kde|, which is a representation of a kernel-density estimate using Gaussian kernels (i.e., a KDE using a mixture model with each point represented as its own Gaussian component). Additionally, in these figures we compare AGNs with the aforementioned radio properties to the two comparison samples described in Sect. \ref{sect:comparison_samples}, showing the smoothed KDE distribution of the \xmm\ comparison sample and the smoothed KDE distribution of the BASS comparison sample. 

In Figs. \ref{fig:hid_fr}, \ref{fig:hid_hl}, \ref{fig:hid_fr_hl}, and \ref{fig:hid_ext} we also display the mean error in the lower-left-hand corner of the plots. This uncertainty reflects the propagation of the error of the X-ray and UV fluxes. We note that this error does not include uncertainty in $\Gamma$. The mean error excludes those that do not have errors in the X-ray flux in the original radio catalog.

\subsection{Morphology: FR I and II}\label{sect:hid_fr}
First, we explore whether FR Is and IIs lie in distinct areas of the HID. The FR I and II sample contains a total of 64 unique sources compiled from the FRXCAT, GRG\_catalog, Gendre+10, Macconi+20, Mingo+19, Miraghaei+17, and ROGUE I catalogs (see Table \ref{tab:classes_tbl}). The redshift distribution of these sources is shown in the upper left of Fig. \ref{fig:z_dist}. It should be noted that we exclude any sources that were marked as ``Small'' in \cite{Mingo19} since the FR classifications are not reliable for these sources. We do however include sources from the GRG\_catalog as the FR classifications for this catalog are reliable.

In Fig. \ref{fig:hid_fr} we show FR Is and FR IIs, with the corresponding colored histogram in \logL{} and \logh{} shown in the side panels. From Fig. \ref{fig:hid_fr}, we see that FR Is populate a different area of the HID than the FR IIs. In this plot, FR Is have a range of hardness values and generally have lower luminosity values, whereas the FR IIs tend to have harder values and are more luminous. Though there is a region of overlap between the contours of the two classes, we clearly see a separation between the two classes. A Kolmogorov-Smirnov (KS) test on the distribution of $\log(\mathrm{hardness})$ and $\log(L_{\mathrm{tot}})$ values individually, both give a probability lower than 0.1\% (p-value < 0.001; see Table \ref{tab:stats}) and we can reject the null hypothesis that the two samples were drawn from the same parent sample (as 0.01\% is much lower than the cut-off of 5\% typically quoted for statistical significance). To test the effects induced on the KS test from the weights of individual distribution tails, we perform an Anderson-Darling (AD) test, which yields consistent results. Both tests thus indicate that the distributions are different.

The distinction between the two populations is clear in luminosity (see Table \ref{tab:stats}), but the results for hardness are more complicated. Low-luminosity sources are more prone to host-galaxy contamination in the soft (E $\lesssim$ 3 keV) band with the net effect of appearing softer than the true intrinsic accretion state. A full discussion of this is provided in Sect. \ref{sect:disc}. 

There are many factors that could affect the placement of these radio AGNs on the HID. To test the relation with black hole mass, we investigate the HID with the Eddington ratio on the ordinate instead of luminosity. The Eddington ratio is a mass normalized luminosity and it thus more appropriate for comparing to XRBs where the black hole mass range is typically much smaller than that of AGNs. Thus, we selected the 36 sources that have black hole mass measurements (either from the original catalog or matched with \citealt{Rakshit20} within 5\as), calculated the Eddington luminosity via the standard equation $L_{\mathrm{Edd}}=1.26\times 10^{38}\times (M/M_{\mathrm{BH}})$ erg/s, and calculated the Eddington ratio using the total luminosity ($L_{\mathrm{tot}}$/$L_{\mathrm{Edd}}$). In Fig. \ref{fig:edd_fr}, we find that the populations still occupy different areas of the HID and thus this separation is not simply due to the difference in black hole mass or observational biases. Again, a KS and AD tests on the distribution of $\log(\mathrm{hardness})$ and $\log(\mathrm{Eddington~Ratio})$ values individually, both give a probability $\sim$1\% for the FR Is and FR IIs and we can reject the null hypothesis that the two samples were drawn from the same parent sample. 

\subsection{Excitation class: HERG and LERG}\label{sect:hid_hl}
Next, we explored whether HERGs and LERGs lie in distinct areas of the HID. Our final sample of HERGs and LERGs has 120 sources from the Best+12, Ching+17, GRG\_catalog, Liao+20\_I, and Macconi+20 catalogs (see Table \ref{tab:classes_tbl}). The redshift distribution of these sources is shown in the upper right of Fig. \ref{fig:z_dist}. It is to be expected that there are fewer HERGs as it is well known that HERGs are rarer \citep[5-15\% compared to the 85-95\% LERGs;][]{Best12,Ching17}. In Fig. \ref{fig:hid_hl}, HERGs and LERGs are shown with the correspondingly colored histogram distributions in \logL{} and \logh{} shown in the side panels. 

From Fig. \ref{fig:hid_hl}, we see that the HERGs and LERGs do occupy different areas of the HID. The LERGs are more widely distributed in both hardness and $L_{\mathrm{tot}}$, whereas the HERGs are harder and more luminous. A \kst\ shows that the HERGs and LERGs are statistically different populations in both \logh\ and \logL\ (p-value < 0.005; see Table \ref{tab:stats}) and an AD test results in a consistent outcome. However, it can be seen by eye in Fig. \ref{fig:hid_fr} that the difference between the areas that each population occupies is not as distinct as the difference between the FR I and II populations. This can also be quantitatively seen in Table \ref{tab:stats} as the difference in both hardness and $L_{\mathrm{tot}}$ between the HERGs and LERGs is smaller than that of the FR I and II populations. Both populations occupy the higher-luminosity and harder portion of the HID, but only the LERGs occupy the lower-luminosity and softer portion of the HID.

As with the FR morphology sources, we explore the effect of Eddington ratio on the ordinate instead of luminosity. For the 28 HERG--LERG sources that have black hole mass measurements, in Fig. \ref{fig:edd_hl} we find that there is stronger visual evidence for these populations occupying different areas of the HID. A \kst\ reveals that the HERGs and LERGs in the Eddington ratio versus hardness plot are not from the same parent populations and thus statistically different populations (p-value < 0.01). An AD test results in a consistent outcome. However, we acknowledge that we are working with small number statistics with this Eddington ratio analysis.

\begin{figure*}
    \centering
    \includegraphics[width=0.8\linewidth]{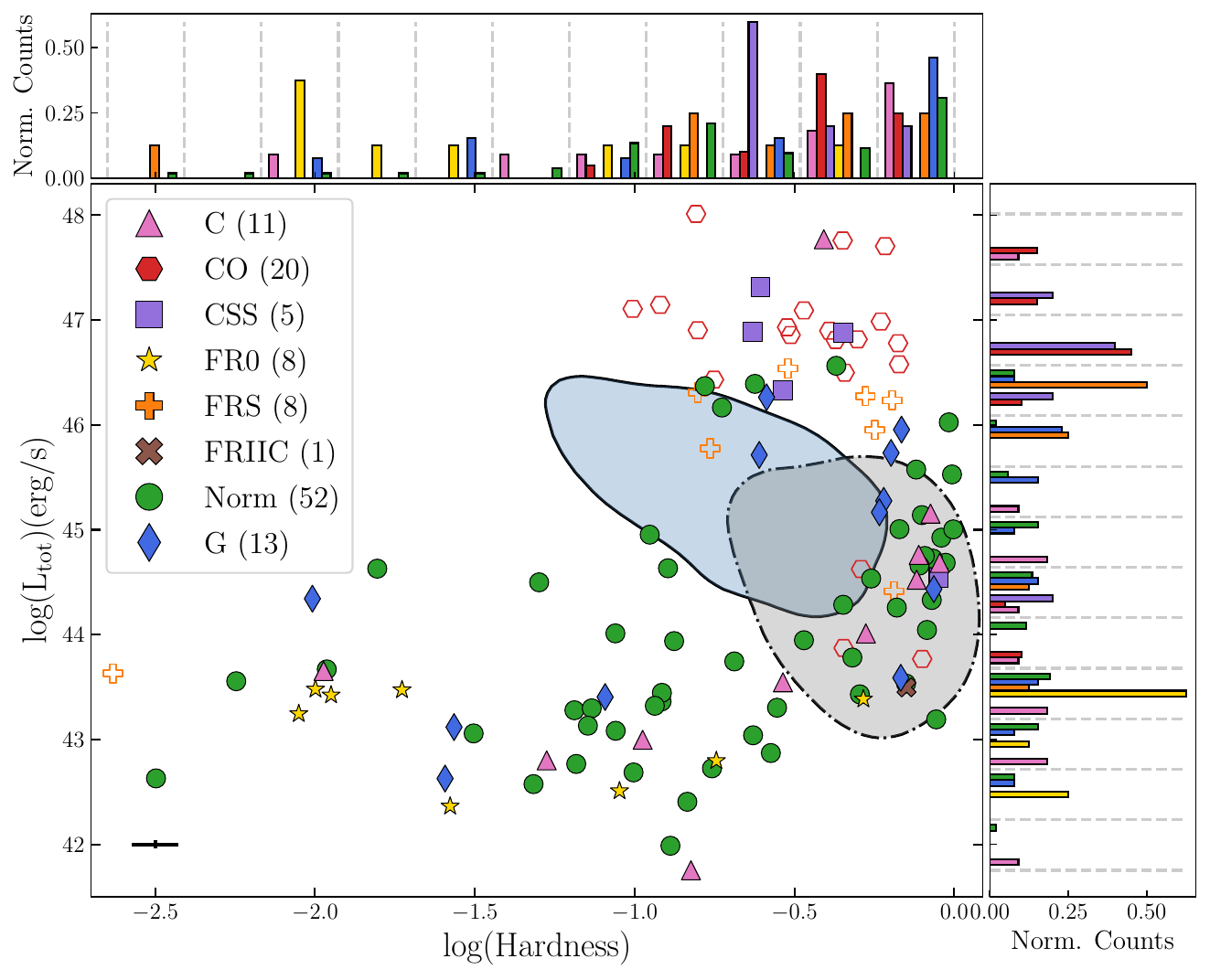}
    \caption{Investigating the placement of radio AGNs with various linear, radio extents on the HID. The color reflects the linear extent classification (see Sect. \ref{sect:hid_ext}): C, CO, CSS, FR0, FRS, and FRIIC are all compact classes, ``G'' stands for GRGs, and ``Norm.'' indicates those with a linear extent between compact and giant. The hollow points are those that could be blazars (see Sect. \ref{sect:disc} for more details). The histograms show the distribution of the normalized number of counts for different classes in hardness and $L_{\mathrm{tot}}$. The mean error is shown in the lower-left-hand corner. The two comparison samples are included and shown as in Figs. \ref{fig:hid_fr_a} and \ref{fig:hid_hl_a}. We see no conclusive evidence for a separation between the various populations.}
    \label{fig:hid_ext}
\end{figure*}

\subsection{Morphology and excitation class: FR I and II HERG and LERG}\label{sect:hid_fr_hl}
Radio AGNs do not have to be classified only according to {either} their morphology {or} excitation class, but instead they can be classified according to both: FR II-HERG, FR II-LERG, and FR I-LERG (FR I-HERGs are extremely rare). We investigate the distribution of FR I and IIs that also have a HERG or LERG classification on the HID. The final sample of sources with FR+HERG--LERG classifications has 39 sources in total from FRXCAT, GRG\_catalog, Macconi+20, and Miraghaei+17. The redshift distribution is shown in the lower left of Fig. \ref{fig:z_dist}. In Fig. \ref{fig:hid_fr_hl}, FRI-LERGs, FR II-LERGs, and FR II-HERGs are shown with the correspondingly colored histogram distributions in \logL{} and \logh{} shown in the side panels. 

From Fig. \ref{fig:hid_fr_hl}, we see that the FR II-HERGs clearly separate from FR I-LERGs. A \kst\ shows that the FR II-HERGs and FR I-LERGs are statistically different populations in both \logh\ and \logL\ (p-value < 0.0002; see Table \ref{tab:stats}) and an AD test results in a consistent outcome. Additionally, the FR II-HERGs and FR I-LERGs have a larger mean difference in \logh\ (0.71) and \logL\ (1.47) than the FR Is and IIs (\logh\ 0.46 and \logL\ 0.97; see Table \ref{tab:stats}). And in fact, the FR II-HERGs and FR I-LERGs have the largest average separation of the populations listed in Table \ref{tab:stats}. We conclude then that combining both FR and excitation class seems to differentiate populations of radio AGNs the most in the HID. FR II-LERGs seem to be an ``intermediate'' population that has overlapping values with both FR II-HERGs and FR I-LERGs but mainly overlaps with FR I-LERGs. 

\bgroup
\def\arraystretch{1.25}
\begin{table*}
\centering
\caption{Results of KS statistical tests between various radio characteristic populations.}
\label{tab:stats}
\begin{tabular}{c|ccccc}
    \hline\hline 
    \textbf{Pop.+value} & $\bm{\langle \mathrm{Pop.~A} \rangle}$ &  $\bm{\langle \mathrm{Pop.~B} \rangle}$ & $\bm{|\langle \bm{\mathrm{A}} \rangle - \langle \bm{\mathrm{B}} \rangle|}$ & \textbf{KS} & \textbf{p-value} \\
    \hline
    FR+$\log(\mathrm{hardness})$ &      FRI:-1.02 &      FRII:-0.56 &       0.46 &     0.49 & 0.000642 \\
     FR+$\log(L_{\mathrm{tot}})$ &      FRI:43.54 &      FRII:44.51 &       0.97 &     0.50 & 0.000510 \\
    HL+$\log(\mathrm{hardness})$ &     HERG:-0.42 &       LERG:-0.8 &       0.38 &     0.40 & 0.002018 \\
     HL+$\log(L_{\mathrm{tot}})$ &     HERG:44.65 &       LERG:43.8 &       0.85 &     0.50 & 0.000043 \\
  FRHL+$\log(\mathrm{hardness})$ &  FRI LERG:-1.1 & FRII HERG:-0.39 &       0.71 &     0.76 & 0.000124 \\
   FRHL+$\log(L_{\mathrm{tot}})$ & FRI LERG:43.16 & FRII HERG:44.63 &       1.47 &     0.76 & 0.000124 \\
    \hline
\end{tabular}
\tablefoot{\bi{Column 1:}  Population name and the value that is put into a two-sample statistical test. FR = morphology classes FR I and FR II. HL = HERG and LERG. FRHL = morphology classes FR I and II along with HERG and LERG classifications. \bi{Column 2:} Population classification for population A and the mean $\log(\mathrm{hardness})$ or $\log(L_{\mathrm{tot}})$ value for that population. Population A classification: mean value. \bi{Column 3:} Population classification for population B and the mean $\log(\mathrm{hardness})$ or $\log(L_{\mathrm{tot}})$ value for that population. Population B classification: mean value. \bi{Column 4:} Absolute value of the difference between the means of populations A and B. \bi{Column 5:} KS statistic for populations A and B. \bi{Column 6:} p-value for the KS test.}
\end{table*}
\egroup

\subsection{Radio extent: Compact to giant}\label{sect:hid_ext}
Lastly, we investigated whether populations of radio sources with different linear extents lie in distinct areas of the HID. We separated the sources into several classes: compact, ``normal,'' and giant. ``Compact'' encompasses many different types of radio sources, and thus we have several subclasses. We define a ``C'' class (compact) for sources that are officially labeled as a CSO, GPS, or HFP and have jets whose linear size is $<$ 1 kpc as identified by \cite{Chandola20}, \cite{Kosmaczewski20}, \cite{Liao20a}, \cite{Liao20b}, and \cite{Sobolewska19}. Then we defined sources that are identified in the parent catalog as compact (C or C$^*$ in Gendre+10 or C in Miraghaei+17) but do not have a detailed classification (e.g., CSO, CSS, GPS, or HFP from \citealt{Sobolewska19}, \citealt{Kosmaczewski20}, and  \citealt{Liao20a}) as ``CO'' (``compact other''). We kept CSS sources separate as they have been suggested to be young FR II sources \citep{Odea98}. Most of the CSS sources are from Liao+20\_I or Liao+20\_II. We kept all FR0s that are all from FRXCAT in a class on their own. We defined the sources in \cite{Mingo14} that had an FR classification but were labeled as ``small''  to be ``FRS'' (``FR small''). We defined any source from the FRXCAT that was originally from the COMP2CAT \citep{JM19} and are FR IIs that have extents that do not exceed 60 kpc as ``FRIIC'' (``FR II compact''). Then we used the classification ``Norm'' (``normal'') to indicate radio AGNs that have an FR classification and have not been identified as either compact or giant. These ``Norm'' sources are from the FRXCAT, Gendre+10, Macconi+20, Mingo+19, Miraghaei+17, and ROGUE I catalogs. And lastly, we use ``G'' (``giant'') to define the GRGs from the GRG\_catalog whose linear extent is $>$ 0.7 Mpc. Generally, the compact sources have linear extents $\lesssim$5 kpc, giant sources have extents greater than 0.7 Mpc, and normal sources are in between. These classifications (``C,'' ``CO,'' ``CSS,'' ``FR0,'' ``FRS,'' ``FRIIC,'' ``Norm,'' and ``G'') are all shown in Fig. \ref{fig:hid_ext} with the correspondingly colored histogram distributions in \logL{} and \logh{} shown in the side panels.

If any source had a compact classification (``C'' or ``CSS'') alongside a normal classification, the final classification was normal. If there were sources with other classifications that matched with the ``CO'' sources, the other classification is chosen as it is more robust. And the ``FR0'' classification is chosen over ``FRS.'' The redshift distribution of all these sources is shown in the lower right of Fig. \ref{fig:z_dist} and one can see that we have a few higher redshift sources when compared to the previously analyzed samples in this work.

In Fig. \ref{fig:hid_ext}, we see no conclusive evidence for a separation between the various populations. One might think that there is a gathering of the compact sources to the upper right-hand corner of the diagram, but some of the sources in the ``CO'' and ``FRC'' classes could be blazars which complicates the conclusions (see Sect. \ref{sect:disc}). We note that there seems to be two different subpopulations of the compact and giant populations; the high-luminosity sources and the low-luminosity sources.

\section{Discussion}\label{sect:disc}

\subsection{Discussion of distinct populations and relation of radio AGN properties to XRB spectral states}
\subsubsection{FR morphology}
When XRBs begin an outburst, they start in the low-hard state with a weaker jet. Then, as the outburst continues the luminosity increases, the jet strengthens, and the source is then in the hard state. If AGNs have analogous spectral states to XRBs, our results from Figs. \ref{fig:hid_fr} and \ref{fig:edd_fr} indicate that FR Is are located in the AGN state diagram where weaker (low-hard) states of XRBs are present and FR II sources are located where stronger (hard) states of XRBs are present. Thus, the FR Is could be the analogs of the early stages of the outburst (with a weaker jet) and the FR IIs could be analogous to the later stages of the outburst (with a brighter, more powerful jet). The strongest similarity between FR morphology and XRB outburst evolution is seen in Fig. \ref{fig:edd_fr} as XRBs are known to go from low Eddington ratio to higher Eddington ratio during an outburst. Our results may suggest that not only environment plays a role in radio-AGN jet morphology, but on average, the effect of the central engine might be significant as well. 

\subsubsection{Excitation class (HERG--LERG)}
A property of radio AGNs that we might expect to reflect different XRB spectral states is excitation class (HERGs and LERGs). HERGs are efficient accretors that accrete between one per cent and ten per cent of their Eddington rate whereas LERGs are inefficient accretors that accrete at a rate below one per cent of their Eddington rate \citep{Best12}. \cite{Best12} suggest that the population dichotomy is caused by a switch between radiatively efficient and radiatively inefficient accretion modes at low accretion rate, which is consistent with synthesis models for AGN evolution \citep{Merloni08}. They show that although LERGs dominate at low radio luminosity and HERGs begin to take over at $L_{\mathrm{1.4\,GHz}}\approx10^{26}$~W/Hz, examples of both classes are found at all radio luminosities.

The placement of the LERGs and HERGs in the HID presented in Fig. \ref{fig:hid_hl} is roughly in line with what we expected in an AGN-XRB analog. When XRBs start an outburst in the low-hard state, there is not much radiation from the central engine. But during the transitions to high-hard state and to the soft state, the amount of radiation from the central engine increases. In the transition to the hard state this is seen as an increase in total luminosity, and in the transition to the soft state this is seen as a redistribution of flux from hard X-rays to soft X-rays, with the soft X-rays corresponding to UV emission in AGNs, which are more efficient in ionizing optical lines. Thus, the HERGs could be the analog of radio AGNs ramping up in accretion power and could be the analog of  XRBs transitioning to either the high-hard or soft state. Additionally, the LERGs that overlap with the HERGs could be sources that are also increasing in accretion power and ultimately X-ray luminosity and the lower-luminosity LERGs could be the initial state. We might have expected there to be a more prominent difference in the areas occupied by the HERG--LERGs in the HID than the FR classifications as the latter can be influenced by the large-scale environment \citep{Kaiser07}. 

Both \xmmcat\ and \swiftcat\ use an absorbed power law to approximate the spectrum in flux calculation (see Sects. \ref{sect:xmm_simult} and \ref{sect:swift_UX} for details). The used values ($N_H\sim$10$^{20}$ and $\Gamma\sim$ 1.7) are representative values for AGNs. Together with our exclusion of the most absorbed sources, this approximation should be sufficient for our entire sample. However, some LERGs might deviate from the spectral shape built into these assumptions, as LERGs mostly show soft X-ray emission arising from the jet \citep[see, e.g.,][]{Hardcastle99,Mingo14}. Their hard X-ray fluxes, and thus also hardness, might be slightly overestimated. Any improvement of their count-to-flux conversion would thus separate them more from HERGs, and thus this uncertainty will not affect our main conclusions.

We note that the widely spread distribution of LERGs compared to HERGs could be due to the fact that there are more LERGs in our sample than HERGs. However, there are typically many more LERGs than HERGs in the larger LERG--HERG parent samples in this work (95\% LERGs and 5\% HERGs \citealt{Best12}; 88\% LERGs and 13\% HERGs \citealt{Ching17}) and our ratio of HERGs to LERGs reflects this (78\% LERGs and 22\% HERGs). Additionally, we note that because we are working with catalog values and not SED fits, we have an upper limit on the LERGs placement in the HID. We would in particular expect a host galaxy correction to move the LERGs down in luminosity and to softer values, and, due to their low luminosity, we would expect a detailed X-ray analysis to move the sources even further down in luminosity and to harder values. We already see evidence for a separation and would thus expect this separation to increase more with detailed spectral analysis.

We also note that there are no HERGs in our sample with Eddington ratios  > 10\%   (Fig. \ref{fig:edd_hl}). To investigate whether we have not underestimated the total luminosity, we compared our $L_{\mathrm{tot}}$ values with the radiatively efficient bolometric luminosities of these sources to determine if the paucity of sources at higher Eddington ratios is due to underestimating the source luminosity. We cross-matched the HERG--LERG sources with SDSS DR16 to obtain [OIII] fluxes for 85 sources. We then calculated the radiatively efficient bolometric luminosity using the [OIII] relation from \cite{Heckman04}, namely $L_{\rm rad,\,bol} = 3500 \times L_{\rm [OIII]}$. We find that for these 85 HERG--LERG sources the $L{tot}$ and radiatively efficient bolometric luminosity measurements agree within 1\%, indicating the lack of highly accreting HERGs is not due to an underestimation of the radiative luminosity. An alternative possibility is that one of the major parent samples of our HERG--LERG sample \citep{Best12} also contained this deficit of high-Eddington-ratio HERGs, which may have been inherited by our sample. Thus we cannot make any conclusions on the relative proportions of high-Eddington-ratio HERGs in the global population from our sample. Future large spectroscopic surveys (e.g., 4MOST) may be able to determine if this is an underlying intrinsic feature of the population or not.

\subsubsection{FR + HERG--LERG}
Radio AGNs can be classified according to both their morphology and excitation class. These classifications and investigations into them can provide insights into the jet and accretion interplay in radio AGNs. Historically, HERGs are associated with FR IIs and LERGs are associated with FR Is. However, there exists an interesting population of FR II-LERGs that broke this historical understanding. Recently, several observational works have investigated the properties of these classes to determine how fundamentally different they are and if any difference is intrinsic or extrinsic \citep{Macconi20,Grandi21}. It is thought that FR II LERGs are a complex population, including at least some evolved or fading FR II HERGs and FR I-like jets in lower-mass hosts \citep[][Mingo et al. in prep.]{Mingo19,Macconi20,Grandi21}. \cite{Garofalo10} presents a model in which FR II-LERGs are aged FR II-HERGs and \cite{Grandi21} finds observational evidence to support this. And \cite{Macconi20} finds that FR II-LERGs have intermediate properties between FR Is and FR II-HERGs.

Next we wanted to determine which classification matters more, the FR or HERG--LERG classification, for the placement of a radio AGNs on the HID. In Sects. \ref{sect:hid_fr} and. \ref{sect:hid_hl} we find that the FR classification seems to have a stronger influence on source placement than excitation class. But from Fig. \ref{fig:hid_fr_hl}, we see that combining these classifications reveals the strongest separation and strongest analog with the XRB HID. It seems that the picture is not particularly clear when one looks simply at one morphology or excitation classes, but when both characteristics are included, the population separation becomes more distinct. 

From Fig. \ref{fig:hid_fr_hl}, we also see that FR II-LERGs seem to have overlapping and intermediate properties between FR I-LERGs and FR II-HERGs similar to what \cite{Macconi20} find. However, the overlap of FR II-LERGs is more with FR I-LERGs than FR II-HERGs. This overlap seems to indicate that the HERG--LERG classification matters more for their placement on the HID, which is contrary to what is seen in Sects. \ref{sect:hid_fr} and \ref{sect:hid_hl}, where FR morphology seemed to distinguish the location of the sources in the HID more prominently. This muddles the answer to the question of which characteristic matters more for the placement of a source in the HID. But it is quite clear from Fig. \ref{fig:hid_fr_hl} that combining morphology and excitation class creates the most distinct populations in the HID. However, we are aware that in Sect. \ref{sect:hid_fr_hl} we are working with small number statistics. It is necessary to increase the sample size to draw definite conclusions. 

\subsubsection{Extent}
It has been suggested that the linear extent of radio jets in radio AGNs can be a sign of evolutionary progression; compact radio galaxies indicate young sources and GRGs indicate evolved, older sources \citep{Subrahmanyan96,Odea21}. However, the size of compact radio AGNs could be due to confinement or of transient nature and the size of GRGs could be attributed to extraordinarily powerful jets, a low-density environment, an old source, or restarted AGN activity. There is still uncertainty as to the origin of the linear extent of these different classes (compact and giant) and is an active area of research. In order to investigate this further, we separated the radio AGNs into categories based on their radio jet extent, placed them on the HID, and compared them to XRBs.

In Fig. \ref{fig:hid_ext}, we find no clear evidence for a separation between different populations of radio AGNs with different linear extents. There might be two different populations of compact sources; the high-luminosity sources and the low-luminosity sources. However, some of the high-luminosity compact sources could be blazars (indicated as hollow shapes in Fig. \ref{fig:hid_ext}). Some of the radio catalogs in this work do explicitly remove blazars. \cite{Liao20a} and \cite{Liao20b} explicitly remove blazars through variability and spectral shape analysis. The CSOs in \cite{Sobolewska11} and \cite{Kosmaczewski20} are not expected to be blazars since typical edge-on configurations of such objects prohibit a face-on configuration \citep{Sobolewska19}. The multiwavelength properties of FR0s provide evidence that FR0s are a stand alone class of objects beyond FR Is, FR IIs, and blazars \citep{Baldi18}. However, the small/compact sources from \cite{Gendre10,Miraghaei17,Mingo19} could indeed be blazars indicated in Fig. \ref{fig:hid_ext} in as these compact objects were selected solely from their size. Thus, some of the sources in the high-luminosity compact sources could be blazars, which would mean that the $L_{P}$ would likely have contamination from the jets. 

If many of the high-luminosity compact sources are not blazars, we speculate that the high-luminosity sources could be the equivalent of XRB ballistic jets. When XRBs move from the hard state to the soft state in the HID, they are evolving from a source that has persistent jets in the hard state to having no radio jet production in the soft state. A source can sometimes then go back to an intermediate state and produce ballistic jets (when XRBs can occasionally cross back over the jet line in small cycles and produce temporary blob injections). The ballistic jets in an intermediate state can sometimes cause the source to go to even higher total luminosities. If radio AGNs are analogous to XRBs, we would expect to see the ballistic jet equivalent (compact) sources overlap with FR II sources as XRB sources that have ballistic jets overlap with sources with strong jets in the XRB HID. In Fig. \ref{fig:hid_ext}, we do in fact see that compact sources overlap with ``normal'' sources (which are usually FR IIs) with some of the overlapping compact sources being CSS sources many of which are known to be small FR II sources. On a related note, we speculate that the low-luminosity sources could be the equivalent of sources that are in the initial states of the outburst that then may evolve into an FR I or II source. 

Pertaining to other extent populations in the HID in Fig. \ref{fig:hid_ext}, we see two populations of GRGs. This could be explained by lower-luminosity GRG radio observations tracing more relic radio emission than the high-luminosity GRGs, whereas UV and X-ray observations trace more recent accretion activity which could be a restarted activity \citep{Bruni20}.

In conclusion, due to the uncertainty in the nature of the high-luminosity sources and the mixing of all the populations, we cannot say with certainty whether or not populations of radio extent are separated or not in the HID.

\subsection{Synthesis and comparison to previous studies}
It is already known that black holes in XRBs and AGNs have similar properties such as: the same fundamental plane of black hole activity \citep{Merloni03,Falcke04,Kording06a}, characteristic timescales \citep{McHardy06}, a common black hole accretion scheme \citep{Arcodia20}, and evolving jets \citep{Zhu20}. However, one of the main current research questions for those studying XRBs and AGNs is whether AGNs have spectral states that are analogous to those of XRBs. As stated in the introduction, \cite{Kording06b} and \cite{Sobolewska11} suggest that different classes of AGNs might correspond to specific spectral states of XRBs. Recent work by \cite{Fernandez21} has found evidence for this. They find that the (a) broad-line Seyferts and about half of the Seyfert 2 population, showing highly excited gas and RQ cores consistent with disk-dominated nuclei, are associated with the soft state of XRBs and (b) the remaining half of Seyfert 2 nuclei and the bright LINERs are associated with the bright hard and intermediate states of XRBs. 

Further, both systems have radio jets, and recently more works have been investigating whether the radio properties of AGNs correlate with specific XRB spectral states. \cite{Svoboda17} found that RL sources have on average higher hardness. This suggests that the AGN radio dichotomy of RL and RQ sources could indeed possibly be explained by the evolution of the accretion states. Similarly, \cite{Zhu20} found that jetted RL quasars are harder than non-jetted RQ quasars. \cite{Fernandez21} find that RL sources are associated with the hard and RQ sources are associated with the soft state. Overall, these recent works provide evidence that AGN radio-loudness does trace different AGN states.

Motivated by these findings, for the first time we investigate the placement on the HID of sources with specific radio properties of the jet and central engine of radio AGNs have been investigated. We explore whether properties of the jet (morphology and extent) and the accretion system (excitation class -- HERG or LERG) reflect discrete XRB accretion states. We find that FR jet morphologies (FR I and FR II) are distinct populations on the HID, and we find evidence that not only environment affects the jet morphologies. We find that HERGs and LERGs also occupy distinct areas of the HID, but they are not as distinct as FR morphology. However, when one considers both FR and excitation class two populations separate out clearly: FR II-HERGs and FRI-LERGs. FR II-HERG are harder and more luminous whereas FRI-LERGs are softer and less luminous. Lastly, we explore whether radio AGNs with different linear radio extents reflect specific XRB spectral states. Overall, we find that specific properties of RL AGNs do indeed correlate with the general nature of XRB spectral states.

In conclusion, it seems that AGNs do have spectral states as XRBs do to some degree and that the radio properties of AGNs do reflect the evolution of XRB states to some extent. {However, it is difficult to differentiate the AGN states and directly correlate them with exact and specific XRB states.} With jet morphology, we find that FR Is are located in the AGN state diagram where weaker (low-hard) states of XRB are present and FR II sources are located where stronger (hard) states of XRB are present. And this difference becomes more distinguished when excitation class is also taken into account. This separation between populations with specific radio properties in the AGN HID argues for a similarity with XRBs HID and thus allows us to draw parallels between XRB and AGN spectral state evolution.

\subsection{Sources of uncertainties}\label{sect:errors}
In this study there are several characteristics and quantities that could effect our results. First, we acknowledge that it is difficult to determine the true placement of low-luminosity sources ($\log(L_{\mathrm{tot}})\lesssim~43$) . \cite{Svoboda17} compared the placement of AGNs and non-active galaxies on the HID and found that the level of mixing of AGNs with non-active galaxies was generally very low, but some non-active galaxies in the sample did exhibit a luminosity exceeding $\log(L_{\mathrm{tot}})\approx43$, where low-luminosity AGNs appear. This implies that the apparent softness of low-luminosity AGNs ($\log(L_{\mathrm{tot}})<44$) may indeed be due to the contribution of the host galaxy. In the absence of extremely high spatial resolution observations, which are only available for nearby galaxies, the only reliable deconvolution of the host galaxy contamination is obtained by a detailed modeling of the AGN SED. This has been done so far only for nearby low-luminosity AGNs and there is not yet a fully accepted consensus on the level of UV host galaxy contamination to the low-luminosity AGN UV measurements \citep{Maoz07,Ho08}. A detailed analysis of the host-galaxy contribution of all low-luminosity sources in our sample is beyond the scope of this work and is the subject of other current work. \cite{Svoboda17} used empirical relations to estimate its contribution and found that when the host-galaxy contribution is corrected for, the low-luminosity sources move rightward on the HID (toward harder values). A host galaxy correction to the UV luminosity will bring the sources downward also in the HID. For simplicity, and to avoid introducing unknown systematic uncertainties, we chose not to apply a similar host galaxy correction.

The second source of uncertainty that could affect the location of the sources on the HID is uncertainty of the photon index $\Gamma$. In the entire sample of all radio galaxies in this work, the mean $\Gamma$ value  is 1.97 with a standard deviation of 0.38. The $\Gamma$ value affects the extrapolation of the 2-10 keV X-ray flux to obtain the power-law luminosity $L_P$ value. Thus, the uncertainty of $\Gamma$ translates to an uncertainty in hardness. However, we do not expect this uncertainty to significantly affect the position in the hardness-luminosity diagram. The extrapolation is done in both directions to high as well as to soft energies. An overestimated value of $\Gamma$ would add more to the soft X-ray band and less to the hard band, and vice versa for an underestimated value of $\Gamma$. Nevertheless, to minimize the uncertainty, we took values from the X-ray spectral fitting when available in the literature references (either from the radio catalog papers or from the \swift\ catalog\footnote{See \href{https://www.swift.ac.uk/2SXPS/docs.php\#sources\_flux}{https://www.swift.ac.uk/2SXPS/docs.php\#sources\_flux} for more details on the $\Gamma$ values calculated for the \swift\ catalog.}) instead of the values derived from the \xmmcat\ by comparing the flux in subsequent energy bands.

We also note that the uncertainty of $\Gamma$ constrained from the catalogs will not be larger than fixing the value of $\Gamma$ to a mean value. \cite{Macconi20} tested the effect of fixing $\Gamma$=1.7 on $L_X$ in the cases that there were poor statistics an/or the emission was complex. \cite{Macconi20} found that varying $\Gamma$ from 1.4 to 2.0 resulted in $\approx$15\% and 40\% change in the $L_X$ values, which represents the maximum uncertainty in case the value constrained from catalog fluxes is largely inaccurate.

A third source of uncertainty is intrinsic obscuration. At X-ray wavelengths, significant obscuration (e.g., $\log(N_{\rm H}) \gtrsim$~22) along the line of sight is known to exist for the majority of AGNs (e.g., \citealt{Ueda14,Buchner14,Ricci15,Buchner17}), as well as more specifically the RL population (e.g., \citealt{Wilkes13,Panessa16,Macconi20,Kuraszkiewicz21}). We already selected sources to have 2-10 keV X-ray photon indices in the range 1.5 < $\Gamma$ < 3.5 (see Sect.~\ref{sect:quality_cuts}), aiming to reject any sources with a flat spectral slope that is typical for the dominant reflection spectrum in heavily obscured AGNs (e.g., \citealt{Murphy09}). However, the possibility still exists that our sample contains obscured AGNs in which other competing spectral components are present $<$\,10\,keV (e.g., photo-ionized gas, collisionally ionized gas, bremsstrahlung emission) that act to disguise the reflection features in the observed spectrum. To completely account for this would require sensitive broadband X-ray spectral fitting of each target with physically motivated torus models, which is outside the scope of this work. We note that any correction for extra obscuration along the line of sight would act to reduce the hardness of our targets, and increase the observed bolometric luminosity -- a net correction that would move points to the upper left of the HID.

For the UV fluxes, although we correct the observed UV fluxes for reddening from the Milky Way, this will not necessarily account for any reddening arising from within the host galaxy or AGN torus of each target. The reddening (and hence intrinsic accretion disk flux) can be approximated from SED decomposition paired with photometric coverage from the far-UV to far-infrared, but such an analysis is outside the scope of our work. Similarly to X-rays, any additional UV correction would likely move points to the upper left of the HID.

A fourth source of uncertainty is the UV and X-ray variability of the AGN. In this work, we have chosen simultaneous UV and X-ray observations when possible. However, due to AGN variability, multiple pointings of simultaneous observations for the same source could result in a different location on the HID for each set of simultaneous observations. A detailed investigation into the effect of variability on the placement of sources in the HID is provided in Fig. 11 of \cite{Svoboda17} and the corresponding analysis. The variability does affect the placement of the sources in Fig. 11 of \citealt{Svoboda17}, but the average placement of the various observations still differentiates the sources from one another. Similarly with a sufficiently large sample of radio galaxies of a particular type, we expect to on average obtain the representative position in the HID of the different classes of radio galaxies and still be able to compare the different populations.

The fifth source of uncertainty is that we use a compilation of 15 different catalogs of radio sources since there is not one comprehensive catalog of all radio galaxies. Each catalog came from a different study that each had its own goals, and thus the sample is not homogeneous in nature. However, the FR, HERG--LERG, and extent definitions are independent, so these classifications are reliable. We also note that the FR classification can be different depending on the radio wavelength, though all works used either 1.4 GHz or 150 MHz. Though the selection of sources may be rather inhomogeneous, it is sufficient for this pilot work to study general trends of the placement of radio galaxies with certain radio characteristics on the HID.

We also note that the X-ray luminosity could be contaminated by contributions from the jet and hot gas (either from the intracluster medium or in the host galaxy). To determine if the X-ray emission is solely from the corona would require extensive X-ray spectral fitting and image analysis of each target, which is beyond the scope of this work.

\section{Conclusions}\label{sect:concl}
In recent years there has been mounting evidence that AGNs may have spectral states that are analogous to those of XRBs. In this work we follow the methodology of \cite{Svoboda17} and investigate whether specific properties of RL AGNs correlate with XRB spectral states. We explore if populations of radio jet morphologies (FR I and II), excitation classes (HERG and LERG), and radio jet linear extents (compact to giant) separate out in the AGN HID (total luminosity vs. hardness). We do this by cross-correlating 15 catalogs of radio galaxies with the desired characteristics from the literature with \xmm\ and \swift\ X-ray and UV source catalogs (see Sects. \ref{sect:cats} and \ref{sect:methods}). We calculated the total luminosity ($L_{\mathrm{tot}}$) and the hardness from the X-ray and UV fluxes (see Sect. \ref{sect:methods}), placed the sources on the AGN HID, and searched for the separation of populations and analogies with the XRB spectral state HID. In investigating if specific radio properties reflect distinct XRB spectral states, we find the following:
\begin{enumerate}
  \item \textbf{Radio morphology (FR I and II):} From a sample of 64 sources (26 FR Is and 38 FR IIs), there is a separation of FR Is from FR IIs at a statistically significant level (p-value < 0.05): FR Is have lower total intensity and softer hardness values, and FR IIs have higher total intensity and harder values (see Sect. \ref{sect:hid_fr}). These populations are separated in an Eddington ratio versus hardness diagram as well. Because FR Is are located in the AGN state diagram, where weaker (low-hard) states of XRBs are present and FR II sources are located where stronger (hard) states of XRBs are present, the FR Is could be the analogs of the early stages of the outburst (with a weaker jet) and the FR IIs could be analogous to the later stages of the outburst (with a brighter, more powerful jet).
  \item \textbf{Excitation class (HERG and LERG):} From a sample of 120 sources (94 LERGs and 26 HERGs), we find that the HERGs and LERGs do occupy different areas of the HID at a statistically significant level, with the LERGs being more widely distributed in both hardness and $L_{\mathrm{tot}}$ and HERGs being harder and more luminous (see Sect. \ref{sect:hid_hl}). This difference becomes more prominent with Eddington ratio versus hardness. However, the difference between the areas that each population occupies is not as distinct as the difference between the FR I and II populations. In an XRB analogy, the HERGs could be the analog of the radio AGNs ramping up in accretion power and the LERGs in the initial state.
  \item \textbf{Radio morphology (FR I and II) plus excitation class (HERG and LERG):} In a sample of 39 sources (12 FR I-LERGs, 8 FR II-LERGs, and 19 FR II-HERGs), there is a clear separation between FR I-LERGs and FR II-HERGs, with FR I-LERGs being less luminous and softer and FR II-HERGs having a higher total luminosity and being harder. The separation between FR I-LERGs and FR II-HERGs is the strongest in this work (at a statistically significant level), which indicates that combining both FR and excitation class seems to differentiate populations of radio AGNs the most in the HID. Thus, both the jet morphology and central engine need to be considered in comparisons to XRBs.
  \item \textbf{Radio extent (compact, normal, giant):} In a sample of 118 sources (53 compact sources, 52 normal, and 13 giant), we do not see any strong evidence for a separation in terms of radio linear extent between the different populations. If the compact sources are not blazars and AGNs have analogous evolutionary states to XRBs, the compact sources could be the equivalent of the ballistic jets found in XRBs.
\end{enumerate}
In conclusion, we find evidence that AGNs with specific radio properties occupy distinct areas of the HID, similar to XRBs. This adds to evidence already found by \cite{Falcke04}, \cite{Nipoti05}, \cite{Kording06b}, \cite{Merloni08}, \cite{Svoboda17}, and \cite{Fernandez21} that different RL AGN populations may follow an evolutionary track similar to the tracks followed by individual XRBs during an outburst. However, it is difficult to perfectly differentiate the AGN states and directly correlate them with specific XRB states. That said, the general separation between populations with specific radio properties in the AGN HID indicates that there is a similarity with XRBs HID and thus allows us to draw parallels between XRB and AGN evolution.

 \textit{Software used in this work:} Astropy \citep{astropy13,astropy18}, TOPCAT \citep{topcat}, Matplotlib \citep{matplotlib}, pandas \citep{pandas10,pandasZendo}, and SciPy \citep{scipy}.

\begin{acknowledgements}
E.M. would like to thank Y. Chandola, P. Dabhade, E. Kosmaczewski, M. Liao, D. Macconi, B. Mingo, and N. \.{Z}ywucka for providing tables of and information about their radio catalogs. E.M. would like to thank D. Garofalo and D. Macconi for useful discussion concerning this work. We would like to thank the anonymous referee for suggestions that added to the depth to this work.

E.M., J.S., A.B., D.K. and P.B. acknowledge financial support from the Czech Science Foundation project No.19-05599Y. This work was supported by the EU-ARC.CZ Large Research Infrastructure grant project LM2018106 of the Ministry of Education, Youth and Sports of the Czech Republic. JS also acknowledges MEYS project LTAUSA17095 for the travel support to USA and useful discussions with Erin Kara.

This research has made use of data obtained from the 4XMM XMM-Newton serendipitous source catalogue compiled by the 10 institutes of the XMM-Newton Survey Science Centre selected by ESA. This work made use of data supplied by the UK Swift Science Data Centre at the University of Leicester.

This work has made use of Astropy, pandas, and SciPy.
\end{acknowledgements}

\bibliographystyle{aa}
\bibliography{agn_states_rad_morph} 

\begin{thebibliography}{148}
\expandafter\ifx\csname natexlab\endcsname\relax\def\natexlab#1{#1}\fi

\bibitem[{{Abramowicz} {et~al.}(1988){Abramowicz}, {Czerny}, {Lasota}, \&
  {Szuszkiewicz}}]{Abramowicz88}
{Abramowicz}, M.~A., {Czerny}, B., {Lasota}, J.~P., \& {Szuszkiewicz}, E. 1988,
  \apj, 332, 646

\bibitem[{{Anderson} \& {Kraft}(1971)}]{Anderson71}
{Anderson}, K.~S. \& {Kraft}, R.~P. 1971, \apjl, 165, L3

\bibitem[{{Arcodia} {et~al.}(2020){Arcodia}, {Ponti}, {Merloni}, \&
  {Nandra}}]{Arcodia20}
{Arcodia}, R., {Ponti}, G., {Merloni}, A., \& {Nandra}, K. 2020, \aap, 638,
  A100

\bibitem[{{Astropy Collaboration} {et~al.}(2018){Astropy Collaboration},
  {Price-Whelan}, {Sip{\H{o}}cz}, {G{\"u}nther}, {Lim}, {Crawford}, {Conseil},
  {Shupe}, {Craig}, {Dencheva}, {Ginsburg}, {Vand erPlas}, {Bradley},
  {P{\'e}rez-Su{\'a}rez}, {de Val-Borro}, {Aldcroft}, {Cruz}, {Robitaille},
  {Tollerud}, {Ardelean}, {Babej}, {Bach}, {Bachetti}, {Bakanov}, {Bamford},
  {Barentsen}, {Barmby}, {Baumbach}, {Berry}, {Biscani}, {Boquien}, {Bostroem},
  {Bouma}, {Brammer}, {Bray}, {Breytenbach}, {Buddelmeijer}, {Burke},
  {Calderone}, {Cano Rodr{\'\i}guez}, {Cara}, {Cardoso}, {Cheedella}, {Copin},
  {Corrales}, {Crichton}, {D'Avella}, {Deil}, {Depagne}, {Dietrich}, {Donath},
  {Droettboom}, {Earl}, {Erben}, {Fabbro}, {Ferreira}, {Finethy}, {Fox},
  {Garrison}, {Gibbons}, {Goldstein}, {Gommers}, {Greco}, {Greenfield},
  {Groener}, {Grollier}, {Hagen}, {Hirst}, {Homeier}, {Horton}, {Hosseinzadeh},
  {Hu}, {Hunkeler}, {Ivezi{\'c}}, {Jain}, {Jenness}, {Kanarek}, {Kendrew},
  {Kern}, {Kerzendorf}, {Khvalko}, {King}, {Kirkby}, {Kulkarni}, {Kumar},
  {Lee}, {Lenz}, {Littlefair}, {Ma}, {Macleod}, {Mastropietro}, {McCully},
  {Montagnac}, {Morris}, {Mueller}, {Mumford}, {Muna}, {Murphy}, {Nelson},
  {Nguyen}, {Ninan}, {N{\"o}the}, {Ogaz}, {Oh}, {Parejko}, {Parley}, {Pascual},
  {Patil}, {Patil}, {Plunkett}, {Prochaska}, {Rastogi}, {Reddy Janga},
  {Sabater}, {Sakurikar}, {Seifert}, {Sherbert}, {Sherwood-Taylor}, {Shih},
  {Sick}, {Silbiger}, {Singanamalla}, {Singer}, {Sladen}, {Sooley},
  {Sornarajah}, {Streicher}, {Teuben}, {Thomas}, {Tremblay}, {Turner},
  {Terr{\'o}n}, {van Kerkwijk}, {de la Vega}, {Watkins}, {Weaver}, {Whitmore},
  {Woillez}, {Zabalza}, \& {Astropy Contributors}}]{astropy18}
{Astropy Collaboration}, {Price-Whelan}, A.~M., {Sip{\H{o}}cz}, B.~M., {et~al.}
  2018, \aj, 156, 123

\bibitem[{{Astropy Collaboration} {et~al.}(2013){Astropy Collaboration},
  {Robitaille}, {Tollerud}, {Greenfield}, {Droettboom}, {Bray}, {Aldcroft},
  {Davis}, {Ginsburg}, {Price-Whelan}, {Kerzendorf}, {Conley}, {Crighton},
  {Barbary}, {Muna}, {Ferguson}, {Grollier}, {Parikh}, {Nair}, {Unther},
  {Deil}, {Woillez}, {Conseil}, {Kramer}, {Turner}, {Singer}, {Fox}, {Weaver},
  {Zabalza}, {Edwards}, {Azalee Bostroem}, {Burke}, {Casey}, {Crawford},
  {Dencheva}, {Ely}, {Jenness}, {Labrie}, {Lim}, {Pierfederici}, {Pontzen},
  {Ptak}, {Refsdal}, {Servillat}, \& {Streicher}}]{astropy13}
{Astropy Collaboration}, {Robitaille}, T.~P., {Tollerud}, E.~J., {et~al.} 2013,
  \aap, 558, A33

\bibitem[{{Baldi} \& {Capetti}(2009)}]{Baldi09}
{Baldi}, R.~D. \& {Capetti}, A. 2009, \aap, 508, 603

\bibitem[{{Baldi} {et~al.}(2015){Baldi}, {Capetti}, \& {Giovannini}}]{Baldi15}
{Baldi}, R.~D., {Capetti}, A., \& {Giovannini}, G. 2015, \aap, 576, A38

\bibitem[{{Baldi} {et~al.}(2018){Baldi}, {Capetti}, \& {Massaro}}]{Baldi18}
{Baldi}, R.~D., {Capetti}, A., \& {Massaro}, F. 2018, \aap, 609, A1

\bibitem[{{Balmaverde} {et~al.}(2006){Balmaverde}, {Capetti}, \&
  {Grandi}}]{Balmaverde06}
{Balmaverde}, B., {Capetti}, A., \& {Grandi}, P. 2006, \aap, 451, 35

\bibitem[{{Barnier} {et~al.}(2022){Barnier}, {Petrucci}, {Ferreira}, {Marcel},
  {Belmont}, {Clavel}, {Corbel}, {Coriat}, {Espinasse}, {Henri}, {Malzac}, \&
  {Rodriguez}}]{Barnier22}
{Barnier}, S., {Petrucci}, P.~O., {Ferreira}, J., {et~al.} 2022, \aap, 657, A11

\bibitem[{{Begelman}(1978)}]{Begelman78}
{Begelman}, M.~C. 1978, \mnras, 184, 53

\bibitem[{{Belloni} {et~al.}(2005){Belloni}, {Homan}, {Casella}, {van der
  Klis}, {Nespoli}, {Lewin}, {Miller}, \& {M{\'e}ndez}}]{Belloni05}
{Belloni}, T., {Homan}, J., {Casella}, P., {et~al.} 2005, \aap, 440, 207

\bibitem[{{Best} \& {Heckman}(2012)}]{Best12}
{Best}, P.~N. \& {Heckman}, T.~M. 2012, \mnras, 421, 1569

\bibitem[{{Best} {et~al.}(2005){Best}, {Kauffmann}, {Heckman}, {Brinchmann},
  {Charlot}, {Ivezi{\'c}}, \& {White}}]{Best05}
{Best}, P.~N., {Kauffmann}, G., {Heckman}, T.~M., {et~al.} 2005, \mnras, 362,
  25

\bibitem[{{Bicknell}(1995)}]{Bicknell95}
{Bicknell}, G.~V. 1995, \apjs, 101, 29

\bibitem[{{Brienza} {et~al.}(2020){Brienza}, {Morganti}, {Harwood}, {Duchet},
  {Rajpurohit}, {Shulevski}, {Hardcastle}, {Mahatma}, {Godfrey}, {Prandoni},
  {Shimwell}, \& {Intema}}]{Brienza20}
{Brienza}, M., {Morganti}, R., {Harwood}, J., {et~al.} 2020, \aap, 638, A29

\bibitem[{{Bruni} {et~al.}(2020){Bruni}, {Panessa}, {Bassani}, {Dallacasa},
  {Venturi}, {Saripalli}, {Brienza}, {Hern{\'a}ndez-Garc{\'\i}a}, {Chiaraluce},
  {Ursini}, {Bazzano}, {Malizia}, \& {Ubertini}}]{Bruni20}
{Bruni}, G., {Panessa}, F., {Bassani}, L., {et~al.} 2020, \mnras, 494, 902

\bibitem[{{Buchner} {et~al.}(2014){Buchner}, {Georgakakis}, {Nandra}, {Hsu},
  {Rangel}, {Brightman}, {Merloni}, {Salvato}, {Donley}, \&
  {Kocevski}}]{Buchner14}
{Buchner}, J., {Georgakakis}, A., {Nandra}, K., {et~al.} 2014, \aap, 564, A125

\bibitem[{{Buchner} {et~al.}(2017){Buchner}, {Schulze}, \& {Bauer}}]{Buchner17}
{Buchner}, J., {Schulze}, S., \& {Bauer}, F.~E. 2017, \mnras, 464, 4545

\bibitem[{{Buttiglione} {et~al.}(2010){Buttiglione}, {Capetti}, {Celotti},
  {Axon}, {Chiaberge}, {Macchetto}, \& {Sparks}}]{Buttiglione10}
{Buttiglione}, S., {Capetti}, A., {Celotti}, A., {et~al.} 2010, \aap, 509, A6

\bibitem[{{Capetti} {et~al.}(2017{\natexlab{a}}){Capetti}, {Massaro}, \&
  {Baldi}}]{FRICAT}
{Capetti}, A., {Massaro}, F., \& {Baldi}, R.~D. 2017{\natexlab{a}}, \aap, 598,
  A49

\bibitem[{{Capetti} {et~al.}(2017{\natexlab{b}}){Capetti}, {Massaro}, \&
  {Baldi}}]{FRIICAT}
{Capetti}, A., {Massaro}, F., \& {Baldi}, R.~D. 2017{\natexlab{b}}, \aap, 601,
  A81

\bibitem[{{\v C}echura {et~al.}(2015){\v C}echura, Vrtilek, \&
  Hadrava}]{Cechura15}
{\v C}echura, J., Vrtilek, S.~D., \& Hadrava, P. 2015, \mnras, 450, 2410

\bibitem[{{Chandola} {et~al.}(2020){Chandola}, {Saikia}, \& {Li}}]{Chandola20}
{Chandola}, Y., {Saikia}, D.~J., \& {Li}, D. 2020, \mnras, 494, 5161

\bibitem[{{Ching} {et~al.}(2017){Ching}, {Sadler}, {Croom}, {Johnston},
  {Pracy}, {Couch}, {Hopkins}, {Jurek}, \& {Pimbblet}}]{Ching17}
{Ching}, J. H.~Y., {Sadler}, E.~M., {Croom}, S.~M., {et~al.} 2017, \mnras, 464,
  1306

\bibitem[{{Corbel} {et~al.}(2013){Corbel}, {Coriat}, {Brocksopp}, {Tzioumis},
  {Fender}, {Tomsick}, {Buxton}, \& {Bailyn}}]{Corbel13}
{Corbel}, S., {Coriat}, M., {Brocksopp}, C., {et~al.} 2013, \mnras, 428, 2500

\bibitem[{{Corbel} {et~al.}(2000){Corbel}, {Fender}, {Tzioumis}, {Nowak},
  {McIntyre}, {Durouchoux}, \& {Sood}}]{Corbel00}
{Corbel}, S., {Fender}, R.~P., {Tzioumis}, A.~K., {et~al.} 2000, \aap, 359, 251

\bibitem[{{Cromwell} \& {Weymann}(1970)}]{Cromwell70}
{Cromwell}, R. \& {Weymann}, R. 1970, \apjl, 159, L147

\bibitem[{{Dabhade} {et~al.}(2020){Dabhade}, {Mahato}, {Bagchi}, {Saikia},
  {Combes}, {Sankhyayan}, {R{\"o}ttgering}, {Ho}, {Gaikwad}, {Raychaudhury},
  {Vaidya}, \& {Guiderdoni}}]{Dabhade20}
{Dabhade}, P., {Mahato}, M., {Bagchi}, J., {et~al.} 2020, \aap, 642, A153

\bibitem[{{Dallacasa} {et~al.}(2000){Dallacasa}, {Stanghellini}, {Centonza}, \&
  {Fanti}}]{Dallacasa00}
{Dallacasa}, D., {Stanghellini}, C., {Centonza}, M., \& {Fanti}, R. 2000, \aap,
  363, 887

\bibitem[{{Denney} {et~al.}(2014){Denney}, {De Rosa}, {Croxall}, {Gupta},
  {Bentz}, {Fausnaugh}, {Grier}, {Martini}, {Mathur}, {Peterson}, {Pogge}, \&
  {Shappee}}]{Denney14}
{Denney}, K.~D., {De Rosa}, G., {Croxall}, K., {et~al.} 2014, \apj, 796, 134

\bibitem[{{Done} {et~al.}(2007){Done}, {Gierli{\'n}ski}, \& {Kubota}}]{Done07}
{Done}, C., {Gierli{\'n}ski}, M., \& {Kubota}, A. 2007, \aapr, 15, 1

\bibitem[{{Duncan} {et~al.}(2019){Duncan}, {Sabater}, {R{\"o}ttgering},
  {Jarvis}, {Smith}, {Best}, {Callingham}, {Cochrane}, {Croston}, {Hardcastle},
  {Mingo}, {Morabito}, {Nisbet}, {Prandoni}, {Shimwell}, {Tasse}, {White},
  {Williams}, {Alegre}, {Chy{\.z}y}, {G{\"u}rkan}, {Hoeft}, {Kondapally},
  {Mechev}, {Miley}, {Schwarz}, \& {van Weeren}}]{Duncan19}
{Duncan}, K.~J., {Sabater}, J., {R{\"o}ttgering}, H.~J.~A., {et~al.} 2019,
  \aap, 622, A3

\bibitem[{{Dunn} {et~al.}(2010){Dunn}, {Fender}, {K{\"o}rding}, {Belloni}, \&
  {Cabanac}}]{Dunn10}
{Dunn}, R.~J.~H., {Fender}, R.~P., {K{\"o}rding}, E.~G., {Belloni}, T., \&
  {Cabanac}, C. 2010, \mnras, 403, 61

\bibitem[{{Esin} {et~al.}(1998){Esin}, {Narayan}, {Cui}, {Grove}, \&
  {Zhang}}]{Esin1998}
{Esin}, A.~A., {Narayan}, R., {Cui}, W., {Grove}, J.~E., \& {Zhang}, S.-N.
  1998, \apj, 505, 854

\bibitem[{{Evans} {et~al.}(2020){Evans}, {Page}, {Osborne}, {Beardmore},
  {Willingale}, {Burrows}, {Kennea}, {Perri}, {Capalbi}, {Tagliaferri}, \&
  {Cenko}}]{Evans20}
{Evans}, P.~A., {Page}, K.~L., {Osborne}, J.~P., {et~al.} 2020, \apjs, 247, 54

\bibitem[{{Falcke} {et~al.}(2004){Falcke}, {K{\"o}rding}, \&
  {Markoff}}]{Falcke04}
{Falcke}, H., {K{\"o}rding}, E., \& {Markoff}, S. 2004, \aap, 414, 895

\bibitem[{{Fanaroff} \& {Riley}(1974)}]{FR74}
{Fanaroff}, B.~L. \& {Riley}, J.~M. 1974, \mnras, 167, 31P

\bibitem[{{Fanti} {et~al.}(1990){Fanti}, {Fanti}, {Schilizzi}, {Spencer}, {Nan
  Rendong}, {Parma}, {van Breugel}, \& {Venturi}}]{Fanti90}
{Fanti}, R., {Fanti}, C., {Schilizzi}, R.~T., {et~al.} 1990, \aap, 231, 333

\bibitem[{{Fender} \& {Belloni}(2012)}]{Fender12}
{Fender}, R. \& {Belloni}, T. 2012, Science, 337, 540

\bibitem[{{Fender} \& {Mu{\~n}oz-Darias}(2016)}]{Fender16}
{Fender}, R. \& {Mu{\~n}oz-Darias}, T. 2016, {The Balance of Power: Accretion
  and Feedback in Stellar Mass Black Holes}, ed. F.~{Haardt}, V.~{Gorini},
  U.~{Moschella}, A.~{Treves}, \& M.~{Colpi}, Vol. 905 (Springer, Cham), 65

\bibitem[{{Fender} {et~al.}(2004){Fender}, {Belloni}, \& {Gallo}}]{Fender04}
{Fender}, R.~P., {Belloni}, T.~M., \& {Gallo}, E. 2004, \mnras, 355, 1105

\bibitem[{{Fern{\'a}ndez-Ontiveros} \& {Mu{\~n}oz-Darias}(2021)}]{Fernandez21}
{Fern{\'a}ndez-Ontiveros}, J.~A. \& {Mu{\~n}oz-Darias}, T. 2021, \mnras, 504,
  5726

\bibitem[{{Gallo} {et~al.}(2003){Gallo}, {Fender}, \& {Pooley}}]{Gallo03}
{Gallo}, E., {Fender}, R.~P., \& {Pooley}, G.~G. 2003, \mnras, 344, 60

\bibitem[{{Garofalo} {et~al.}(2010){Garofalo}, {Evans}, \&
  {Sambruna}}]{Garofalo10}
{Garofalo}, D., {Evans}, D.~A., \& {Sambruna}, R.~M. 2010, \mnras, 406, 975

\bibitem[{{Gendre} {et~al.}(2010){Gendre}, {Best}, \& {Wall}}]{Gendre10}
{Gendre}, M.~A., {Best}, P.~N., \& {Wall}, J.~V. 2010, \mnras, 404, 1719

\bibitem[{{Grandi} {et~al.}(2021){Grandi}, {Torresi}, {Macconi}, {Boccardi}, \&
  {Capetti}}]{Grandi21}
{Grandi}, P., {Torresi}, E., {Macconi}, D., {Boccardi}, B., \& {Capetti}, A.
  2021, \apj, 911, 17

\bibitem[{{Grinberg} {et~al.}(2014){Grinberg}, {Pottschmidt}, {B{\"o}ck},
  {Schmid}, {Nowak}, {Uttley}, {Tomsick}, {Rodriguez}, {Hell}, {Markowitz},
  {Bodaghee}, {Cadolle Bel}, {Rothschild}, \& {Wilms}}]{Grinberg14}
{Grinberg}, V., {Pottschmidt}, K., {B{\"o}ck}, M., {et~al.} 2014, \aap, 565, A1

\bibitem[{{Guainazzi} {et~al.}(2002){Guainazzi}, {Matt}, {Fiore}, \&
  {Perola}}]{Guainazzi02}
{Guainazzi}, M., {Matt}, G., {Fiore}, F., \& {Perola}, G.~C. 2002, \aap, 388,
  787

\bibitem[{{G{\"u}ver} \& {{\"O}zel}(2009)}]{Guver09}
{G{\"u}ver}, T. \& {{\"O}zel}, F. 2009, \mnras, 400, 2050

\bibitem[{{Hardcastle}(2018)}]{Hardcastle18}
{Hardcastle}, M. 2018, Nature Astronomy, 2, 273

\bibitem[{{Hardcastle} {et~al.}(2007){Hardcastle}, {Evans}, \&
  {Croston}}]{Hardcastle07}
{Hardcastle}, M.~J., {Evans}, D.~A., \& {Croston}, J.~H. 2007, \mnras, 376,
  1849

\bibitem[{{Hardcastle} \& {Worrall}(1999)}]{Hardcastle99}
{Hardcastle}, M.~J. \& {Worrall}, D.~M. 1999, \mnras, 309, 969

\bibitem[{{Harrison} {et~al.}(2015){Harrison}, {Thomson}, {Alexander}, {Bauer},
  {Edge}, {Hogan}, {Mullaney}, \& {Swinbank}}]{Harrison15}
{Harrison}, C.~M., {Thomson}, A.~P., {Alexander}, D.~M., {et~al.} 2015, \apj,
  800, 45

\bibitem[{{Heckman} \& {Best}(2014)}]{Heckman14}
{Heckman}, T.~M. \& {Best}, P.~N. 2014, \araa, 52, 589

\bibitem[{{Heckman} {et~al.}(2004){Heckman}, {Kauffmann}, {Brinchmann},
  {Charlot}, {Tremonti}, \& {White}}]{Heckman04}
{Heckman}, T.~M., {Kauffmann}, G., {Brinchmann}, J., {et~al.} 2004, \apj, 613,
  109

\bibitem[{{Ho}(2008)}]{Ho08}
{Ho}, L.~C. 2008, \araa, 46, 475

\bibitem[{{Homan} {et~al.}(2005){Homan}, {Buxton}, {Markoff}, {Bailyn},
  {Nespoli}, \& {Belloni}}]{Homan05}
{Homan}, J., {Buxton}, M., {Markoff}, S., {et~al.} 2005, \apj, 624, 295

\bibitem[{Hunter(2007)}]{matplotlib}
Hunter, J.~D. 2007, Computing in Science Engineering, 9, 90

\bibitem[{{Ineson} {et~al.}(2015){Ineson}, {Croston}, {Hardcastle}, {Kraft},
  {Evans}, \& {Jarvis}}]{Ineson15}
{Ineson}, J., {Croston}, J.~H., {Hardcastle}, M.~J., {et~al.} 2015, \mnras,
  453, 2682

\bibitem[{{Jarvis} {et~al.}(2019){Jarvis}, {Harrison}, {Thomson}, {Circosta},
  {Mainieri}, {Alexander}, {Edge}, {Lansbury}, {Molyneux}, \&
  {Mullaney}}]{Jarvis19}
{Jarvis}, M.~E., {Harrison}, C.~M., {Thomson}, A.~P., {et~al.} 2019, \mnras,
  485, 2710

\bibitem[{{Jimenez-Gallardo} {et~al.}(2019){Jimenez-Gallardo}, {Massaro},
  {Capetti}, {Prieto}, {Paggi}, {Baldi}, {Grossova}, {Ostorero},
  {Siemiginowska}, \& {Viada}}]{JM19}
{Jimenez-Gallardo}, A., {Massaro}, F., {Capetti}, A., {et~al.} 2019, \aap, 627,
  A108

\bibitem[{{Kaiser} \& {Best}(2007)}]{Kaiser07}
{Kaiser}, C.~R. \& {Best}, P.~N. 2007, \mnras, 381, 1548

\bibitem[{{Kalberla} {et~al.}(2005){Kalberla}, {Burton}, {Hartmann}, {Arnal},
  {Bajaja}, {Morras}, \& {P{\"o}ppel}}]{Kalberla05}
{Kalberla}, P.~M.~W., {Burton}, W.~B., {Hartmann}, D., {et~al.} 2005, \aap,
  440, 775

\bibitem[{{Kellermann} {et~al.}(2016){Kellermann}, {Condon}, {Kimball},
  {Perley}, \& {Ivezi{\'c}}}]{Kellermann16}
{Kellermann}, K.~I., {Condon}, J.~J., {Kimball}, A.~E., {Perley}, R.~A., \&
  {Ivezi{\'c}}, {\v{Z}}. 2016, \apj, 831, 168

\bibitem[{{Kellermann} {et~al.}(1989){Kellermann}, {Sramek}, {Schmidt},
  {Shaffer}, \& {Green}}]{Kellermann89}
{Kellermann}, K.~I., {Sramek}, R., {Schmidt}, M., {Shaffer}, D.~B., \& {Green},
  R. 1989, \aj, 98, 1195

\bibitem[{{Konar} {et~al.}(2013){Konar}, {Hardcastle}, {Jamrozy}, \&
  {Croston}}]{Konar13}
{Konar}, C., {Hardcastle}, M.~J., {Jamrozy}, M., \& {Croston}, J.~H. 2013,
  \mnras, 430, 2137

\bibitem[{{K{\"o}rding} {et~al.}(2006{\natexlab{a}}){K{\"o}rding}, {Falcke}, \&
  {Corbel}}]{Kording06a}
{K{\"o}rding}, E., {Falcke}, H., \& {Corbel}, S. 2006{\natexlab{a}}, \aap, 456,
  439

\bibitem[{{K{\"o}rding} {et~al.}(2006{\natexlab{b}}){K{\"o}rding}, {Jester}, \&
  {Fender}}]{Kording06b}
{K{\"o}rding}, E.~G., {Jester}, S., \& {Fender}, R. 2006{\natexlab{b}}, \mnras,
  372, 1366

\bibitem[{{Kosmaczewski} {et~al.}(2020){Kosmaczewski}, {Stawarz},
  {Siemiginowska}, {Cheung}, {Ostorero}, {Sobolewska}, {Kozie{\l}-Wierzbowska},
  {W{\'o}jtowicz}, \& {Marchenko}}]{Kosmaczewski20}
{Kosmaczewski}, E., {Stawarz}, {\L}., {Siemiginowska}, A., {et~al.} 2020, \apj,
  897, 164

\bibitem[{{Koss} {et~al.}(2017){Koss}, {Trakhtenbrot}, {Ricci}, {Lamperti},
  {Oh}, {Berney}, {Schawinski}, {Balokovi{\'c}}, {Baronchelli}, {Crenshaw},
  {Fischer}, {Gehrels}, {Harrison}, {Hashimoto}, {Hogg}, {Ichikawa}, {Masetti},
  {Mushotzky}, {Sartori}, {Stern}, {Treister}, {Ueda}, {Veilleux}, \&
  {Winter}}]{Koss17}
{Koss}, M., {Trakhtenbrot}, B., {Ricci}, C., {et~al.} 2017, \apj, 850, 74

\bibitem[{{Kozie{\l}-Wierzbowska} {et~al.}(2020){Kozie{\l}-Wierzbowska},
  {Goyal}, \& {{\.Z}ywucka}}]{ROGUEI20}
{Kozie{\l}-Wierzbowska}, D., {Goyal}, A., \& {{\.Z}ywucka}, N. 2020, \apjs,
  247, 53

\bibitem[{{Kuraszkiewicz} {et~al.}(2021){Kuraszkiewicz}, {Wilkes}, {Atanas},
  {Buchner}, {McDowell}, {Willner}, {Ashby}, {Azadi}, {Barthel}, {Haas},
  {Worrall}, {Birkinshaw}, {Antonucci}, {Chini}, {Fazio}, {Lawrence}, \&
  {Ogle}}]{Kuraszkiewicz21}
{Kuraszkiewicz}, J., {Wilkes}, B.~J., {Atanas}, A., {et~al.} 2021, \apj, 913,
  134

\bibitem[{{LaMassa} {et~al.}(2015){LaMassa}, {Cales}, {Moran}, {Myers},
  {Richards}, {Eracleous}, {Heckman}, {Gallo}, \& {Urry}}]{LaMassa15}
{LaMassa}, S.~M., {Cales}, S., {Moran}, E.~C., {et~al.} 2015, \apj, 800, 144

\bibitem[{{Ledlow} \& {Owen}(1996)}]{Ledlow96}
{Ledlow}, M.~J. \& {Owen}, F.~N. 1996, \aj, 112, 9

\bibitem[{{Liao} \& {Gu}(2020)}]{Liao20a}
{Liao}, M. \& {Gu}, M. 2020, \mnras, 491, 92

\bibitem[{{Liao} {et~al.}(2020){Liao}, {Gu}, {Zhou}, \& {Chen}}]{Liao20b}
{Liao}, M., {Gu}, M., {Zhou}, M., \& {Chen}, L. 2020, \mnras, 497, 482

\bibitem[{{Maccagni} {et~al.}(2020){Maccagni}, {Murgia}, {Serra}, {Govoni},
  {Morokuma-Matsui}, {Kleiner}, {Buchner}, {J{\'o}zsa}, {Kamphuis},
  {Makhathini}, {Moln{\'a}r}, {Prokhorov}, {Ramaila}, {Ramatsoku}, {Thorat}, \&
  {Smirnov}}]{Maccagni20}
{Maccagni}, F.~M., {Murgia}, M., {Serra}, P., {et~al.} 2020, \aap, 634, A9

\bibitem[{{Maccarone} {et~al.}(2003){Maccarone}, {Gallo}, \&
  {Fender}}]{Maccarone03}
{Maccarone}, T.~J., {Gallo}, E., \& {Fender}, R. 2003, \mnras, 345, L19

\bibitem[{{Macconi} {et~al.}(2020){Macconi}, {Torresi}, {Grandi}, {Boccardi},
  \& {Vignali}}]{Macconi20}
{Macconi}, D., {Torresi}, E., {Grandi}, P., {Boccardi}, B., \& {Vignali}, C.
  2020, \mnras, 493, 4355

\bibitem[{{MacLeod} {et~al.}(2019){MacLeod}, {Green}, {Anderson}, {Bruce},
  {Eracleous}, {Graham}, {Homan}, {Lawrence}, {LeBleu}, {Ross}, {Ruan},
  {Runnoe}, {Stern}, {Burgett}, {Chambers}, {Kaiser}, {Magnier}, \&
  {Metcalfe}}]{MacLeod19}
{MacLeod}, C.~L., {Green}, P.~J., {Anderson}, S.~F., {et~al.} 2019, \apj, 874,
  8

\bibitem[{{Maoz}(2007)}]{Maoz07}
{Maoz}, D. 2007, \mnras, 377, 1696

\bibitem[{{Matt} {et~al.}(2003){Matt}, {Guainazzi}, \& {Maiolino}}]{Matt03}
{Matt}, G., {Guainazzi}, M., \& {Maiolino}, R. 2003, \mnras, 342, 422

\bibitem[{{McClintock} \& {Remillard}(2006)}]{McClintock06}
{McClintock}, J.~E. \& {Remillard}, R.~A. 2006, in Compact stellar X-ray
  sources, Vol.~39, 157--213

\bibitem[{{McHardy} {et~al.}(2006){McHardy}, {Koerding}, {Knigge}, {Uttley}, \&
  {Fender}}]{McHardy06}
{McHardy}, I.~M., {Koerding}, E., {Knigge}, C., {Uttley}, P., \& {Fender},
  R.~P. 2006, \nat, 444, 730

\bibitem[{{Merloni} \& {Heinz}(2008)}]{Merloni08}
{Merloni}, A. \& {Heinz}, S. 2008, \mnras, 388, 1011

\bibitem[{{Merloni} {et~al.}(2003){Merloni}, {Heinz}, \& {di
  Matteo}}]{Merloni03}
{Merloni}, A., {Heinz}, S., \& {di Matteo}, T. 2003, \mnras, 345, 1057

\bibitem[{{Mingo} {et~al.}(2019){Mingo}, {Croston}, {Hardcastle}, {Best},
  {Duncan}, {Morganti}, {Rottgering}, {Sabater}, {Shimwell}, {Williams},
  {Brienza}, {Gurkan}, {Mahatma}, {Morabito}, {Prandoni}, {Bondi}, {Ineson}, \&
  {Mooney}}]{Mingo19}
{Mingo}, B., {Croston}, J.~H., {Hardcastle}, M.~J., {et~al.} 2019, \mnras, 488,
  2701

\bibitem[{{Mingo} {et~al.}(2014){Mingo}, {Hardcastle}, {Croston}, {Dicken},
  {Evans}, {Morganti}, \& {Tadhunter}}]{Mingo14}
{Mingo}, B., {Hardcastle}, M.~J., {Croston}, J.~H., {et~al.} 2014, \mnras, 440,
  269

\bibitem[{{Mirabel} \& {Rodr{\'\i}guez}(1994)}]{Mirabel94}
{Mirabel}, I.~F. \& {Rodr{\'\i}guez}, L.~F. 1994, \nat, 371, 46

\bibitem[{{Miraghaei} \& {Best}(2017)}]{Miraghaei17}
{Miraghaei}, H. \& {Best}, P.~N. 2017, \mnras, 466, 4346

\bibitem[{{Murphy} \& {Yaqoob}(2009)}]{Murphy09}
{Murphy}, K.~D. \& {Yaqoob}, T. 2009, \mnras, 397, 1549

\bibitem[{{Narayan} {et~al.}(2012){Narayan}, {S{\"A} dowski}, {Penna}, \&
  {Kulkarni}}]{Narayan12}
{Narayan}, R., {S{\"A} dowski}, A., {Penna}, R.~F., \& {Kulkarni}, A.~K. 2012,
  \mnras, 426, 3241

\bibitem[{{Netzer}(2019)}]{Netzer19}
{Netzer}, H. 2019, \mnras, 488, 5185

\bibitem[{{Nipoti} {et~al.}(2005){Nipoti}, {Blundell}, \& {Binney}}]{Nipoti05}
{Nipoti}, C., {Blundell}, K.~M., \& {Binney}, J. 2005, \mnras, 361, 633

\bibitem[{{O'Dea}(1998)}]{Odea98}
{O'Dea}, C.~P. 1998, \pasp, 110, 493

\bibitem[{{O'Dea} \& {Saikia}(2021)}]{Odea21}
{O'Dea}, C.~P. \& {Saikia}, D.~J. 2021, \aapr, 29, 3

\bibitem[{{Page} {et~al.}(2012){Page}, {Brindle}, {Talavera}, {Still}, {Rosen},
  {Yershov}, {Ziaeepour}, {Mason}, {Cropper}, {Breeveld}, {Loiseau}, {Mignani},
  {Smith}, \& {Murdin}}]{Page12}
{Page}, M.~J., {Brindle}, C., {Talavera}, A., {et~al.} 2012, \mnras, 426, 903

\bibitem[{{Page} {et~al.}(2014){Page}, {Yershov}, {Breeveld}, {Kuin},
  {Mignani}, {Smith}, {Rawlings}, {Oates}, {Siegel}, \& {Roming}}]{Page14}
{Page}, M.~J., {Yershov}, V., {Breeveld}, A., {et~al.} 2014, in Proceedings of
  Swift: 10 Years of Discovery (SWIFT 10, 37

\bibitem[{pandas~development team(2020)}]{pandasZendo}
pandas~development team, T. 2020, pandas-dev/pandas: Pandas

\bibitem[{{Panessa} {et~al.}(2019){Panessa}, {Baldi}, {Laor}, {Padovani},
  {Behar}, \& {McHardy}}]{Panessa19}
{Panessa}, F., {Baldi}, R.~D., {Laor}, A., {et~al.} 2019, Nature Astronomy, 3,
  387

\bibitem[{{Panessa} {et~al.}(2016){Panessa}, {Bassani}, {Landi}, {Bazzano},
  {Dallacasa}, {La Franca}, {Malizia}, {Venturi}, \& {Ubertini}}]{Panessa16}
{Panessa}, F., {Bassani}, L., {Landi}, R., {et~al.} 2016, \mnras, 461, 3153

\bibitem[{{Panessa} \& {Giroletti}(2013)}]{Panessa13}
{Panessa}, F. \& {Giroletti}, M. 2013, \mnras, 432, 1138

\bibitem[{{Panessa} {et~al.}(2015){Panessa}, {Tarchi}, {Castangia}, {Maiorano},
  {Bassani}, {Bicknell}, {Bazzano}, {Bird}, {Malizia}, \&
  {Ubertini}}]{Panessa15}
{Panessa}, F., {Tarchi}, A., {Castangia}, P., {et~al.} 2015, \mnras, 447, 1289

\bibitem[{{P{\^a}ris} {et~al.}(2018){P{\^a}ris}, {Petitjean}, {Aubourg},
  {Myers}, {Streblyanska}, {Lyke}, {Anderson}, {Armengaud}, {Bautista},
  {Blanton}, {Blomqvist}, {Brinkmann}, {Brownstein}, {Brandt}, {Burtin},
  {Dawson}, {de la Torre}, {Georgakakis}, {Gil-Mar{\'\i}n}, {Green}, {Hall},
  {Kneib}, {LaMassa}, {Le Goff}, {MacLeod}, {Mariappan}, {McGreer}, {Merloni},
  {Noterdaeme}, {Palanque-Delabrouille}, {Percival}, {Ross}, {Rossi},
  {Schneider}, {Seo}, {Tojeiro}, {Weaver}, {Weijmans}, {Y{\`e}che}, {Zarrouk},
  \& {Zhao}}]{Paris18}
{P{\^a}ris}, I., {Petitjean}, P., {Aubourg}, {\'E}., {et~al.} 2018, \aap, 613,
  A51

\bibitem[{{Parma} {et~al.}(1999){Parma}, {Murgia}, {Morganti}, {Capetti}, {de
  Ruiter}, \& {Fanti}}]{Parma99}
{Parma}, P., {Murgia}, M., {Morganti}, R., {et~al.} 1999, \aap, 344, 7

\bibitem[{{Planck Collaboration} {et~al.}(2020){Planck Collaboration},
  {Aghanim}, {Akrami}, {Ashdown}, {Aumont}, {Baccigalupi}, {Ballardini},
  {Banday}, {Barreiro}, {Bartolo}, {Basak}, {Battye}, {Benabed}, {Bernard},
  {Bersanelli}, {Bielewicz}, {Bock}, {Bond}, {Borrill}, {Bouchet}, {Boulanger},
  {Bucher}, {Burigana}, {Butler}, {Calabrese}, {Cardoso}, {Carron},
  {Challinor}, {Chiang}, {Chluba}, {Colombo}, {Combet}, {Contreras}, {Crill},
  {Cuttaia}, {de Bernardis}, {de Zotti}, {Delabrouille}, {Delouis}, {Di
  Valentino}, {Diego}, {Dor{\'e}}, {Douspis}, {Ducout}, {Dupac}, {Dusini},
  {Efstathiou}, {Elsner}, {En{\ss}lin}, {Eriksen}, {Fantaye}, {Farhang},
  {Fergusson}, {Fernandez-Cobos}, {Finelli}, {Forastieri}, {Frailis},
  {Fraisse}, {Franceschi}, {Frolov}, {Galeotta}, {Galli}, {Ganga},
  {G{\'e}nova-Santos}, {Gerbino}, {Ghosh}, {Gonz{\'a}lez-Nuevo}, {G{\'o}rski},
  {Gratton}, {Gruppuso}, {Gudmundsson}, {Hamann}, {Handley}, {Hansen},
  {Herranz}, {Hildebrandt}, {Hivon}, {Huang}, {Jaffe}, {Jones}, {Karakci},
  {Keih{\"a}nen}, {Keskitalo}, {Kiiveri}, {Kim}, {Kisner}, {Knox},
  {Krachmalnicoff}, {Kunz}, {Kurki-Suonio}, {Lagache}, {Lamarre}, {Lasenby},
  {Lattanzi}, {Lawrence}, {Le Jeune}, {Lemos}, {Lesgourgues}, {Levrier},
  {Lewis}, {Liguori}, {Lilje}, {Lilley}, {Lindholm}, {L{\'o}pez-Caniego},
  {Lubin}, {Ma}, {Mac{\'\i}as-P{\'e}rez}, {Maggio}, {Maino}, {Mandolesi},
  {Mangilli}, {Marcos-Caballero}, {Maris}, {Martin}, {Martinelli},
  {Mart{\'\i}nez-Gonz{\'a}lez}, {Matarrese}, {Mauri}, {McEwen}, {Meinhold},
  {Melchiorri}, {Mennella}, {Migliaccio}, {Millea}, {Mitra},
  {Miville-Desch{\^e}nes}, {Molinari}, {Montier}, {Morgante}, {Moss}, {Natoli},
  {N{\o}rgaard-Nielsen}, {Pagano}, {Paoletti}, {Partridge}, {Patanchon},
  {Peiris}, {Perrotta}, {Pettorino}, {Piacentini}, {Polastri}, {Polenta},
  {Puget}, {Rachen}, {Reinecke}, {Remazeilles}, {Renzi}, {Rocha}, {Rosset},
  {Roudier}, {Rubi{\~n}o-Mart{\'\i}n}, {Ruiz-Granados}, {Salvati}, {Sandri},
  {Savelainen}, {Scott}, {Shellard}, {Sirignano}, {Sirri}, {Spencer},
  {Sunyaev}, {Suur-Uski}, {Tauber}, {Tavagnacco}, {Tenti}, {Toffolatti},
  {Tomasi}, {Trombetti}, {Valenziano}, {Valiviita}, {Van Tent}, {Vibert},
  {Vielva}, {Villa}, {Vittorio}, {Wandelt}, {Wehus}, {White}, {White},
  {Zacchei}, \& {Zonca}}]{Planck18}
{Planck Collaboration}, {Aghanim}, N., {Akrami}, Y., {et~al.} 2020, \aap, 641,
  A6

\bibitem[{{Rakshit} {et~al.}(2020){Rakshit}, {Stalin}, \&
  {Kotilainen}}]{Rakshit20}
{Rakshit}, S., {Stalin}, C.~S., \& {Kotilainen}, J. 2020, \apjs, 249, 17

\bibitem[{{Remillard} \& {McClintock}(2006)}]{Remillard06}
{Remillard}, R.~A. \& {McClintock}, J.~E. 2006, \araa, 44, 49

\bibitem[{{Ricci} {et~al.}(2017){Ricci}, {Trakhtenbrot}, {Koss}, {Ueda}, {Del
  Vecchio}, {Treister}, {Schawinski}, {Paltani}, {Oh}, {Lamperti}, {Berney},
  {Gandhi}, {Ichikawa}, {Bauer}, {Ho}, {Asmus}, {Beckmann}, {Soldi},
  {Balokovi{\'c}}, {Gehrels}, \& {Markwardt}}]{Ricci17}
{Ricci}, C., {Trakhtenbrot}, B., {Koss}, M.~J., {et~al.} 2017, \apjs, 233, 17

\bibitem[{{Ricci} {et~al.}(2015){Ricci}, {Ueda}, {Koss}, {Trakhtenbrot},
  {Bauer}, \& {Gandhi}}]{Ricci15}
{Ricci}, C., {Ueda}, Y., {Koss}, M.~J., {et~al.} 2015, \apjl, 815, L13

\bibitem[{{Richards} {et~al.}(2006){Richards}, {Lacy}, {Storrie-Lombardi},
  {Hall}, {Gallagher}, {Hines}, {Fan}, {Papovich}, {Vanden Berk}, {Trammell},
  {Schneider}, {Vestergaard}, {York}, {Jester}, {Anderson}, {Budav{\'a}ri}, \&
  {Szalay}}]{Richards06}
{Richards}, G.~T., {Lacy}, M., {Storrie-Lombardi}, L.~J., {et~al.} 2006, \apjs,
  166, 470

\bibitem[{{Risaliti} {et~al.}(2009){Risaliti}, {Miniutti}, {Elvis}, {Fabbiano},
  {Salvati}, {Baldi}, {Braito}, {Bianchi}, {Matt}, {Reeves}, {Soria}, \&
  {Zezas}}]{Risaliti09}
{Risaliti}, G., {Miniutti}, G., {Elvis}, M., {et~al.} 2009, \apj, 696, 160

\bibitem[{{Rosen} {et~al.}(2016){Rosen}, {Webb}, {Watson}, {Ballet}, {Barret},
  {Braito}, {Carrera}, {Ceballos}, {Coriat}, {Della Ceca}, {Denkinson},
  {Esquej}, {Farrell}, {Freyberg}, {Gris{\'e}}, {Guillout}, {Heil},
  {Koliopanos}, {Law-Green}, {Lamer}, {Lin}, {Martino}, {Michel}, {Motch},
  {Nebot Gomez-Moran}, {Page}, {Page}, {Page}, {Pakull}, {Pye}, {Read},
  {Rodriguez}, {Sakano}, {Saxton}, {Schwope}, {Scott}, {Sturm}, {Traulsen},
  {Yershov}, \& {Zolotukhin}}]{Rosen16}
{Rosen}, S.~R., {Webb}, N.~A., {Watson}, M.~G., {et~al.} 2016, \aap, 590, A1

\bibitem[{{Ruan} {et~al.}(2019){Ruan}, {Anderson}, {Eracleous}, {Green},
  {Haggard}, {MacLeod}, {Runnoe}, \& {Sobolewska}}]{Ruan19}
{Ruan}, J.~J., {Anderson}, S.~F., {Eracleous}, M., {et~al.} 2019, \apj, 883, 76

\bibitem[{{Sabater} {et~al.}(2019){Sabater}, {Best}, {Hardcastle}, {Shimwell},
  {Tasse}, {Williams}, {Br{\"u}ggen}, {Cochrane}, {Croston}, {de Gasperin},
  {Duncan}, {G{\"u}rkan}, {Mechev}, {Morabito}, {Prandoni}, {R{\"o}ttgering},
  {Smith}, {Harwood}, {Mingo}, {Mooney}, \& {Saxena}}]{Sabater19}
{Sabater}, J., {Best}, P.~N., {Hardcastle}, M.~J., {et~al.} 2019, \aap, 622,
  A17

\bibitem[{{Sadler} {et~al.}(2014){Sadler}, {Ekers}, {Mahony}, {Mauch}, \&
  {Murphy}}]{Sadler14}
{Sadler}, E.~M., {Ekers}, R.~D., {Mahony}, E.~K., {Mauch}, T., \& {Murphy}, T.
  2014, \mnras, 438, 796

\bibitem[{{Schawinski} {et~al.}(2015){Schawinski}, {Koss}, {Berney}, \&
  {Sartori}}]{Schawinski15}
{Schawinski}, K., {Koss}, M., {Berney}, S., \& {Sartori}, L.~F. 2015, \mnras,
  451, 2517

\bibitem[{{Scott} {et~al.}(2004){Scott}, {Kriss}, {Brotherton}, {Green},
  {Hutchings}, {Shull}, \& {Zheng}}]{Scott04}
{Scott}, J.~E., {Kriss}, G.~A., {Brotherton}, M., {et~al.} 2004, \apj, 615, 135

\bibitem[{{Shabala} {et~al.}(2008){Shabala}, {Ash}, {Alexander}, \&
  {Riley}}]{Shabala08}
{Shabala}, S.~S., {Ash}, S., {Alexander}, P., \& {Riley}, J.~M. 2008, \mnras,
  388, 625

\bibitem[{{Shabala} {et~al.}(2020){Shabala}, {Jurlin}, {Morganti}, {Brienza},
  {Hardcastle}, {Godfrey}, {Krause}, \& {Turner}}]{Shabala20}
{Shabala}, S.~S., {Jurlin}, N., {Morganti}, R., {et~al.} 2020, \mnras, 496,
  1706

\bibitem[{{Shappee} {et~al.}(2014){Shappee}, {Prieto}, {Grupe}, {Kochanek},
  {Stanek}, {De Rosa}, {Mathur}, {Zu}, {Peterson}, {Pogge}, {Komossa}, {Im},
  {Jencson}, {Holoien}, {Basu}, {Beacom}, {Szczygie{\l}}, {Brimacombe},
  {Adams}, {Campillay}, {Choi}, {Contreras}, {Dietrich}, {Dubberley},
  {Elphick}, {Foale}, {Giustini}, {Gonzalez}, {Hawkins}, {Howell}, {Hsiao},
  {Koss}, {Leighly}, {Morrell}, {Mudd}, {Mullins}, {Nugent}, {Parrent},
  {Phillips}, {Pojmanski}, {Rosing}, {Ross}, {Sand}, {Terndrup}, {Valenti},
  {Walker}, \& {Yoon}}]{Shappee14}
{Shappee}, B.~J., {Prieto}, J.~L., {Grupe}, D., {et~al.} 2014, \apj, 788, 48

\bibitem[{{Shimwell} {et~al.}(2019){Shimwell}, {Tasse}, {Hardcastle}, {Mechev},
  {Williams}, {Best}, {R{\"o}ttgering}, {Callingham}, {Dijkema}, {de Gasperin},
  {Hoang}, {Hugo}, {Mirmont}, {Oonk}, {Prandoni}, {Rafferty}, {Sabater},
  {Smirnov}, {van Weeren}, {White}, {Atemkeng}, {Bester}, {Bonnassieux},
  {Br{\"u}ggen}, {Brunetti}, {Chy{\.z}y}, {Cochrane}, {Conway}, {Croston},
  {Danezi}, {Duncan}, {Haverkorn}, {Heald}, {Iacobelli}, {Intema}, {Jackson},
  {Jamrozy}, {Jarvis}, {Lakhoo}, {Mevius}, {Miley}, {Morabito}, {Morganti},
  {Nisbet}, {Orr{\'u}}, {Perkins}, {Pizzo}, {Schrijvers}, {Smith}, {Vermeulen},
  {Wise}, {Alegre}, {Bacon}, {van Bemmel}, {Beswick}, {Bonafede}, {Botteon},
  {Bourke}, {Brienza}, {Calistro Rivera}, {Cassano}, {Clarke}, {Conselice},
  {Dettmar}, {Drabent}, {Dumba}, {Emig}, {En{\ss}lin}, {Ferrari}, {Garrett},
  {G{\'e}nova-Santos}, {Goyal}, {G{\"u}rkan}, {Hale}, {Harwood}, {Heesen},
  {Hoeft}, {Horellou}, {Jackson}, {Kokotanekov}, {Kondapally},
  {Kunert-Bajraszewska}, {Mahatma}, {Mahony}, {Mandal}, {McKean}, {Merloni},
  {Mingo}, {Miskolczi}, {Mooney}, {Nikiel-Wroczy{\'n}ski}, {O'Sullivan},
  {Quinn}, {Reich}, {Roskowi{\'n}ski}, {Rowlinson}, {Savini}, {Saxena},
  {Schwarz}, {Shulevski}, {Sridhar}, {Stacey}, {Urquhart}, {van der Wiel},
  {Varenius}, {Webster}, \& {Wilber}}]{Shimwell19}
{Shimwell}, T.~W., {Tasse}, C., {Hardcastle}, M.~J., {et~al.} 2019, \aap, 622,
  A1

\bibitem[{{S{\k{a}}dowski} {et~al.}(2014){S{\k{a}}dowski}, {Narayan},
  {McKinney}, \& {Tchekhovskoy}}]{Sadowski14}
{S{\k{a}}dowski}, A., {Narayan}, R., {McKinney}, J.~C., \& {Tchekhovskoy}, A.
  2014, \mnras, 439, 503

\bibitem[{{Sobolewska} {et~al.}(2019){Sobolewska}, {Siemiginowska},
  {Guainazzi}, {Hardcastle}, {Migliori}, {Ostorero}, \&
  {Stawarz}}]{Sobolewska19}
{Sobolewska}, M., {Siemiginowska}, A., {Guainazzi}, M., {et~al.} 2019, \apj,
  871, 71

\bibitem[{{Sobolewska} {et~al.}(2011){Sobolewska}, {Siemiginowska}, \&
  {Gierli{\'n}ski}}]{Sobolewska11}
{Sobolewska}, M.~A., {Siemiginowska}, A., \& {Gierli{\'n}ski}, M. 2011, \mnras,
  413, 2259

\bibitem[{{Subrahmanyan} {et~al.}(1996){Subrahmanyan}, {Saripalli}, \&
  {Hunstead}}]{Subrahmanyan96}
{Subrahmanyan}, R., {Saripalli}, L., \& {Hunstead}, R.~W. 1996, \mnras, 279,
  257

\bibitem[{{Svoboda} {et~al.}(2017){Svoboda}, {Guainazzi}, \&
  {Merloni}}]{Svoboda17}
{Svoboda}, J., {Guainazzi}, M., \& {Merloni}, A. 2017, \aap, 603, A127

\bibitem[{{Taylor}(2005)}]{topcat}
{Taylor}, M.~B. 2005, in Astronomical Society of the Pacific Conference Series,
  Vol. 347, Astronomical Data Analysis Software and Systems XIV, ed.
  P.~{Shopbell}, M.~{Britton}, \& R.~{Ebert}, 29

\bibitem[{{Tchekhovskoy} \& {Bromberg}(2016)}]{Tchekhovskoy16}
{Tchekhovskoy}, A. \& {Bromberg}, O. 2016, \mnras, 461, L46

\bibitem[{{Tohline} \& {Osterbrock}(1976)}]{Tohline76}
{Tohline}, J.~E. \& {Osterbrock}, D.~E. 1976, \apjl, 210, L117

\bibitem[{{Turner}(2018)}]{Turner18}
{Turner}, R.~J. 2018, \mnras, 476, 2522

\bibitem[{{Ueda} {et~al.}(2014){Ueda}, {Akiyama}, {Hasinger}, {Miyaji}, \&
  {Watson}}]{Ueda14}
{Ueda}, Y., {Akiyama}, M., {Hasinger}, G., {Miyaji}, T., \& {Watson}, M.~G.
  2014, \apj, 786, 104

\bibitem[{{Vantyghem} {et~al.}(2014){Vantyghem}, {McNamara}, {Russell}, {Main},
  {Nulsen}, {Wise}, {Hoekstra}, \& {Gitti}}]{Vantyghem14}
{Vantyghem}, A.~N., {McNamara}, B.~R., {Russell}, H.~R., {et~al.} 2014, \mnras,
  442, 3192

\bibitem[{{Vasudevan} \& {Fabian}(2009)}]{Vasudevan09}
{Vasudevan}, R.~V. \& {Fabian}, A.~C. 2009, \mnras, 392, 1124

\bibitem[{{V{\'e}ron-Cetty} \& {V{\'e}ron}(2010)}]{Veron10}
{V{\'e}ron-Cetty}, M.~P. \& {V{\'e}ron}, P. 2010, \aap, 518, A10

\bibitem[{Virtanen {et~al.}(2020)Virtanen, Gommers, Oliphant, Haberland, Reddy,
  Cournapeau, Burovski, Peterson, Weckesser, Bright, {van der Walt}, Brett,
  Wilson, Millman, Mayorov, Nelson, Jones, Kern, Larson, Carey, Polat, Feng,
  Moore, {VanderPlas}, Laxalde, Perktold, Cimrman, Henriksen, Quintero, Harris,
  Archibald, Ribeiro, Pedregosa, {van Mulbregt}, \& {SciPy 1.0
  Contributors}}]{scipy}
Virtanen, P., Gommers, R., Oliphant, T.~E., {et~al.} 2020, Nature Methods, 17,
  261

\bibitem[{{Webb} {et~al.}(2020){Webb}, {Coriat}, {Traulsen}, {Ballet}, {Motch},
  {Carrera}, {Koliopanos}, {Authier}, {de la Calle}, {Ceballos}, {Colomo},
  {Chuard}, {Freyberg}, {Garcia}, {Kolehmainen}, {Lamer}, {Lin}, {Maggi},
  {Michel}, {Page}, {Page}, {Perea-Calderon}, {Pineau}, {Rodriguez}, {Rosen},
  {Santos Lleo}, {Saxton}, {Schwope}, {Tom{\'a}s}, {Watson}, \&
  {Zakardjian}}]{Webb20}
{Webb}, N.~A., {Coriat}, M., {Traulsen}, I., {et~al.} 2020, \aap, 641, A136

\bibitem[{{W}es {M}c{K}inney(2010)}]{pandas10}
{W}es {M}c{K}inney. 2010, in {P}roceedings of the 9th {P}ython in {S}cience
  {C}onference, ed. {S}t\'efan van~der {W}alt \& {J}arrod {M}illman, 56 -- 61

\bibitem[{{Whittam} {et~al.}(2016){Whittam}, {Riley}, {Green}, \&
  {Jarvis}}]{Whittam16}
{Whittam}, I.~H., {Riley}, J.~M., {Green}, D.~A., \& {Jarvis}, M.~J. 2016,
  \mnras, 462, 2122

\bibitem[{{Wilkes} {et~al.}(2013){Wilkes}, {Kuraszkiewicz}, {Haas}, {Barthel},
  {Leipski}, {Willner}, {Worrall}, {Birkinshaw}, {Antonucci}, {Ashby}, {Chini},
  {Fazio}, {Lawrence}, {Ogle}, \& {Schulz}}]{Wilkes13}
{Wilkes}, B.~J., {Kuraszkiewicz}, J., {Haas}, M., {et~al.} 2013, \apj, 773, 15

\bibitem[{{Williams} {et~al.}(2019){Williams}, {Hardcastle}, {Best}, {Sabater},
  {Croston}, {Duncan}, {Shimwell}, {R{\"o}ttgering}, {Nisbet}, {G{\"u}rkan},
  {Alegre}, {Cochrane}, {Goyal}, {Hale}, {Jackson}, {Jamrozy}, {Kondapally},
  {Kunert-Bajraszewska}, {Mahatma}, {Mingo}, {Morabito}, {Prandoni},
  {Roskowinski}, {Shulevski}, {Smith}, {Tasse}, {Urquhart}, {Webster}, {White},
  {Beswick}, {Callingham}, {Chy{\.z}y}, {de Gasperin}, {Harwood}, {Hoeft},
  {Iacobelli}, {McKean}, {Mechev}, {Miley}, {Schwarz}, \& {van
  Weeren}}]{Williams19}
{Williams}, W.~L., {Hardcastle}, M.~J., {Best}, P.~N., {et~al.} 2019, \aap,
  622, A2

\bibitem[{{Willingale} {et~al.}(2013){Willingale}, {Starling}, {Beardmore},
  {Tanvir}, \& {O'Brien}}]{Willingale13}
{Willingale}, R., {Starling}, R.~L.~C., {Beardmore}, A.~P., {Tanvir}, N.~R., \&
  {O'Brien}, P.~T. 2013, \mnras, 431, 394

\bibitem[{{Yershov}(2014)}]{Yershov14}
{Yershov}, V.~N. 2014, \apss, 354, 97

\bibitem[{{Zdziarski}(1985)}]{Zdziarski1985}
{Zdziarski}, A.~A. 1985, \apj, 289, 514

\bibitem[{{Zdziarski} {et~al.}(2004){Zdziarski}, {Gierli{\'n}ski},
  {Miko{\l}ajewska}, {Wardzi{\'n}ski}, {Smith}, {Harmon}, \&
  {Kitamoto}}]{Zdziarski04}
{Zdziarski}, A.~A., {Gierli{\'n}ski}, M., {Miko{\l}ajewska}, J., {et~al.} 2004,
  \mnras, 351, 791

\bibitem[{{Zhang}(2013)}]{Zhang13}
{Zhang}, S.-N. 2013, Frontiers of Physics, 8, 630

\bibitem[{{Zhu} {et~al.}(2020){Zhu}, {Brandt}, {Luo}, {Wu}, {Xue}, \&
  {Yang}}]{Zhu20}
{Zhu}, S.~F., {Brandt}, W.~N., {Luo}, B., {et~al.} 2020, \mnras, 496, 245

\end{thebibliography}

\begin{appendix}

\section{Rejected source counts}\label{A:rejects}

\bgroup
\def\arraystretch{1.25}
\setlength{\tabcolsep}{4pt}
\begin{table*}
\centering
\caption{Number of sources rejected in the data quality cut process.}
\label{tab:rejects}
\begin{tabular}{c|cccccc|cccccc}
    \hline\hline 
     & \multicolumn{6}{c|}{\xmm} & \multicolumn{6}{c}{\swift\ Only} \\
    \textbf{Catalog} & \textbf{N$_\mathrm{\bf{start}}$} & \textbf{<3$\mathbf{\sigma}$} &  \textbf{Underexp.} &  \textbf{$\mathbf{\Gamma\leq}$1.5} &  \textbf{{$\mathbf{\Gamma\geq}$3.0}} &  \textbf{N$_\mathrm{\bf{end}}$} &  \textbf{N$_\mathrm{\bf{start}}$} & \textbf{<3$\mathbf{\sigma}$} &  \textbf{Underexp.} &  \textbf{$\mathbf{\Gamma\leq}$1.5} &  \textbf{{$\mathbf{\Gamma\geq}$3.0}} & \textbf{N$_\mathrm{\bf{end}}$}\\
    \hline
    Best+12 &        110 &           0 &                  36 &               19 &                2 &       58 &           59 &             0 &                37 &                  3 &                  3 &         18 \\
    Chandola+20 &          1 &           0 &                   0 &                0 &                0 &        1 &            3 &             0 &                 2 &                  1 &                  0 &          0 \\
    Ching+17 &         38 &           0 &                  13 &               10 &                1 &       16 &           18 &             1 &                12 &                  1 &                  1 &          4 \\
    FRXCAT &         26 &           0 &                   9 &                4 &                2 &       14 &            7 &             0 &                 2 &                  1 &                  0 &          4 \\
    Gendre+10 &         40 &           0 &                   9 &               13 &                0 &       20 &           35 &             5 &                 4 &                 11 &                  0 &         19 \\
    GRG\_catalog &         23 &           0 &                   5 &               12 &                0 &        8 &           10 &             0 &                 2 &                  3 &                  0 &          5 \\
    Kosmaczewski+20 &          7 &           0 &                   0 &                6 &                0 &        1 &            5 &             0 &                 0 &                  2 &                  0 &          3 \\
    Liao+20\_I &          9 &           0 &                   2 &                4 &                0 &        1 &            4 &             0 &                 0 &                  1 &                  0 &          3 \\
    Liao+20\_II &         16 &           0 &                   0 &                4 &                0 &        9 &            7 &             0 &                 0 &                  3 &                  0 &          4 \\
    Macconi+20 &         26 &           0 &                   0 &                3 &                0 &       23 &            2 &             0 &                 0 &                  0 &                  0 &          2 \\
Mingo+19 &         15 &           0 &                   4 &                3 &                0 &        8 &           11 &             0 &                 6 &                  0 &                  0 &          5 \\
    Miraghaei+17\_FR &         16 &           0 &                   5 &                5 &                0 &        7 &            9 &             0 &                 4 &                  2 &                  0 &          3 \\
Miraghaei+17\_FR\_HL &          7 &           0 &                   3 &                2 &                0 &        3 &            6 &             0 &                 4 &                  1 &                  0 &          1 \\
    ROGUEI &         34 &           0 &                   8 &                5 &                1 &       21 &           17 &             0 &                 8 &                  1 &                  0 &          8 \\
    Sobolewska+19a &         11 &           0 &                   2 &                7 &                0 &        2 &            5 &             0 &                 0 &                  3 &                  0 &          2 \\
    \hline
\end{tabular}
\tablefoot{\bi{Column 1:} Catalog name as defined in Sect. \ref{sect:cats}. \bi{Column 2:} Number of sources that cross-matched with \xmm\ (either X-ray and UV or just UV). \bi{Column 3:} Number of sources that were rejected because the UV detection was < 3$\sigma$. \bi{Column 4:} Number of sources that were rejected due to under exposure because either (a)  the X-ray exposure time is less than 10 ks or (b) the uncertainty in a UV or X-ray flux measurement exceeds 100\%. \bi{Column 5:} Number of sources that had $\Gamma\leq$1.5 and are identified as too obscured to be modeled by a power law. \bi{Column 6:} Number of sources that had $\Gamma\geq$3.0 and are identified as too physically far from what is seen in AGNs. \bi{Column 7:} Number of sources remaining after subtracting the sources that were rejected by the data quality cuts in Cols. 3-6. \bi{Columns 8-13:} Same as Cols. 2-7 but for \swift\ matches instead of \xmm, except that \bi{Col. 10} is only sources that were rejected because the uncertainty in a UV or X-ray flux measurement  exceeds 100\% without an exposure requirement as with \xmm.}
\end{table*}
\egroup

In Sect. \ref{sect:quality_cuts} we apply several data quality cuts to the radio catalogs. In this appendix we provide the number of sources that were rejected with each cut (see Table \ref{tab:rejects}).

\FloatBarrier
 
\section{Extent plot variations}
\begin{figure*}
    \begin{subfigure}{.5\textwidth}
      \centering
      \includegraphics[width=0.99\linewidth]{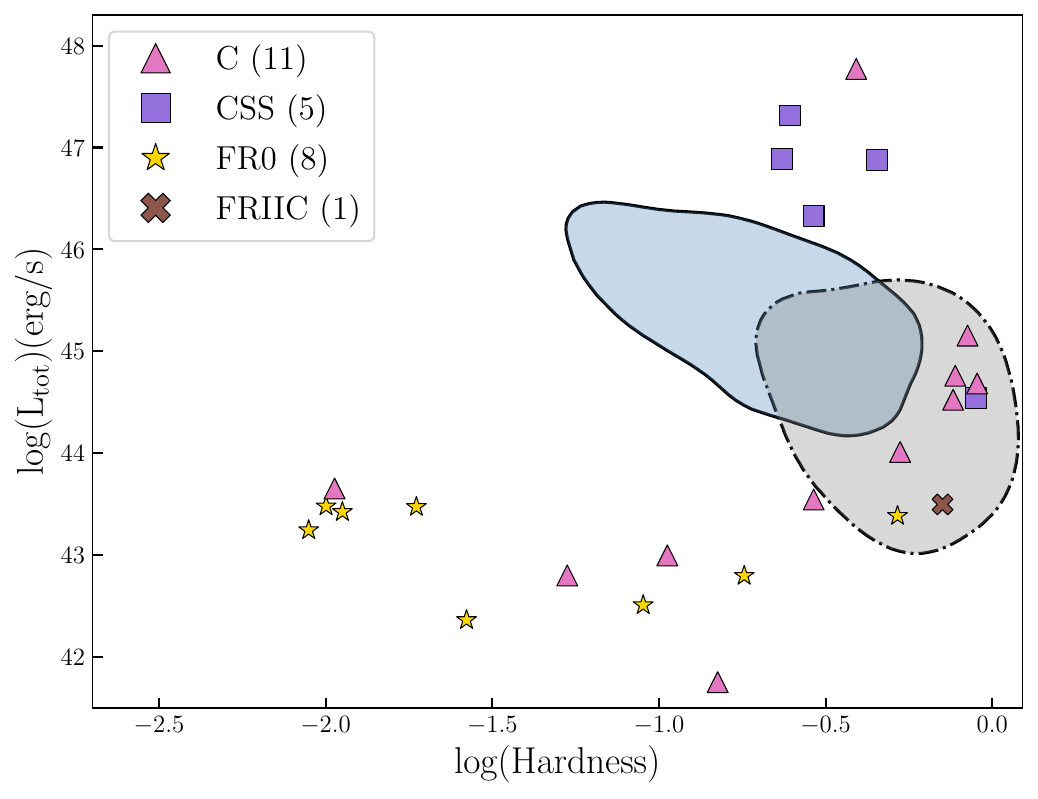}
      \caption{}
    \end{subfigure}
    \begin{subfigure}{.5\textwidth}
      \centering
      \includegraphics[width=0.99\linewidth]{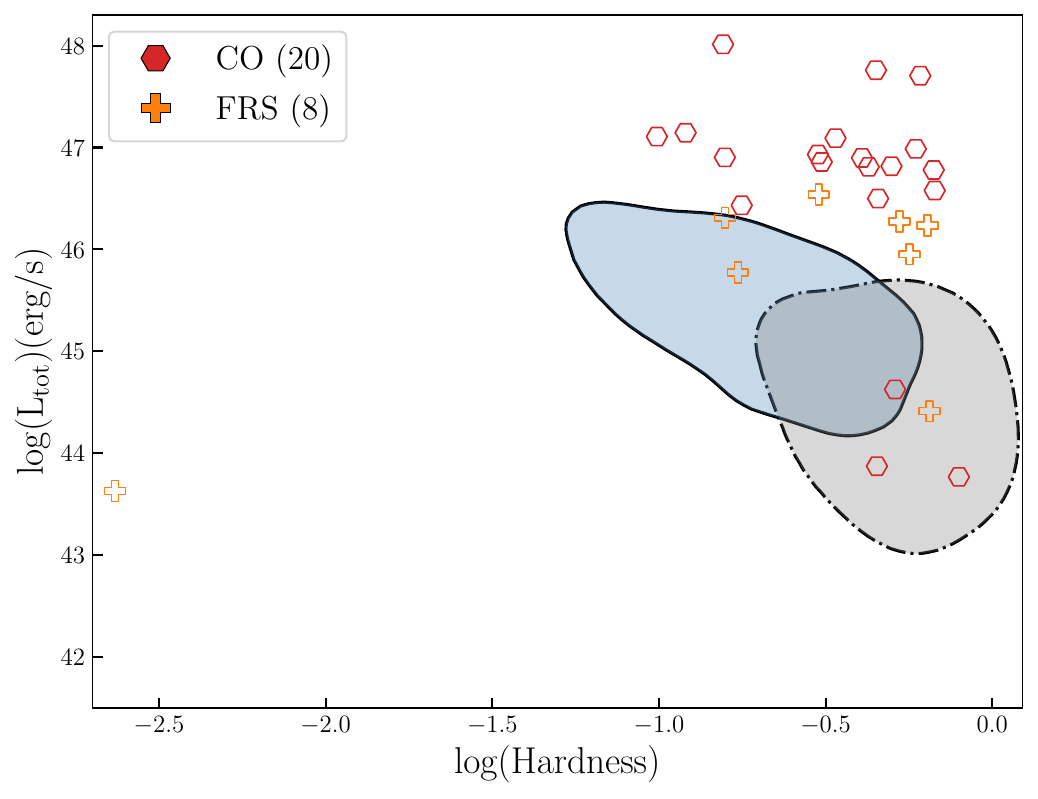}
      \caption{}
    \end{subfigure}
    \begin{subfigure}{.5\textwidth}
      \centering
      \includegraphics[width=0.99\linewidth]{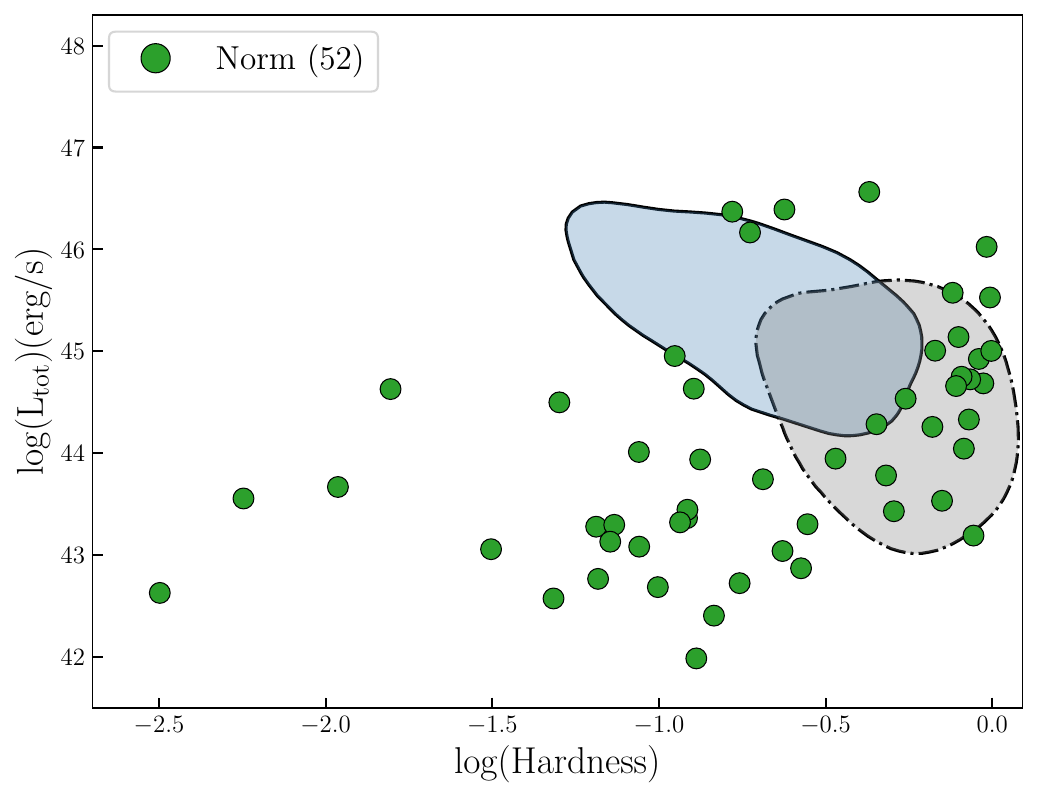}
      \caption{}
    \end{subfigure}
    \begin{subfigure}{.5\textwidth}
      \centering
      \includegraphics[width=0.99\linewidth]{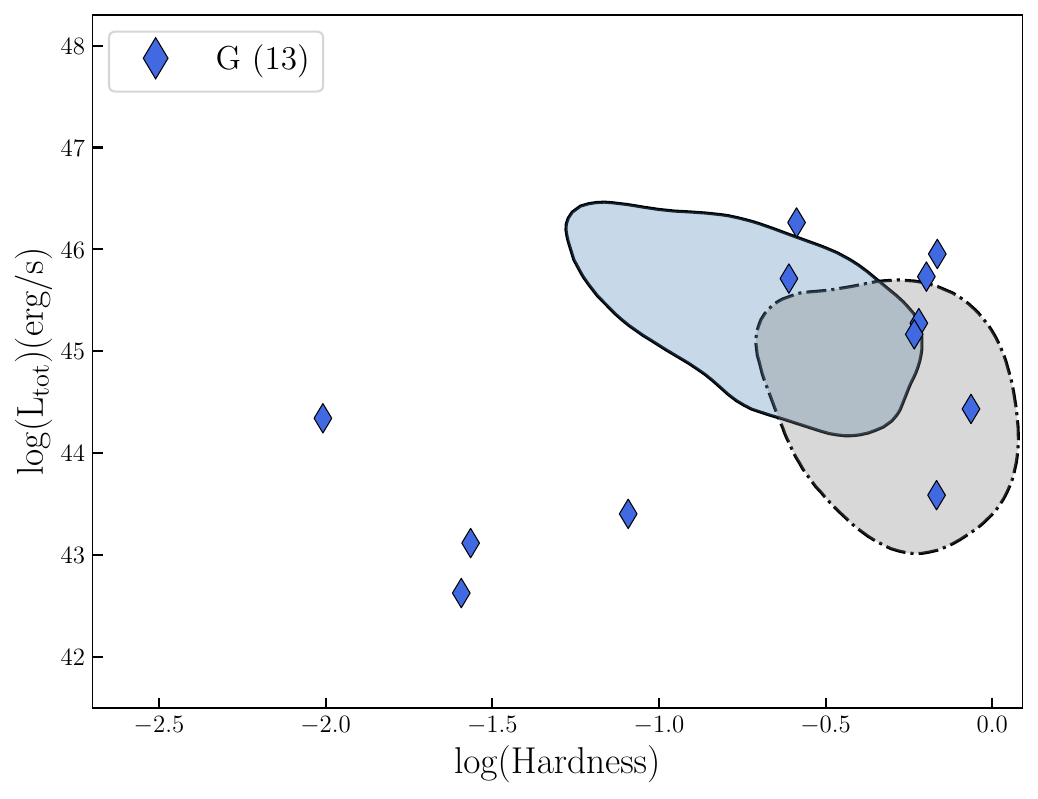}
      \caption{}
    \end{subfigure}
   \caption{Variations in the extent HID (Fig. \ref{fig:hid_ext}) with certain subsets of populations. \textbf{(a):} Sources with a certain compact extent classification: C, CSS, and FR0. \textbf{(b):} Sources that are compact but could possibly be blazars. \textbf{(c):} Sources that have a ``normal'' extent classification -- not compact and not giant. \textbf{(d):} Sources that are GRGs.}
  \label{fig:extent_populations}
\end{figure*}

We provide a suite of extent plots that show the compact, possible blazars, ``normal,'' and giant sources individually to help the reader see the different populations more clearly (see Fig. \ref{fig:extent_populations}).

\end{appendix}

\end{document}